\documentclass[useAMS,usenatbib]{mn2e}

\usepackage{url,times,graphicx,amsmath,amsfonts,amssymb,aas_macros,color}

\usepackage[dvipsnames,svgnames,hyperref]{xcolor}
\usepackage[
    pdftex,
    a4paper=true,
    breaklinks=true,
    bookmarks=true,
    bookmarksopen=false,
    bookmarksopenlevel=2
    bookmarksnumbered=true,
    bookmarkstype=toc,
    colorlinks=true,
    citecolor=NavyBlue,
    linkcolor=black,
    menucolor=green,
    urlcolor=blue,
]{hyperref}
\usepackage{comment}

\newcommand{\galacticus}{\textsc{Galacticus}}
\newcommand{\DLB}{\textsc{DLB07}}
\newcommand{\lgalaxy}{\textsc{Lgalaxies}}
\newcommand{\mice}{\textsc{Mice}}
\newcommand{\morgana}{\textsc{Morgana}}
\newcommand{\pinocchio}{\textsc{Pinocchio}}
\newcommand{\ysam}{\textsc{ySAM}}
\newcommand{\galform}{\textsc{Galform}}
\newcommand{\galformVGP}{\textsc{Galform-gp14}}
\newcommand{\galformBOW}{\textsc{Galform-kb06}}
\newcommand{\galformFONT}{\textsc{Galform-kf08}}

\newcommand{\sag}{\textsc{Sag}}
\newcommand{\sage}{\textsc{Sage}}
\newcommand{\somerville}{\textsc{SantaCruz}}
\newcommand{\galics}{\textsc{GalICS-2.0}}
\newcommand{\skibba}{\textsc{Skibba}}

\newcommand{\ahf}{\textsc{AHF}}

\newcommand{\subfind}{\textsc{SUBFIND}}

\newcommand{\Mbnd}{{\ifmmode{M_{\rm bnd}}\else{$M_{\rm bnd}$}\fi}}
\newcommand{\Mfof}{{\ifmmode{M_{\rm fof}}\else{$M_{\rm fof}$}\fi}}
\newcommand{\Mcrit}{{\ifmmode{M_{\rm 200c}}\else{$M_{\rm 200c}$}\fi}}
\newcommand{\Mmean}{{\ifmmode{M_{\rm 200m}}\else{$M_{\rm 200m}$}\fi}}
\newcommand{\MBN}{{\ifmmode{M_{\rm BN98}}\else{$M_{\rm BN98}$}\fi}}

\newcommand{\Rcrit}{{\ifmmode{R_{\rm 200c}}\else{$R_{\rm 200c}$}\fi}}

\newcommand{\hMpc}{{\ifmmode{h^{-1}{\rm Mpc}}\else{$h^{-1}$Mpc}\fi}}
\newcommand{\hkpc}{{\ifmmode{h^{-1}{\rm kpc}}\else{$h^{-1}$kpc}\fi}}
\newcommand{\hMsun}{{\ifmmode{h^{-1}{\rm {M_{\odot}}}}\else{$h^{-1}{\rm{M_{\odot}}}$}\fi}}
\newcommand{\Mstar}{{\ifmmode{M_{*}}\else{$M_{*}$}\fi}}
\newcommand{\Mhalo}{{\ifmmode{M_{\rm halo}}\else{$M_{\rm halo}$}\fi}}
\newcommand{\Ngal}{{\ifmmode{N_{\rm gal}}\else{$N_{\rm gal}$}\fi}}
\newcommand{\Norph}{{\ifmmode{N_{\rm orphan}}\else{$N_{\rm orphan}$}\fi}}
\newcommand{\Nxorph}{{\ifmmode{N_{\rm non-orphan}}\else{$N_{\rm non-orphan}$}\fi}}
\newcommand{\Zsolar}{{\ifmmode{Z_{\odot}}\else{$Z_{\odot}$}\fi}}
\newcommand{\Msun}{{\ifmmode{{\rm {M_{\odot}}}}\else{${\rm{M_{\odot}}}$}\fi}}
\newcommand{\ltsima}{$\; \buildrel < \over \sim \;$}
\newcommand{\gtsima}{$\; \buildrel > \over \sim \;$}
\newcommand{\lsim}{\lower.5ex\hbox{\ltsima}}
\newcommand{\gsim}{\lower.5ex\hbox{\gtsima}}

\def\sSFR{sSFR}

\def\lesssim{\mathrel{\hbox{\rlap{\hbox{\lower4pt\hbox{$\sim$}}}\hbox{$<$}}}}
\def\gtrsim{\mathrel{\hbox{\rlap{\hbox{\lower4pt\hbox{$\sim$}}}\hbox{$>$}}}}

\def\Nmodels{14}
\def\NSAMs{12}
\def\NHODs{2}

\newcommand{\Tab}[1]{Table~\ref{#1}}
\newcommand{\Sec}[1]{Section~\ref{#1}}
\newcommand{\App}[1]{Appendix~\ref{#1}}
\newcommand{\Eq}[1]{Eq.~(\ref{#1})}
\newcommand{\Fig}[1]{Fig.~\ref{#1}}
\newcommand{\beq}{\begin{equation}}
\newcommand{\eeq}{\end{equation}}
\def\beqa{\begin{eqnarray}}
\def\eeqa{\end{eqnarray}}
\def\hMpc{$h^{-1}\,{\rm Mpc}$}
\def\hkpc{$h^{-1}\,{\rm kpc}$}

\title[nIFTy Galaxies]
{nIFTy Cosmology: Comparison of Galaxy Formation Models}
\author[Knebe et. al]
       {Alexander Knebe,$^{1}$\thanks{E-mail: alexander.knebe@uam.es}
       Frazer R. Pearce,$^{2}$				
       Peter A. Thomas,$^{3}$ 					
       Andrew Benson,$^{4}$   \newauthor 				
       Jeremy Blaizot,$^{5,6,7}$ 			
       Richard Bower,$^{8}$			        
       Jorge Carretero,$^{9,10}$ 				
       Francisco J. Castander,$^{9}$	 \newauthor		
       Andrea Cattaneo,$^{11}$			
       Sofia A. Cora,$^{12,13}$				
       Darren J. Croton,$^{14}$			        
       Weiguang Cui,$^{15}$	 \newauthor			
       Daniel Cunnama,$^{16}$  			
       Gabriella De Lucia,$^{17}$			
       Julien E. Devriendt,$^{18}$			
       Pascal J. Elahi,$^{19}$\newauthor				
       Andreea Font,$^{20}$	  		        
       Fabio Fontanot,$^{17}$			
       Juan Garcia-Bellido,$^{1,21}$			
       Ignacio D. Gargiulo,$^{12,13}$	 \newauthor		
       Violeta Gonzalez-Perez,$^{8}$			
       John Helly,$^{8}$			
       Bruno Henriques,$^{22}$		        
       Michaela Hirschmann,$^{23}$ \newauthor			
       Jaehyun Lee,$^{24}$				
       Gary A. Mamon,$^{23}$			
       Pierluigi Monaco,$^{25,17}$			
       Julian Onions,$^{2}$	\newauthor		
       Nelson D. Padilla,$^{28,29}$ 				
       Chris Power,$^{15}$				
       Arnau Pujol,$^{9}$			
       Ramin A. Skibba,$^{26}$	 \newauthor			
       Rachel S. Somerville,$^{27}$	
       Chaichalit Srisawat,$^{3}$ 	                
       Cristian A. Vega-Mart\'inez,$^{12}$  \newauthor	
       Sukyoung K. Yi$^{24}$			        
\\
\\
$^{1}$Departamento de F\'isica Te\'{o}rica, M\'{o}dulo 15, Facultad de Ciencias, Universidad Aut\'{o}noma de Madrid, 28049 Madrid, Spain\\
$^{2}$School of Physics \& Astronomy, University of Nottingham, Nottingham NG7 2RD, UK\\
$^{3}$Department of Physics \& Astronomy, University of Sussex, Brighton, BN1 9QH, UK\\
$^{4}$Carnegie Observatories, 813 Santa Barbara Street, Pasadena, CA 91101, USA\\
$^{5}$Universit\`e de Lyon, Lyon, F-69003, France\\
$^{6}$Universit\`e Lyon 1, Observatoire de Lyon, 9 avenue Charles Andr\`e, Saint-Genis Laval, F-69230, France\\
$^{7}$CNRS, UMR 5574, Centre de Recherche Astrophysique de Lyon ; Ecole Normale Sup\`erieure de Lyon, Lyon, F-69007, France\\
$^{8}$Institute for Computational Cosmology, Department of Physics, University of Durham, South Road, Durham, DH1 3LE, UK\\
$^{9}$Institut de Ci\`encies de l'Espai, IEEC-CSIC, Campus UAB, 08193 Bellaterra, Barcelona, Spain\\
$^{10}$Port d{'}Informaci\'{o} Cient\'{i}fica (PIC) Edifici D, Universitat Aut\`{o}noma de Barcelona (UAB), E-08193 Bellaterra (Barcelona), Spain.\\
$^{11}$GEPI, Observatoire de Paris, CNRS, 61, Avenue de l'Observatoire 75014, Paris  France\\
$^{12}$Instituto de Astrof\'isica de La Plata (CCT La Plata, CONICET, UNLP), Paseo del Bosque s/n, B1900FWA, La Plata, Argentina.\\
$^{13}$Facultad de Ciencias Astron\'omicas y Geof\'{\i}sicas, Universidad Nacional de La Plata, Paseo del Bosque s/n, B1900FWA, La Plata, Argentina\\
$^{14}$Centre for Astrophysics and Supercomputing, Swinburne University of Technology, Hawthorn, Victoria 3122, Australia\\
$^{15}$International Centre for Radio Astronomy Research, University of Western Australia, 35 Stirling Highway, Crawley, Western Australia 6009, Australia\\
$^{16}$Physics Department, University of Western Cape, Bellville, Cape Town 7535, South Africa\\
$^{17}$INAF - Astronomical Observatory of Trieste, via Tiepolo 11, I-34143 Trieste, Italy\\
$^{18}$Astrophysics, University of Oxford, Denys Wilkinson Building, Keble Road, Oxford, OX1\,3RH, UK\\
$^{19}$Sydney Institute for Astronomy, A28, School of Physics, The University of Sydney, NSW 2006, Australia\\
$^{20}$Astrophysics Research Institute, Liverpool John Moores University,  IC2, Liverpool Science Park, 146 Brownlow Hill,  Liverpool L3 5RF, UK\\
$^{21}$Instituto de F\'isica Te\'{o}rica, Universidad Aut\'{o}noma de Madrid (IFT-UAM/CSIC), 28049 Madrid, Spain\\
$^{22}$Max-Planck-Institut f\"ur Astrophysik, Karl-Schwarzschild-Str. 1, 85741 Garching b. M\"unchen, Germany\\
$^{23}$Institut d'Astrophysique de Paris (UMR 7095: CNRS \& UPMC), 98 bis Bd Arago, F-75014 Paris, France\\
$^{24}$Department of Astronomy and Yonsei University Observatory, Yonsei University, Seoul 120-749, Republic of Korea\\
$^{25}$Dipartimento di Fisica, Universit\`a di Trieste, via Tiepolo 11, 34143 Trieste, Italy\\
$^{26}$Department of Physics, Center for Astrophysics and Space Sciences, University of California, 9500 Gilman Drive, San Diego, CA 92093\\
$^{27}$Department of Physics and Astronomy, Rutgers University, 136 Frelinghuysen Road, Piscataway, NJ 08854, USA\\
$^{28}$Instituto de Astrofisica, Universidad Catolica de Chile, Santiago, Chile\\
$^{29}$Centro de Astro-Ingenieria, Universidad Catolica de Chile, Santiago, Chile\\
 }

\begin{document}

\date{Accepted XXXX . Received XXXX; in original form XXXX}

\pagerange{\pageref{firstpage}--\pageref{lastpage}} \pubyear{2010}

\maketitle

\label{firstpage}

\clearpage

\begin{abstract}

We present a comparison of \Nmodels\ galaxy formation models: \NSAMs\ different
semi-analytical models and \NHODs\ halo-occupation distribution models for
galaxy formation based upon the same cosmological simulation and merger tree information derived from it.

The participating codes have proven to be very successful in their own
right but they have all been calibrated independently using various
observational data sets, stellar models, and merger trees. In this paper we apply
them without recalibration and this leads to a
wide variety of predictions for the stellar mass function, specific
star formation rates, stellar-to-halo mass ratios, and the abundance
of orphan galaxies.  The scatter is much larger than seen
in previous comparison studies primarily because the codes have been
used outside of their native environment within which they are well tested and calibrated.

The purpose of the `nIFTy comparison of galaxy formation models' is to
bring together as many different galaxy formation modellers as
possible and to investigate a common approach to model calibration.  This
paper provides a unified description for all participating models and
presents the initial, uncalibrated comparison as a baseline for our
future studies where we will develop a common
calibration framework and address the extent to which that reduces the
scatter in the model predictions seen here.

\end{abstract}
\noindent

\section{Introduction} \label{sec:introduction}

Understanding the formation and evolution of galaxies within a
self-consistent cosmological context is one of the outstanding and
most challenging topics of astrophysics and cosmology. Over the last
few decades great strides forward have been made along two distinct
lines: on the one hand, through directly accounting for the baryonic
component (gas, stars, supermassive black holes, etc.) in cosmological
simulations that include hydrodynamics and gravity and on the other
hand, through a procedure known as semi-analytic modelling (SAM), in
which a statistical estimate of the distribution of dark matter haloes
and their merger history -- either coming from cosmological
simulations or extended Press-Schechter/Lagrangian methods -- is
combined with simplified yet physically motivated prescriptions to
estimate the distribution of the physical properties of galaxies. To
date the vast computational challenge related to simulating the
baryonic component has made it impractical for a general adoption of
the former approach within the large volumes necessary for galaxy
surveys and hence a lot of effort has been devoted to the latter SAM
strategy. Further, most of the modelling of sub-grid physics in
hydrodynamical simulations relies on schemes akin to the ones used in
the SAM approach, and hence it is more effective to apply them in
post-processing to a simulation where parameter scans are then less
costly.  Further, over this period we have not only witnessed
significant advances in simulation techniques
\citep[e.g.][]{Vogelsberger14,Schaye15}, but the original ideas for
SAMs \citep[cf.][]{White78} have undergone substantial refinement too
\citep[e.g.][and in particular all the people and methods detailed
  below in
  \Sec{sec:models}]{Cole91,white91,Lacey91,Blanchard92,Kauffmann93}.

Some SAMs simply rely on analytical forms for the underlying merger
trees based upon (conditional) mass functions from \citet{Press74} or
Extended Press-Schechter \citep{Bond91} as, for instance, described in
\citet{Kolatt99b}.  Other codes take as input halo merger trees
derived from cosmological $N$-body simulations \citep[see][for the
  historical origin of both techniques]{Lacey93,Roukema97}. While the
former remain a critical and powerful approach, advances in computing
power (especially for dark matter only simulations) have shifted the
focus of SAM developers towards the utilization of $N$-body merger
trees as input to their models as they more reliably capture
non-linear structure formation. For a recent comparison of merger tree
construction methods, the influence of the underlying halo finder and
the impact for (a particular) SAM we refer the reader to results
coming out of a previous comparison project `Sussing Merger
Trees'\footnote{\url{http://popia.ft.uam.es/SussingMergerTrees}}
\citep{Srisawat13,Avila14,Lee14}: \citet{Lee14}, for instance, have
shown that SAM parameters can be re-tuned to overcome differences
between different merger trees; something of great relevance for the
work presented here, as we will see later.

As well as considering the influence of the halo finder and merger
tree construction on any galaxy formation model, it is also important
to consider the different semi-analytical techniques and methods
themselves. Where this strategy has been used it has thus far focused
primarily on comparing the physical details of the model.
For instance, \citet{Somerville99}
\citep[as well as][]{Lu11b} implemented various physical prescriptions
into a single code; by design this tested the underlying physical
assumptions and principles rather than for any code-to-code
(dis-)similarities. \citet{Fontanot09} and \citet{Kimm09} compared
different SAMs, but without using the same merger trees as an
input. Similarly, \citet{Contreras13} compared Durham and Munich SAMs
for the Millennium simulation \citep{Springel05b} measuring and
comparing the halo occupation distribution found within
them. Different channels for bulge formation in two distinct SAMs have
been compared by \citet{DeLucia11} and \citet{Fontanot11}.
\citet[][]{Fontanot12} compared the predictions of three different
SAMs \citep[the same ones as used already in][]{Fontanot09,Kimm09} for
the star formation rate function to observations.

The first move towards using identical inputs was undertaken by
\citet{Maccio10} where three SAMs were compared, this time using the
same merger tree for all of them. However, the emphasis was on
studying (four) Milky Way-sized dark matter haloes.
\citet{DiazGimenez10} have analyzed the properties of compact galaxy
groups in mocks obtained using three SAMs applied to the same merger
trees derived from the Millennium simulation.  They found that the
fraction of compact groups that were not dense quartets in real space
varied from 24\% to 41\% depending on the SAM.  In \citet{Snaith11}
four SAMs (two Durham and two Munich flavours) all based upon trees
extracted from the Millennium simulation have been compared with a
special focus on the luminosity function of galaxy groups --
highlighting differences amongst models, especially for the magnitude
gap distribution between first- and second-ranked group
galaxies. Stripped down versions of three models (again utilising
identical merger trees) have been investigated in great detail by
\citet{DeLucia10}: they studied primarily various assumptions for gas
cooling and galaxy mergers and found that different assumptions in the
modelling of galaxy mergers can result in significant differences in
the timings of mergers, with important consequences for the formation
and evolution of massive galaxies. Most recently \citet{Lu14} used
merger trees extracted from the Bolshoi simulation \citep{Klypin11} as
input to three SAMs. They conclude that in spite of the significantly
different parameterizations for star formation and feedback processes,
the three models yield qualitatively similar predictions for the
assembly histories of galaxy stellar mass and star formation over
cosmic time. Note that all three models in this study were tuned and
calibrated to the same observational data set, i.e. the stellar mass
function of local galaxies. Additionally, it should not go unmentioned
that a lot of effort has gone into comparing SAMs to the results of
cosmological hydrodynamic simulations, either using the same merger
trees for both
\citep{Yoshida02,Helly03,Cattaneo07,Saro10,Hirschmann12b,Monaco14} or
not \citep{Benson01,Lu11b} -- yielding galaxy populations with similar
statistical properties with discrepancies primarily arising for
cooling rates, gas consumption, and star formation efficiencies. For an elaborate
review of semi-analytical models in relation to hydrodynamic simulations please refer to
\citet{Somerville14}.

In addition to the maturation of SAMs over the past ten years, the
field of halo models of galaxy clustering has produced other powerful
techniques for associating dark matter haloes with galaxies: the halo
occupation distribution \citep[HOD;][]{Jing98, Cooray02}, as well as
the complementary models of the conditional luminosity function
\citep{Yang03} and subhalo abundance matching \citep{Conroy06}.  HOD
models have been developed to reproduce the observed real-space and
redshift-space clustering statistics of galaxies as well as their
luminosity, colour, and stellar mass distributions, and they have
clarified the similarities and differences between `central' and
`satellite' galaxies within dark matter haloes.  The major difference
between SAMs and HODs is that the former model physical processes with
merger trees whereas the latter are relating numerical data (for a
given redshift) to observations in a statistical manner, i.e. they
bypass an explicit modelling of the baryonic physics and rely on a
statistical description of the link between dark matter and
galaxies. While SAMs are therefore guaranteed to return a galaxy
population that evolves self-consistently across cosmic time, HOD models --
by construction -- provide an accurate reproduction of the galactic
content of haloes. In what follows, we refer to both of them as galaxy
formation models, unless we want to highlight their differences, and
in those cases we again refer to them as SAM or HOD model.

In this work -- emerging out of the `nIFTy cosmology'
workshop\footnote{\url{http://popia.ft.uam.es/nIFTyCosmology}} -- we
are continuing previous comparison efforts, but substantially
extending the set of galaxy formation models: \Nmodels\ models are
participating this time, \NSAMs\ SAMs and \NHODs\ HOD models. We are
further taking a slightly different approach: we fix the underlying
merger tree and halo catalogue but, for the initial
comparison presented here, we allow each and every model to use its
favourite parameter set, i.e.~we keep the same parameter set as in
their reference model based upon their favourite cosmological
simulation and merger tree realisation.\footnote{Note
  that not all models entering this comparison were designed ab initio
  to work with $N$-body trees.} By doing this we are deliberately not
directly testing for different implementations of the same physics,
but rather we are attempting to gauge the output scatter across models
when given the same cosmological simulation as input. Further, this
particular simulation and its trees are different from the
ones for which the models were originally developed and tested. By
following this strategy,
we aim at testing the variations across models outside their native environment
and {\em without} re-tuning.  In that regard, it also needs to be mentioned that
any galaxy formation model involves a certain level of degeneracy with respect
to its parameters \citep{Henriques09}.  This is further complicated by the fact
that different models are likely using different parameterizations for the
included physical processes.  And while it has been shown in some of the
previously undertaken comparisons that there is a certain level of consistency
\citep{Fontanot09,Lu14}, there nevertheless remains scatter. A study including model
re-calibration will form the next stage of this project and will be presented in
a future work.

The remainder of the paper is structured as follows: in \Sec{sec:data}
we briefly present the underlying simulation, the halo catalogue and
the merger tree; we further summarize five different halo mass
definitions typically used. In \Sec{sec:models} we give a brief
description of galaxy formation models in general; a detailed
description of each individual model is reserved for \App{app:models}
which also serves as a review of the galaxy formation models featured
here. Some general comments on the layout and strategy of the comparison are given in \Sec{sec:comparison}. The results for the stellar mass function -- the key property studied here -- are put forward in \Sec{sec:SMF} whereas all other properties are compared in \Sec{sec:galaxies}. We
present a discussion of the results in \Sec{sec:discussion} and close
with our conclusions in \Sec{sec:conclusions}.

\section{The Provided Data} \label{sec:data}
The halo catalogues used for this paper are extracted from 62
snapshots of a cosmological dark-matter-only simulation undertaken
using the \textsc{Gadget-3} $N$-body code \citep{Springel05} with
initial conditions drawn from the WMAP7 cosmology \citep[][$\Omega_{\rm m}=0.272$, $\Omega_\Lambda=0.728$, $\Omega_{\rm b}=0.0455$, $\sigma_8=0.807$, $h=0.7$, $n_s=0.96$]{Komatsu11}.
We use $270^3$ particles in a box of co-moving width $62.5$\hMpc, with
a dark-matter particle mass of $9.31\times10^8$\hMsun. Haloes were
identified with \subfind\ \citep{Springel01subfind} applying a
Friends-Of-Friends (FOF) linking length of $b=0.2$ for the host
haloes. Only (sub-)haloes with at least 20 particles were kept. The
merger trees were generated with the \textsc{MergerTree} code that
forms part of the publicly available \ahf\ package
\citep{Knollmann09,Gill04a}. For a study of the interplay between halo
finder and merger tree we refer the interested reader to
\cite{Avila14}.

As we run the \Nmodels\ galaxy formation models with their standard calibration, we have to consider five different mass definitions and provide them in the halo catalogues; three of them are based upon a spherical-overdensity assumption \citep{Press74}

\begin{equation} \label{eq:virialradius}
  M_{\rm ref}(<R_{\rm ref}) = \Delta_{\rm ref} \rho_{\rm ref} \frac{4\pi}{3} R_{\rm ref}^3\ ,
\end{equation}

\noindent
and two more on the FOF algorithm \citep{Davis85}. The masses supplied are summarized as follows:\\

\begin{tabular}{lll}
$\bullet$ \Mfof: 		& \multicolumn{2}{l}{the mass of all particles inside the FOF group}\\
$\bullet$ \Mbnd: 	& \multicolumn{2}{l}{the bound mass of the FOF group}\\
$\bullet$ \Mcrit: 	& $\Delta_{\rm ref}=200$,				& $\rho_{\rm ref}=\rho_{\rm c}$\\
$\bullet$ \Mmean: 	& $\Delta_{\rm ref}=200$,				& $\rho_{\rm ref}=\rho_{\rm b}$\\
$\bullet$ \MBN: 	& $\Delta_{\rm ref}=\Delta_{\rm BN98}$,	& $\rho_{\rm ref}=\rho_{\rm c}$\\
\end{tabular}\\

\noindent
where $\rho_{\rm c}$ and $\rho_{\rm b}$ are the critical and background density of the Universe, respectively, both of which are functions of redshift and cosmology. $\Delta_{\rm BN98}$ is the virial factor as given by Eq.(6) in \citet{Bryan98}, and $R_{\rm ref}$ is the corresponding halo radius for which the interior mean density matches the desired value on the right-hand side of \Eq{eq:virialradius}.

Note that \subfind\ returns exclusive masses, i.e. the mass of
host haloes does not include the mass of its subhaloes. Further, the
aforementioned five mass definitions only apply to host haloes; for subhaloes
\subfind\ always returns the mass of particles bound to it.

\begin{table*}
  \caption{Participating galaxy formation models. Alongside the name (first column) and the major reference (last column) we list the applied halo mass definition (second column) and highlight whether the calibration of the model included the stellar mass function (SMF) and/or a luminosity function (LF). Note that the respective model calibration is not necessarily limited to either of these two quantities. We additionally list the assumed initial mass function (IMF). For more details please refer to the individual model descriptions in the \App{app:models}.}
\label{tab:models}
\begin{center}
\begin{tabular}{llllll}
\hline
code name	& mass 		 	& \multicolumn{2}{c}{calibration} 	& IMF &reference\\
\hline
 \galacticus	& \MBN			& SMF	& LF				 	& Chabrier	& \cite{Benson12}		\\
 \galics		& \Mbnd			& SMF	& ---					& Kennicutt	& Cattaneo et al. (in prep.)		\\
 \morgana 	& \Mfof			& SMF	& ---					& Chabrier	& \citet{Monaco07}		\\
 \sag			& \Mbnd			& ---		& LF					& Salpeter		& \citet{gargiulo_2014}	\\
 \somerville 	& \MBN			& SMF	& ---					& Chabrier	& \citet{Somerville08}	\\
 \ysam 		& \Mcrit			& SMF	& ---					& Chabrier	& \citet{lee13}			\\
\\
\underline{Durham flavours:}\\
 \galformVGP	& \Mbnd			& ---		& LF					& Kennicutt	& \citet{gp14}			\\
 \galformBOW	& \Mbnd			& ---		& LF					& Kennicutt	& \citet{Bower06}	 	\\
 \galformFONT	& \Mbnd			& ---		& LF					& Kennicutt	& \citet{Font08}			\\
\\
\underline{Munich flavours:}\\
 \DLB 		& \Mcrit 			& ---		& LF					& Chabrier	& \citet{delucia_sam_2007}	\\
 \lgalaxy		& \Mcrit			& SMF	& LF					& Chabrier	& \citet{Henriques13}\\
 \sage 		& \Mcrit			& SMF	& ---					& Chabrier	& \citet{Croton06}			\\
\\
\underline{HOD models:}\\
 \mice 		& \Mfof			& ---		& LF					& 'diet' Salpeter	& \citet{Carretero14}		\\
 \skibba		& \Mcrit			& --- 		& LF					& Chabrier	& \citet{Skibba09}		\\
\hline
\end{tabular}
\end{center}
\end{table*}

\section{The Galaxy Formation Models} \label{sec:models}
The galaxy formation models used in this paper follow one of two
approaches, which are either `semi-analytic model' (SAM) or `halo occupation
distribution' (HOD) based models. We give a brief outline of both below. A far
more detailed description of the semi-analytic galaxy formation method
can be found in \citet[][]{baugh_review_2006}, whereas \citet[][]{Cooray02} provides
a review of the formalism and applications of the halo-based description of non-linear gravitational clustering forming the basis of HOD models.

Both model techniques require a dark matter halo catalogue, derived
from an $N$-body simulation as described above or produced analytically,
and take the resultant halo properties as input. Furthermore,
SAMs require that the haloes be grouped into merger
trees of common ancestry across cosmic time. For SAMs, the merger
trees describing halo evolution directly affect the evolution of
galaxies that occupy them. HOD models in their standard incarnation
do not make use of the information of the temporal evolution of haloes; HOD models
rather provides a mapping between haloes and galaxies at a given redshift.
However, the utilization of merger trees is in principle possible and there are recent advances in that
direction \citep{Yang12,Behroozi13}.

The models and their respective reference are summarized in
\Tab{tab:models} where we also list their favourite choice for the
mass definition applied for all the results presented in the main body
of the paper. Some of the models also applied one (or more) of the
alternative mass definitions and results for the influence on the results are presented in
\App{app:halomassdefinition}. The table also includes information on
whether the models included stellar masses and/or a luminosity
function during the calibration process. Note that the calibration is
not necessarily limited to either or both of these galaxy properties.
And we also chose to provide the assumed initial mass function (IMF)
in that table which has influences on the stellar masses (and star formation
rates) presented later in the paper. However, for more details about the models please refer
to \App{app:models}.

\subsection{Semi-analytic galaxy formation models}
Semi-analytic models
encapsulate the main physical processes governing galaxy formation and evolution in
a set of coupled
parameterised differential equations. In these equations, the
parameters are not arbitrary but set the efficiency of the various
physical processes being modelled. Given that these processes are
often ill-understood in detail,
the parameters play a dual role of both
balancing efficiencies and accommodating the (sometimes significant)
uncertainties. Their exact values should never be taken too
literally. However, a sensible model is usually one whose parameters
are set at an order-of-magnitude level of reasonableness for the
physics being represented. All models are calibrated against a key set
of observables, however exactly which observables are used changes
from model to model.

In practice, the semi-analytic coupled equations are what describe how
baryons move between different reservoirs of mass. These baryons are treated analytically in the halo merger trees and followed through
time. The primary reservoirs of baryonic mass used in all models in
this paper include the hot halo gas, cold disk gas, stars,
supermassive black holes, and gas ejected from the halo. Additionally,
different models may more finely delineate these mass components of
the halo/galaxy system (e.g. breaking cold gas into HI and H$_2$) or even
add new ones (e.g. the intra-cluster stars).

The physics that a semi-analytic model will try to capture can
typically be broken into the processes below:

\begin{itemize}
	\item \textit{Pristine gas infall:} As a halo grows its bound mass increases. Most SAMs assume that new dark matter also brings with it the cosmic baryon fraction of new baryonic mass in the form of pristine gas. This gas may undergo heating as it falls onto the halo to form a hot halo, or it may sink to the centre along (cold) streams to feed the galaxy. At early times, infall may be substantially reduced by photoionization heating.

	\item \textit{Hot halo gas cooling:} A hot halo of gas around a galaxy will lose energy and the densest gas at the centre will cool and coalesce onto the central galaxy in less than a Hubble time. It is usually assumed that this cooling gas conserves angular momentum, which leads to the formation of a cold galactic disk of gas, within which stars can form.

	\item \textit{Star formation in the cold gas disk:} If the surface density of cold gas in the disk is high enough, molecular clouds will collapse and star forming regions will occur. The observed correlation between star formation rate density and cold gas density or density divided by disk dynamical time can be applied to estimate a star formation rate \citep{Kennicutt98}.

	\item \textit{Supernova feedback and the production of metals:} Very massive new stars have short lifetimes and will become supernova on very short timescales. The assumed initial mass function (IMF) of the stars will determine the rate of these events, which will return mass and energy back into the disk, or even blow gas entirely out into the halo and beyond \citep{Dekel86}. Supernovae are also the primary channel to produce and move metals in and around the galaxy/halo system.

	\item \textit{Disk instabilities:} Massive disks can buckle under their own weight. Simple analytic arguments to estimate the stability of a disk can be applied to modelled galaxies. Unstable disks will form bars that move mass (stars and/or gas) and angular momentum inward. This redistribution of mass facilitates a change in morphology where stars pile up in the centre, leading to the growth of a bulge or pseudo-bulge \citep{Efstathiou82,Mo98}.

	\item \textit{Halo mergers:} With time haloes will merge, and thus their occupant galaxies will as well. Modern cosmological simulations readily resolve structures within structures, the so-called subhalo (and hence satellite) population. However, not all models use this information, instead treating the dynamical evolution of substructures analytically. Subhaloes can either be tidally destroyed or eventually merge. It also happens that subhaloes will be lost below the mass resolution limit of the simulation, and different models treat such occurrences differently. It is common to keep tracking the now `orphan' satellites until they merge with the central galaxy.
	
	\item \textit{Galaxy mergers and morphological transformation:} Galaxy mergers are destructive if the mass ratio of the two galaxies is close to unity (major merger). In this case the disks are usually destroyed to form a spheroid, however, new accreted cooling gas can lead to the formation of a new disk of stars around it. Less significant mergers (minor mergers) will simply cause the larger galaxy to consume the smaller one. Note that this is a simplified picture and the detailed implementation in a SAM might be based upon more refined considerations \citep[e.g.][]{Hopkins09}

	\item \textit{Merger and instability induced starbursts:} Mergers and instabilities can also produce significant bursts of star formation on short timescales \citep{Mihos94a,Mihos94b,Mihos96}. These are often modelled separately from the more quiescent mode of star formation ongoing in the disk.

	\item \textit{Supermassive black holes:} Most modern SAMs model the formation of supermassive black holes at the centres of galaxies. Such black holes typically grow through galaxy merger-induced and/or disk instability-induced cold disk gas accretion and/or by the slow accretion of hot gas out of the halo.

	\item \textit{Active galactic nuclei:} Accretion of gas onto a supermassive black hole will trigger an active phase. For rapid growth this will produce a so-called `quasar mode' event, typically occurring from galaxy mergers. For more quiescent growth the `radio mode' may occur. Feedback resulting from the latter has been used by modellers to shut down the cooling of gas into massive galaxies, effectively ceasing their star formation to produce a `red and dead' population of ellipticals, as observed. Some models also include outflows driven by the `quasar mode'.

\end{itemize}

It is important to note that not all the processes above are
parameterised. Often simple analytic theory will provide a reasonable
approximation for the behaviour of the baryons in that reservoir and
their movement to a different reservoir. In other cases an observed
correlation between two properties already predicted by the model can
be used to predict a new property.  However, sometimes there is little
guidance as to how a physical process should be captured. In such cases a
power-law or other simple relationship is often applied. 
In many cases parameterization of SAM recipes is also based upon results from numerical experiments.
And as we learn more about galaxies and their evolution such prescriptions are updated
and refined, in the hope that the model overall becomes a better
representation of the real Universe.

Each model used in this paper is described in more detail in
\App{app:models}. There the reader can find references that provide a
complete list of the baryonic reservoirs used and a description of the
equations employed to move baryons between them.

\subsection{Halo occupation distribution models}
Halo models of galaxy clustering offer a powerful alternative to the
semi-analytic method to produce large mock galaxy catalogues. Therefore and for
comparison to the SAMs, respectively, we include them in our analysis.  As
described in the introduction, halo models may be of three types: halo
occupation distribution (HOD); conditional luminosity function (CLF); subhalo
abundance matching (SHAM); or some combination or extension of these. Although
these models are not equivalent, they tend to have similar inferences and
predictions for most galaxy properties and distributions as a function of halo
mass and halocentric distance.  The two halo models (i.e. \mice\ and \skibba) in
this paper are best described as HOD models - even though they both also
incorporate subhalo properties and hence feature a SHAM component.

The primary purpose of halo models is to statistically link the properties of
dark matter haloes (at a given redshift) to galaxy properties in a relatively
simple way.  In contrast with SAMs, physical relations and processes are
inferred from halo modeling rather than input in the models.  By utilizing
statistics of observed galaxies and simulated dark matter haloes, halo modelers
describe the abundances and spatial distributions of central and satellite
galaxies in haloes as a function of host halo mass, circular velocity,
concentration, or other properties, and thus provide a guide for and constraints
on the formation and evolution of these galaxies.

Given the halo mass function, halo bias, and halo density profile, HOD models
naturally begin with the halo occupation distribution, $P(N|M,c\backslash s,x)$,
where $M$ is the host halo mass, $c\backslash s$ refers to central or satellite
galaxy status, $x$ refers to some galaxy property such as stellar mass or star
formation rate.  As in SAMs, most models implicitly or explicitly assume that
the central object in a halo is special and is (mostly) more massive than its
satellites.  The mean central galaxy HOD, $\langle N_\mathrm{c}|M \rangle$,
follows an error function, which assumes a lognormal distribution for the
central galaxy luminosity (or stellar mass) function at fixed halo mass.  The
mean satellite galaxy HOD, $\langle N_\mathrm{s}|M \rangle$, approximately
follows a power-law as a function of $(M/M_1)^\alpha$, where the parameter
$M_1\propto M_\mathrm{min}$ and determines the critical mass above which haloes
typically host at least one satellite within the selection limits.  CLF models
have different parameterizations, but the CLFs (or conditional stellar mass
functions) may be integrated to obtain HODs.  All HOD parameters may evolve with
redshift, though in practice, the stellar mass-halo mass relation, for example,
evolves very little especially at the massive end.  However, the occupation number
of satellites as a function of mass evolves significantly.

HOD models are constructed to reproduce conditional distributions and clustering
of galaxies from their halo mass distribution, but modeling choices and choices
of which constraints to use result in different models having different
predictions for relations between galaxies and haloes.  In addition, one must
make decisions about how and whether to model halo exclusion (the fact that
haloes have a finite size and cannot overlap so much so that one halo's center
lies within the radius of another halo), scale-dependent bias, velocity bias,
stripping and disruption of satellites, buildup of the intracluster light, etc.
One can incorporate subhalo properties and distributions, as has been done here
(which makes the \mice\ and \skibba\ models HOD/SHAM hybrids), to model orphan
satellites as well.  For bimodal or more complex galaxy properties, such as
color, star formation rate, and morphology, there is no unique way to model
their distributions: hence models have different predictions for the quenching
and structural evolution of central and satellite galaxies.  Models that
incorporate merger trees may have different predictions for galaxy
growth by merging vis-a-vis in situ star formation; however, the two models
presented here are applied to each snapshot individually.

\subsection{Orphan galaxies}
`Orphan' galaxies \citep{Springel01subfind,Gao04b,Guo10,Frenk12} are galaxies
which, for a variety of reasons (such as a mass resolution limit and tidal
stripping), no longer retain the dark matter halo they formed within. Some of
the processes that create orphans are physical. For instance, when a galaxy and
its halo enter a larger host halo tidal forces from the latter will lead to
stripping of the galaxy's halo. But there are also numerical issues leading to
orphan galaxies: halo finders as applied to the simulation data have intrinsic
problems identifying subhaloes close to the centre of the host halo \citep[see,
  e.g.,][]{Knebe11,Muldrew11,Onions12} and therefore the galaxy's halo might
disappear from the merger tree \citep[see][]{Srisawat13,Avila14}. As, unless
they have merged, galaxies do not simply disappear from one simulation snapshot
to the next, the majority of the SAMs deal with this by keeping the galaxy alive
and in their catalogues. They evolve associated properties such as position and
velocity in different ways: some teams freeze these at the values they had when
the host halo was last present while others integrate their orbits analytically
or tag a background dark matter particle and follow that instead. Note that
information about the motion of dark matter particles was not supplied to the
modellers during the comparison and so this latter method was not available in
this study. For that reason we do not include any analyses which are affected by
the choice of how to assign positions and velocities to orphans.

\subsection{Calibration} 
For the initial comparison presented here we do not require a common
calibration, but rather each semi-analytical model is used `as is'. In
particular this means that each model has been run with its preferred physical
prescriptions and corresponding set of parameters, as detailed in the Appendix,
{\it without specific retuning}.  One of the purposes in doing this is to show
the importance of providing adequate calibraton when applying the models.  We
show in Section~\ref{sec:SMF} that the raw scatter in the stellar mass function
is very large but reduces considerably when models use their preferred merger
tree/halo mass definition and initial-mass-function.  A detailed comparison of
models when retuned to match the nIFTy merger trees and halo catalogues will be
the topic of a future paper.

\section{The Comparison} \label{sec:comparison}

\begin{figure}
 \begin{center}
   \includegraphics[width=0.8\columnwidth]{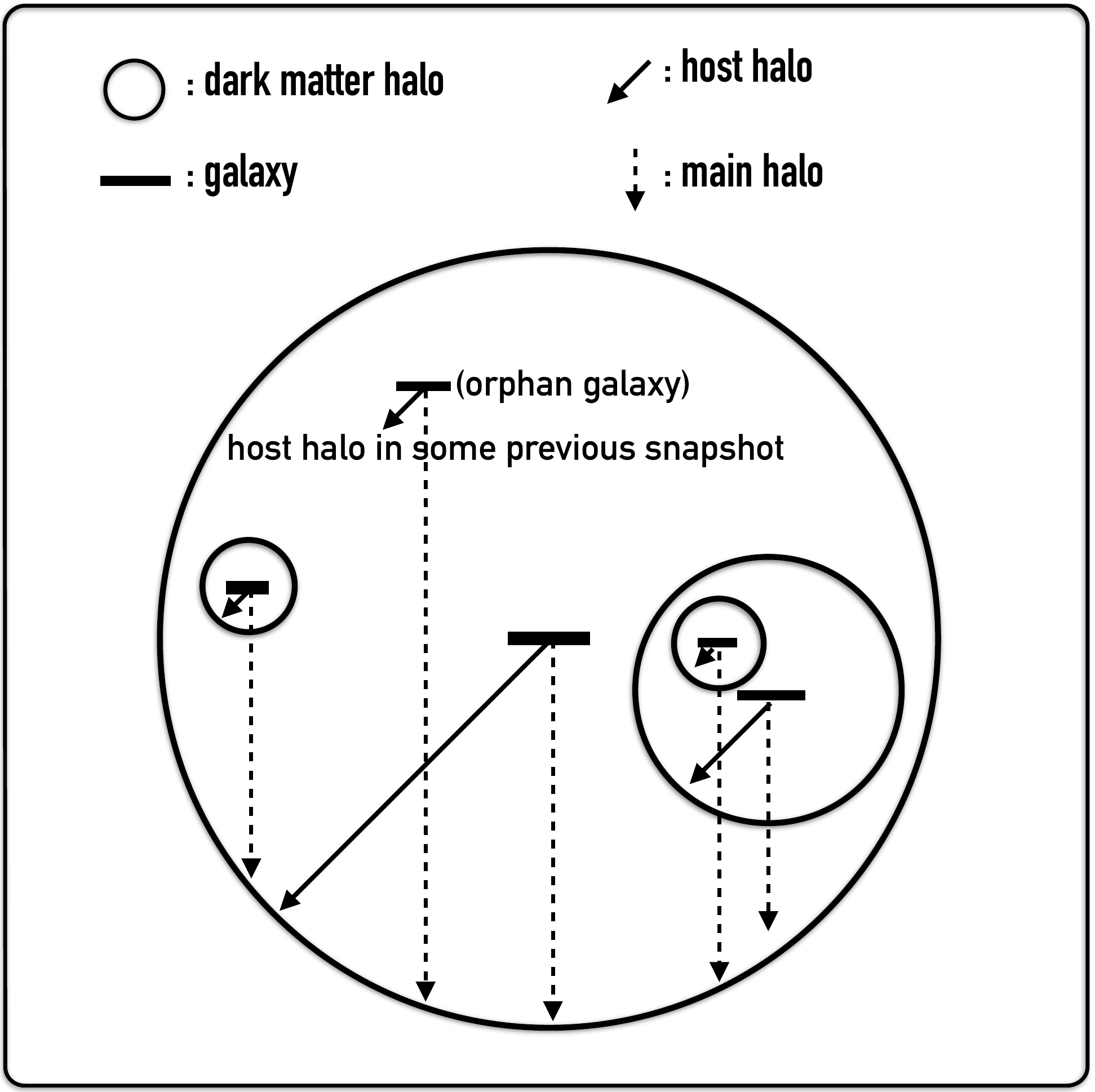}
   \caption{Sketch illustrating the population of galaxies residing inside a
     dark matter `main halo' (outermost circle). Each galaxy can have its own
     dark matter halo which we will refer to as `host halo' (or simply
     `halo'). Galaxies can also be devoid of such a `galaxy host halo' and then
     they are tagged as an `orphan' galaxy.}
 \label{fig:SAMgalaxies}
 \end{center}
 \end{figure}

First we present our methodology and establish the terminology used
throughout the remainder of the paper. We illustrate in
\Fig{fig:SAMgalaxies} how galaxies and the haloes they live within are
connected. A circle represents a dark matter halo, a horizontal line a
galaxy. A solid arrow points to the host halo of the galaxy whereas a
dashed arrow points to the main halo the galaxy orbits within. Note
that any substructure hierarchy has been flattened to one single
level, i.e. sub-subhaloes will point to the highest level main
halo. The sketch also includes orphan galaxies which will have a
pointer to its `last' dark matter host halo.

To facilitate comparison, in all subsequent plots that make use of
halo mass we (arbitrarily) chose $M_{\rm bnd}$ regardless of the mass
definition used internally by each model. Each galaxy in the supplied
galaxy catalogues points to its host halo (either the present
one, if it still exists, or its last one in the case of orphans) and
this link is used to select $M_{\rm bnd}$ from the original halo
catalogue -- irrespective of the mass definition used in the actual
model. This allows us to directly compare properties over a single set
of halo masses without concern that the underlying dark matter
framework has shifted slightly or is evolving differently with
redshift.

We have further limited all of the comparisons presented here to
galaxies with stellar masses $M_{*}>10^{9}$\hMsun\ -- a mass threshold
appropriate for simulations with a resolution comparable to the
Millennium simulation \citep[see][]{Guo11}. Additionally, when
connecting galaxies to their haloes we have applied a mass threshold
for the latter of $M_{\rm halo}>10^{11}$\hMsun, corresponding to
approximately 100 dark matter particles for the simulation used here.

When interpreting the plots we need to bear in mind -- as can be
verified in \App{app:models} -- that a great variety of model
calibrations exist: some models tune to stellar mass functions (SMF),
some models to luminosity functions (LF), and the HOD models to the
clustering properties of galaxies. To facilitate the differentiation
between calibrations in all subsequent plots we chose colours for the
models as follows: models calibrated using only SMFs are presented in
blue, models calibrated using LFs in red, and models using a
combination of both in green; the \NHODs\ HOD models are shown in
black, but with different linestyles.

Note that not all models necessarily appear in all plots. For
instance, the \skibba\ HOD model only provided a galaxy catalogue for
redshift $z=0$ and hence is not shown in evolutionary plots. Some
models do not feature orphans and so do not appear in the
corresponding plots.

For the comparison presented in this first paper we now focus on the stellar mass
function (SMF) to be presented in the following \Sec{sec:SMF}. There we will also investigate
several origins for model-to-model variations seen not only for the SMF, but
also for other galaxy properties to be presented in \Sec{sec:galaxies},
e.g. star formation rates, number density of galaxies, orphan fractions, and the
halo occupation. We deliberately exclude luminosity-based properties as they
introduce another layer of modelling, i.e. the employed stellar population
synthesis and dust model.

\section{Stellar Mass Function} \label{sec:SMF}

 \begin{figure}
   \includegraphics[width=\columnwidth]{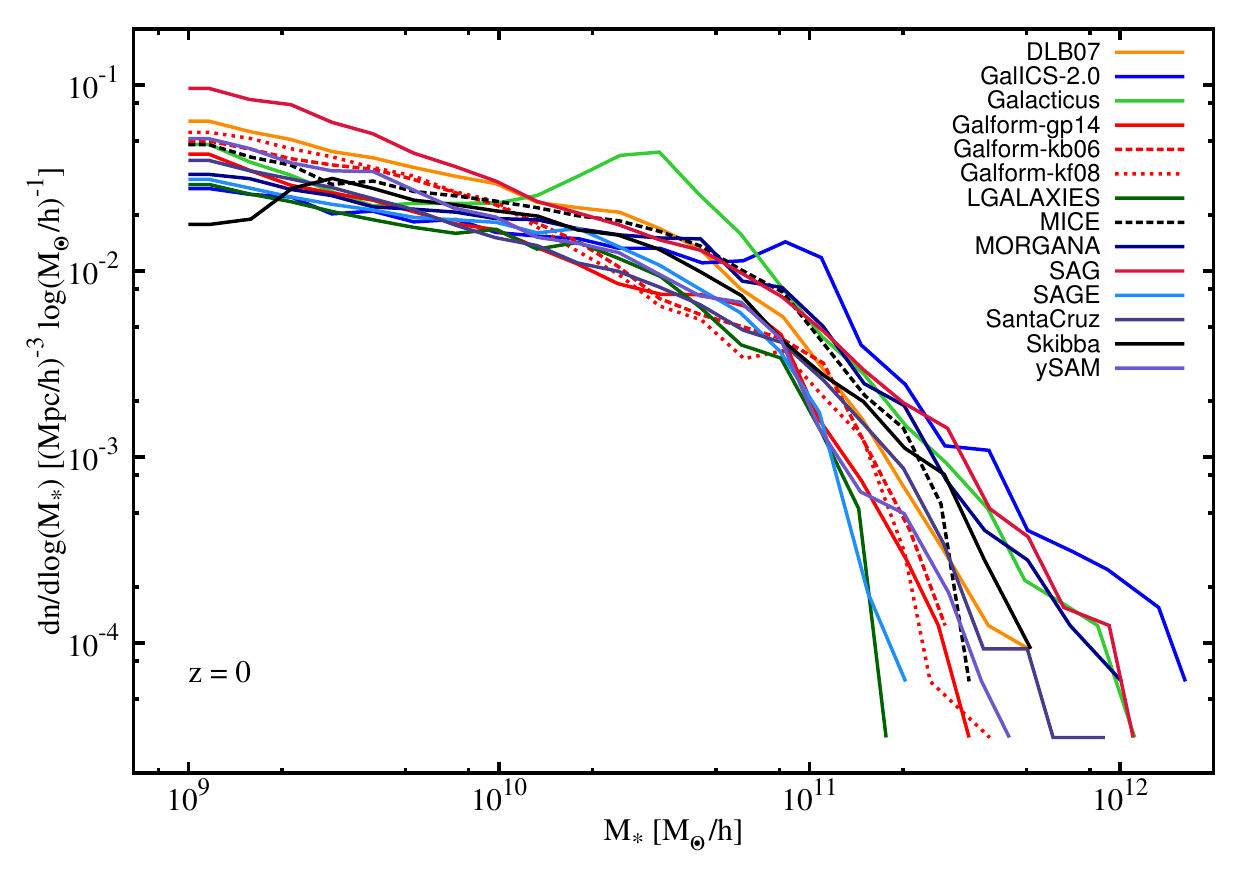}
   \includegraphics[width=\columnwidth]{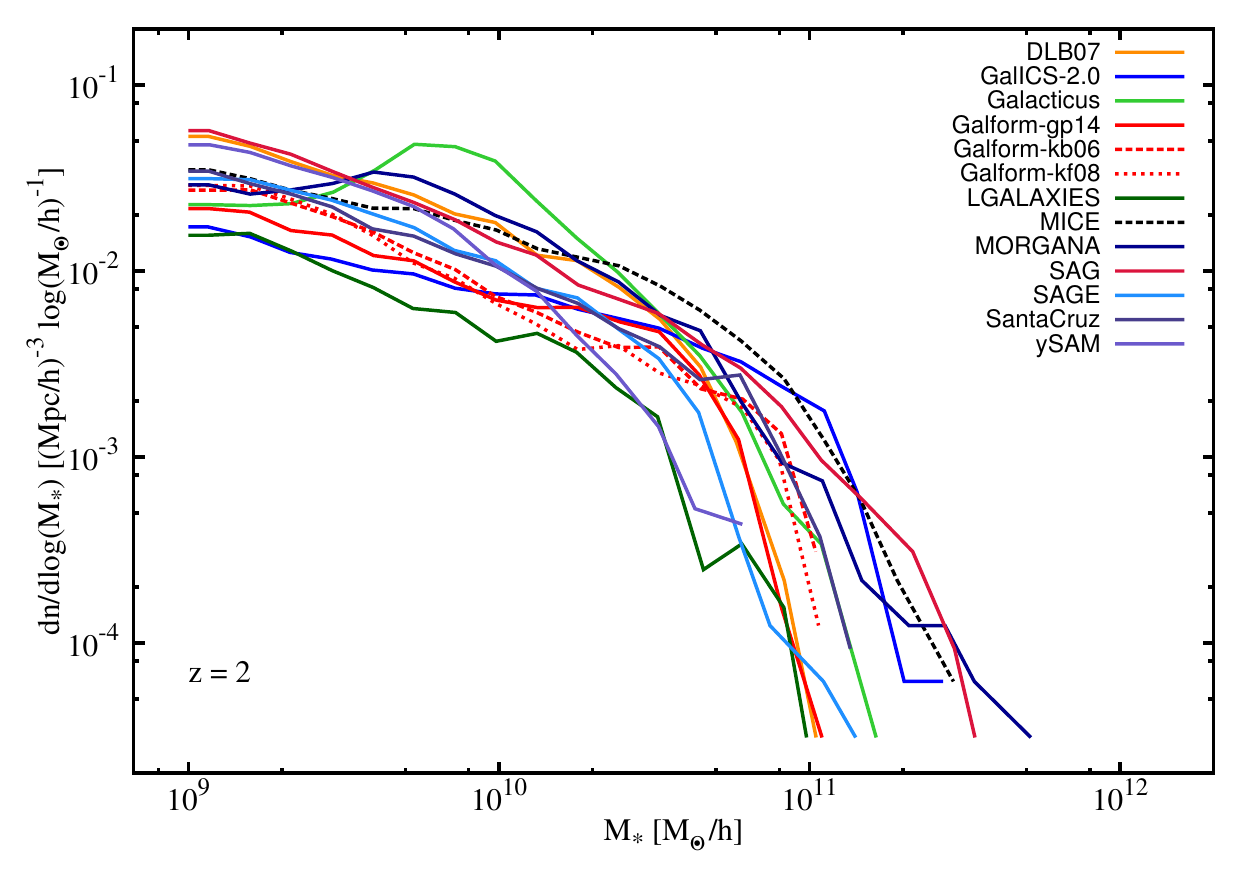}
   \caption{Stellar mass function at redshift $z=0$ (top) and $z=2$ (bottom). Each model used its preferred mass definition and initial stellar mass function.}
 \label{fig:NgalMstar}
 \end{figure}

 \begin{figure}
   \includegraphics[width=\columnwidth]{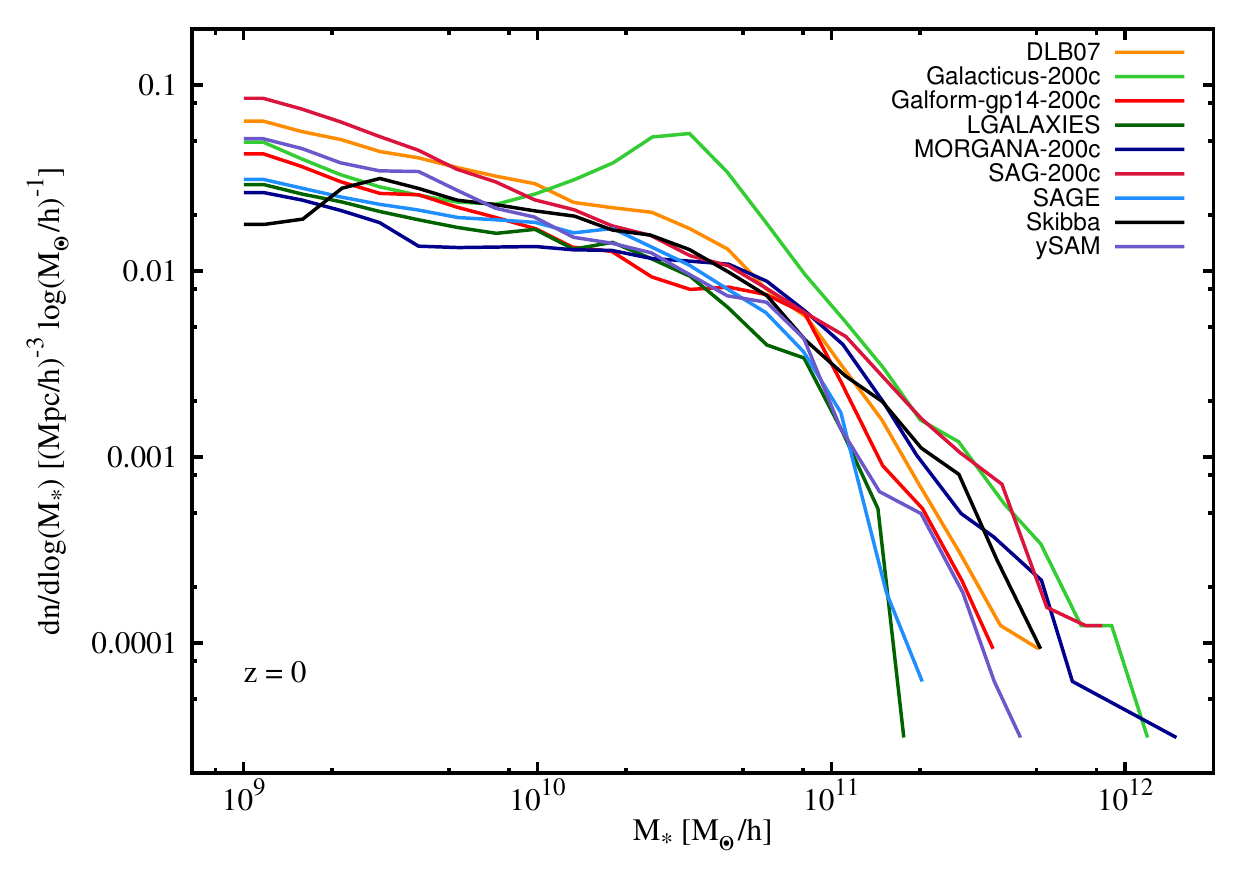}
   \caption{Stellar mass function at redshift $z=0$ for models that (also)
     returned galaxy catalogues using \Mcrit\ as the mass definition. To be
     compared against the upper panel of \Fig{fig:NgalMstar}}
 \label{fig:NgalMstarMassDefinition}
 \end{figure}

A key property, used by many of the models presented here to constrain
their parameters, is the stellar mass function of galaxies: we therefore
dedicate a full section to its presentation and discussion. It is
shown in \Fig{fig:NgalMstar} both at redshift $z=0$ (top panel) and $z=2$
(bottom panel). This plot indicates that there is quite a range in both galaxy
abundance and mass (influenced by star formation rate and star formation history) across the models. These
differences are apparent at $z=0$ where the models are calibrated, but 
are even more pronounced at redshift $z=2$. At $z=0$ the model results
vary in amplitude by around a factor of 3 in the main and exhibit high mass
cut-offs of varying steepness and position; at $z=2$ the differences in
amplitude are even larger, reflecting a broad variation in the location of the
peak in star formation rates (presented in the next section).

The HOD model \mice\ lies within the range of stellar mass functions provided by
the SAMs as does the \skibba\ model above $10^{9.5}\hMsun$. At $z=2$ the
\skibba\ model does not provide a return while the \mice\ model features amongst
the models with the largest number of high mass galaxies, i.e.~the \morgana\ and
\sag\ models.  While in \morgana\ the overproduction of massive galaxies is
connected to the inefficiency of the chosen AGN feedback implementation to
quench cooling in massive haloes, in \sag\ and \mice\ it could additionally be
related to the assumption of a Salpeter IMF which implies a higher mass estimate
than for a Chabrier IMF (see \Sec{sec:IMF} below).  The mass function for
\lgalaxy\ at $z=2$ is lower than any other model due to the delayed
reincorporation of gas ejected from supernova feedback that shifts star
formation in low mass galaxies to later times \citep{Henriques13}. At both
redshifts the \galacticus\ model displays a bump in the stellar mass function
around $10^{10}\hMsun$ due to the matching of feedback from active galactic
nuclei and supernovae.  For completeness we also checked that the scatter seen here basically remains
unchanged when restricting the analysis to (non-)central galaxies and
(non-)orphans, respectively.  

The differences seen here are huge, especially at the high-mass end, even when
models have implemented the same physical phenomena such as supernova and AGN
feedback.  For instance, \lgalaxy\ and \galacticus\ both allow the black hole to
accrete from the hot halo, with associated jets and bubbles producing `radio
mode' feedback: however, the mass of the largest galaxies differs by around an
order of magnitude at redshift $z=0$.  In order to understand how much of this
difference arises from the different physical implementations, we first need to
consider other factors that may influence the results.  For example, the models:
\begin{itemize}
 \item[a)] use a variety of halo mass definitions;
 \item[b)] use different initial-mass functions (IMFs);
 \item[c)] have been taken out of their native environment, i.e. they have been
   applied to a halo catalogue and tree structure that they were not developed or
   tested for;
 \item[d)] have not been re-calibrated to this new setup; and
 \item[e)] have not been tuned to the same observational data.
\end{itemize}
In the following sub-sections we will address points a-c) in more detail.
Points d) and e) are more complex and will be left for a future study.

\subsection{Mass Definition} \label{sec:MassDefinition}
It can be seen from \Tab{tab:models} that the models participating in this
comparison applied a variety of different mass definitions (which were
introduced in \Sec{sec:data}) to define the dark matter haloes that formed their
halo merger tree. But as several of the code representatives also returned
galaxy catalogues using mass definitions other than their default one, we are
able to prepare a plot that shows the stellar mass function for \Mcrit, i.e. the
mass definition for which the maximum number of models exist. We show that plot
as \Fig{fig:NgalMstarMassDefinition} where we see that the effect of changing
the mass definition is smaller than the model-to-model variation and hence not
the primary source of it.

 \begin{figure}
   \includegraphics[width=\columnwidth]{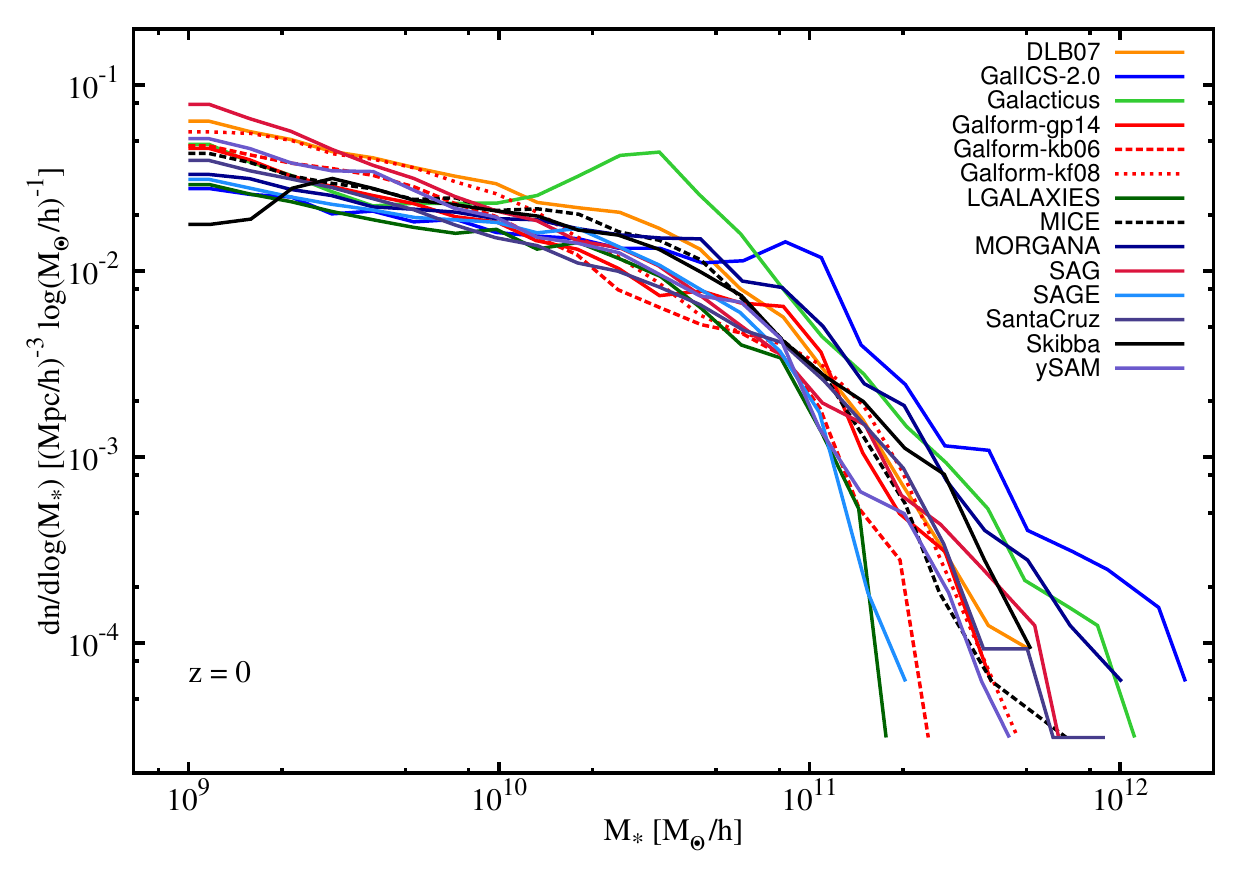}
   \caption{Stellar mass function at redshift $z=0$ after applying a correction
     for the applied IMF, i.e. models have been corrected towards a Chabrier
     IMF. To be compared against the upper panel of \Fig{fig:NgalMstar}}
 \label{fig:NgalMstarIMF}
 \end{figure}

 \begin{figure}
   \includegraphics[width=\columnwidth]{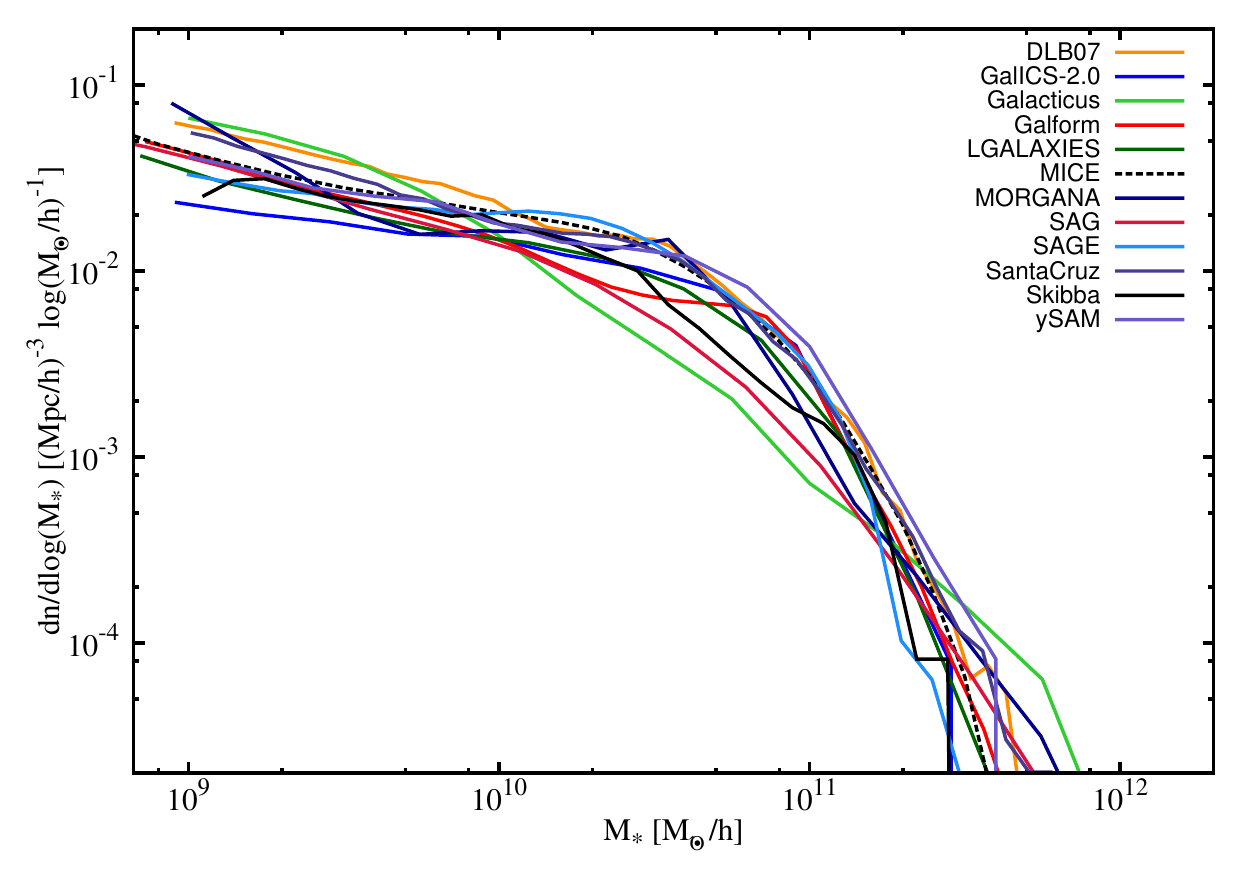}
   \caption{Stellar mass function at redshift $z=0$ for all the models when
     created in the native environment used during the model calibration. All
     curves have been corrected for the IMF towards Chabrier. This figure uses
     the same scale as (and should be compared to) \Fig{fig:NgalMstarIMF}.} 
 \label{fig:SMFnative}
 \end{figure}

 \begin{figure*}
   \includegraphics[width=1.99\columnwidth]{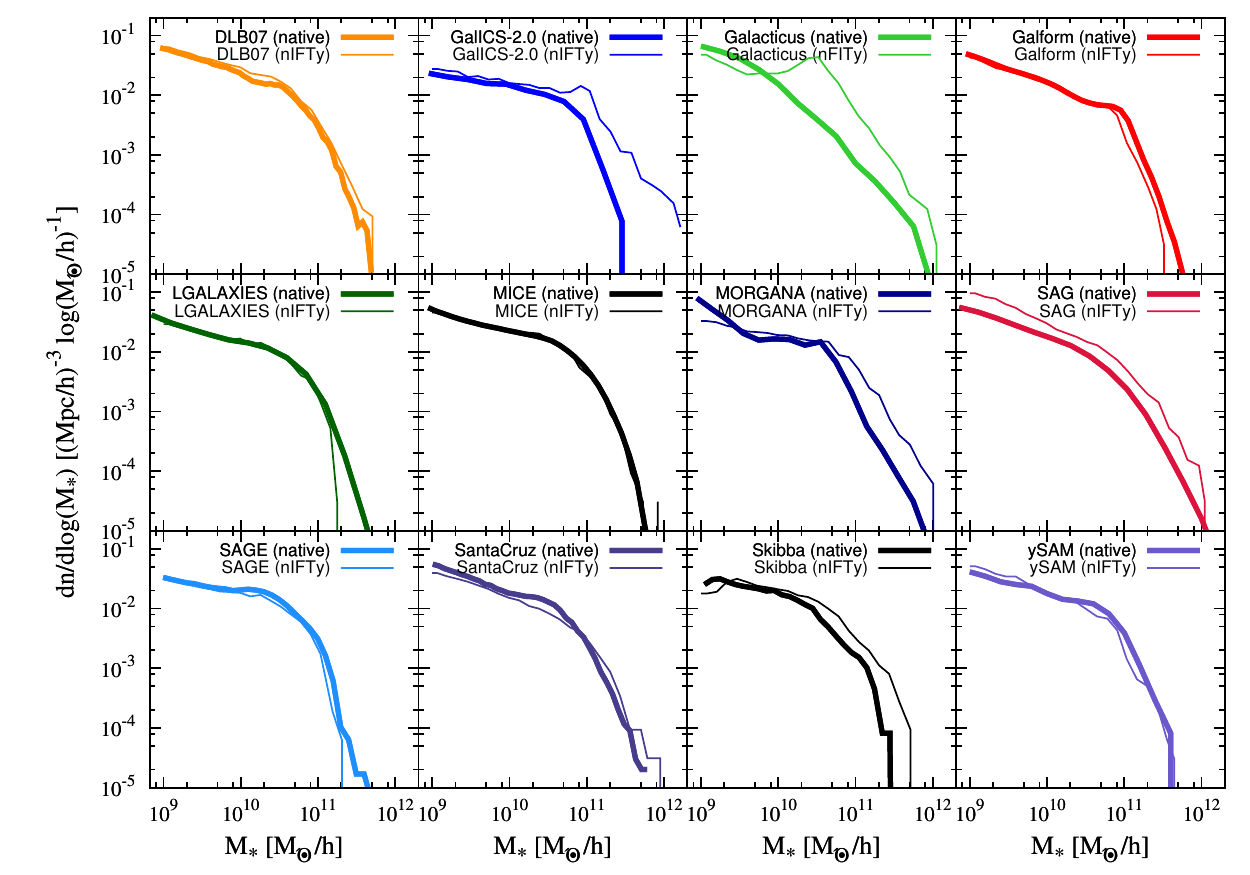}
   \caption{Comparing the stellar mass functions for each model as given in its
     native environment (thick lines) and when applied with the same parameters
     to the nIFTy data (thin lines).}
 \label{fig:SMFnativenifty}
 \end{figure*}

\App{app:halomassdefinition} provides a direct comparison of models for two
different mass definitions (their standard one and \Mcrit). That appendix
further shows its influence on other galaxy properties such as the
stellar-to-halo mass ratio and the number and star formation density evolutions.

\subsection{IMF Correction} \label{sec:IMF}
An additional source of scatter is that the models assumed various initial stellar
mass functions.  Hence we transformed the stellar masses returned by each model
to a unified Chabrier IMF \citep{Chabrier03}. For that we used the following equations \citep{Mitchell13,Bell:01}:
\begin{equation} \label{eq:IMFcorrection}
\begin{array}{lcll}
\log_{10}(M^{\rm Chabrier}_{*}) & = & \log_{10}(M^{\rm Salpeter}_{*})    & - \ 0.240\\
\log_{10}(M^{\rm Chabrier}_{*}) & = & \log_{10}(M^{\rm diet-Salpeter}_{*})    & - \ 0.090\\
\log_{10}(M^{\rm Chabrier}_{*}) & = & \log_{10}(M^{\rm Kennicutt}_{*})  & + \ 0.089\\
\end{array}
\end{equation}
Note that this is only a rough correction, as these numbers depend on the
stellar population synthesis (SPS) model, age and metallicity of the simple
stellar population and on looking to one or several bands when estimating
stellar masses from broad band photometry.

The models have been corrected as follows:
\begin{itemize}
 \item \galics: tuned to observations w/ Chabrier IMF;
 \item \galform: Kennicut $\rightarrow$ Chabrier;
 \item \mice: diet-Salpeter $\rightarrow$ Chabrier;
 \item \sag: Salpeter $\rightarrow$ Chabrier;
\end{itemize}
noting that we left \galics\ untouched because this model tuned its parameters
to an observational data that itself assumed already a Chabrier IMF. 

In \Fig{fig:NgalMstarIMF} we show the resulting stellar mass function for all
models where we notice again that the scatter is only slightly reduced. Some
additional information is again provided in \App{app:IMF}.

\subsection{Model Environment} \label{sec:native}

While the whole idea of the comparison presented in this paper is to
apply galaxy formation models to the same halo catalogues and merger
trees coming from a unique cosmological simulation, we have seen that
there is a non-negligible scatter across properties. This scatter is
larger than in previous comparison projects which encompassed fewer
models \citep[e.g.][]{Kimm09,Fontanot09,Fontanot12,Contreras13,Lu14}. As we could neither attribute the increased variations to halo
mass definitions nor the assumed initial stellar mass functions we are now going to show that
this in part comes from taking models out of their native
environment. The majority of the models have been designed using a
certain simulation and tree structure. However for the comparison
presented here this environment has often been substantially changed,
and model parameters have not been adjusted to reflect this (as
mentioned before, such a recalibration will form part of the follow-up
project). The effect of this approach can be appreciated in the two
following plots.

\Fig{fig:SMFnative} shows the stellar mass functions at redshift $z=0$ for all
models \textit{as given when applied to the simulation and merger trees used for calibration}; we refer to this setup as `native environment'. 
These data points were directly provided by the code representatives, then converted to a common IMF, as in the
previous section, and plotted on the same axes/scale as for
\Fig{fig:NgalMstar}. The agreement between the models is now much improved
indicating that the main part of the scatter seen before is due to models being
applied to simulation data they were not adapted to (and that might even feature a different cosmology). There are still a few
outliers on the low-mass end, e.g. \galics: this model was calibrated on the UltraVISTA SMFs at
different redshifts, not on local mass functions; and models calibrated on
high-$z$ data tend to underestimate the low-mass end of the local SMF.

To underline the influence of the merger trees to the model results we directly compare in \Fig{fig:SMFnativenifty} the native (thick lines) to the nIFTy (thin lines) stellar mass function where each panel represents one model. This plot quantifies the sensitivity of the model to the underlying simulation and merger tree. It shows that \textit{recalibration is required whenever a new simulation is to be used}. While this is common practice within the community, its necessity has been shown here for the first time. A forthcoming companion paper will further address the influence of the applied observational data set to the remaining model-to-model scatter.

\section{Galaxies and their Haloes} \label{sec:galaxies}
In this section we extend the comparison to several additional properties
including star formation rate, the stellar mass fraction, the number density
(evolution) of galaxies, and the relation between galaxies and their dark matter
haloes.

\subsection{Star formation rate} \label{sec:sfr}

\begin{figure}
   \includegraphics[width=\columnwidth]{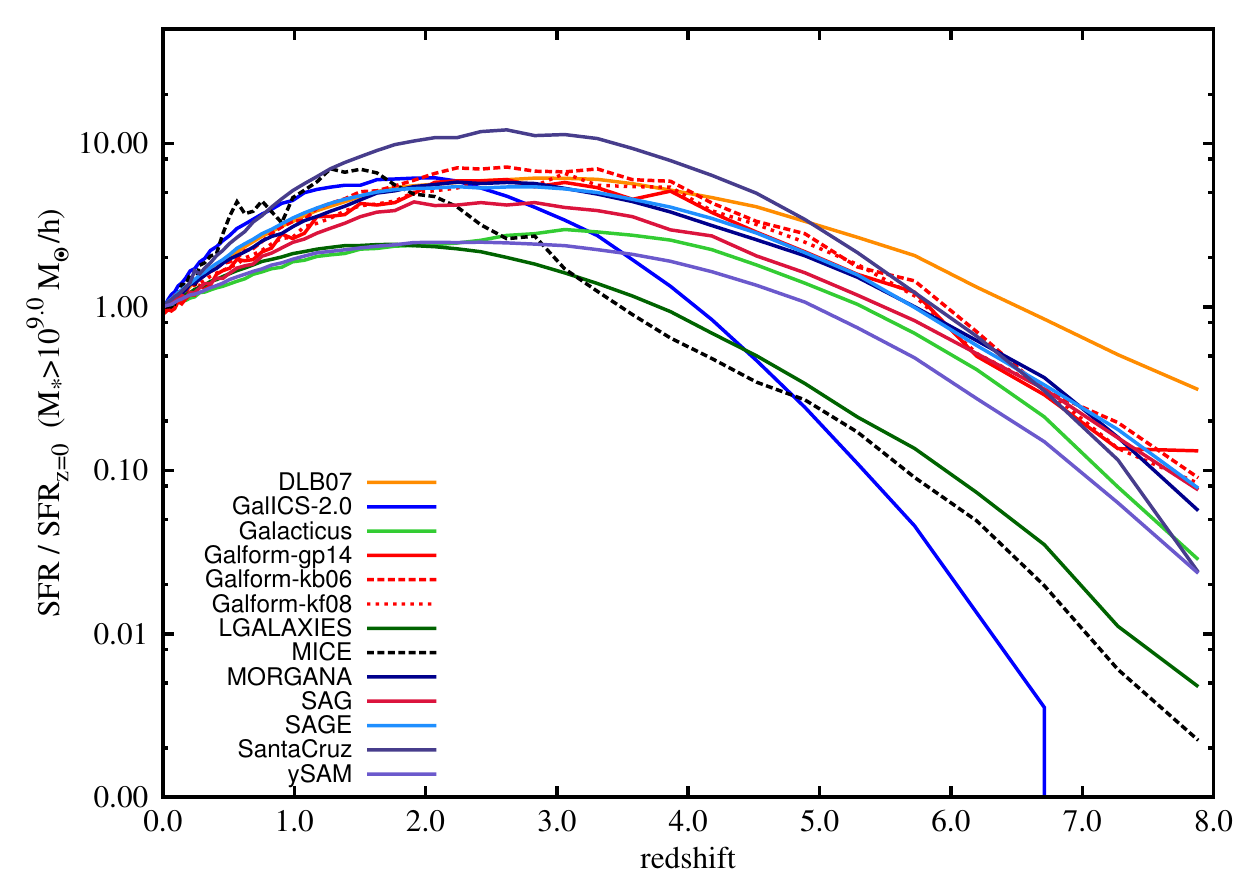}
   \caption{Star formation rate density for galaxies with
     $M_{*}>10^{9}$\hMsun\ as a function of redshift (normalized to the redshift
     $z=0$ values listed in \Tab{tab:sumMstar}).}
 \label{fig:SFRzred}
 \end{figure}

The stellar mass of a galaxy studied in the previous section depends upon the
evolution of its star formation rate (SFR). Therefore we now turn to the history
of the star formation rate across all considered models. In \Fig{fig:SFRzred} we
show the redshift evolution, noting that all the curves in this plot have been
normalized by their redshift $z=0$ values (which are given in the third column
of \Tab{tab:sumMstar}). In this way we separate trends from absolute
differences. An un-normalized version of \Fig{fig:SFRzred} can be found in
\App{app:unnormalized}. Remember that the HOD model \skibba\ does not produce
high redshift outputs and so appears neither in \Fig{fig:SFRzred} nor in the SFR
columns of the accompanying \Tab{tab:sumMstar}.

For the SAMs the peak of star formation is about redshift $z\sim 2-3$ followed
by a rapid decrease at late times -- in agreement with observations and the
uncertainties seen within them \citep[e.g.][]{Madau14}. But amongst these models
there are also differences: in \lgalaxy, for instance, the peak is at smaller
redshifts while for \galacticus\ it is at earlier times; and the HOD model
\mice\ shows a relatively high SFR at low redshifts, i.e. \mice\ stars are
formed preferentially later than in the other models.  These differences are
reflected in the fact that, given the redshift $z=0$ normalisation in the plot,
there are differences in amplitude of an order of magnitude at redshift $z>6$.

\begin{figure}
   \includegraphics[width=\columnwidth]{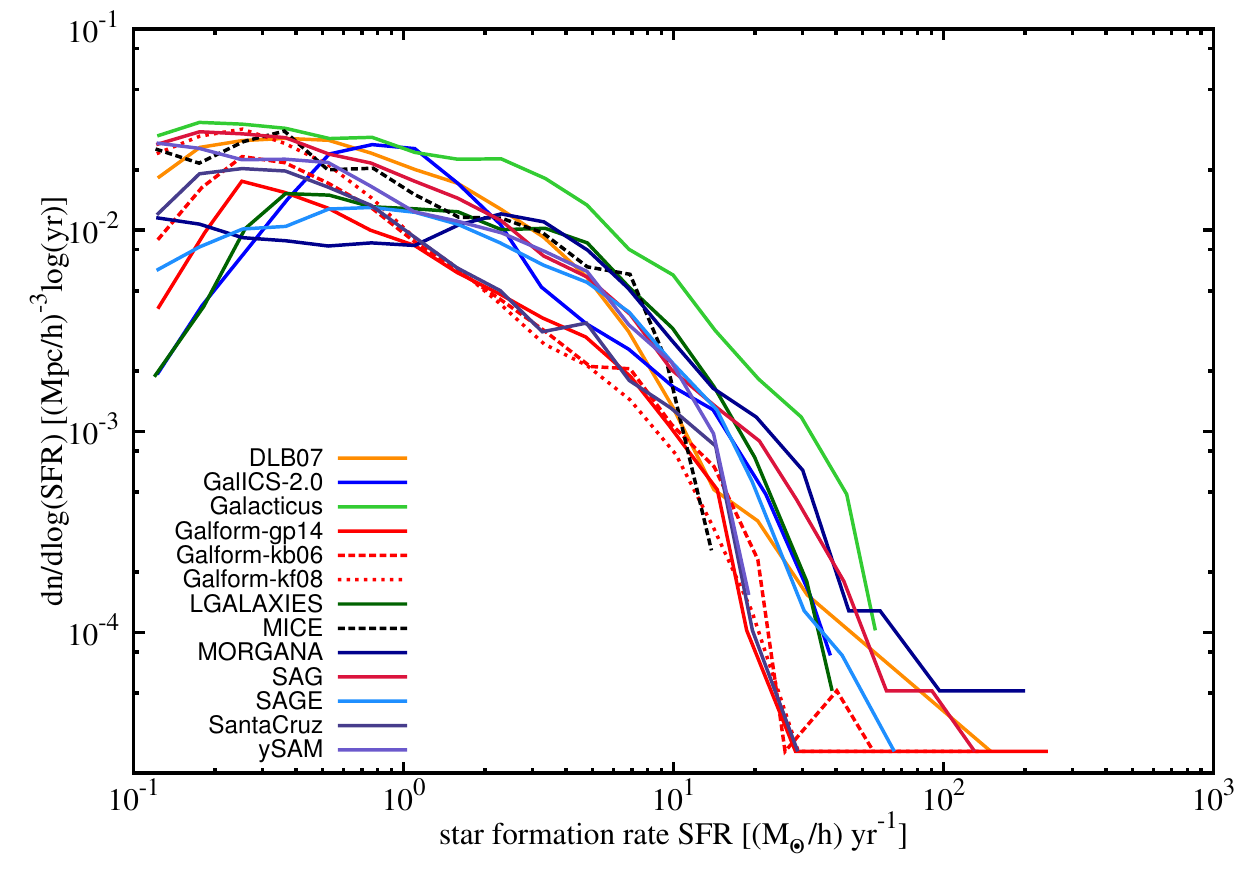}
   \caption{Star formation rate distribution function at redshift $z=0$.}
 \label{fig:NgalSFR}
 \end{figure}

While the previous figure has shown the integrated star formation rate, we
inspect its redshift $z=0$ properties more closely in \Fig{fig:NgalSFR} where we
present the star-formation-rate distribution function, i.e.~the number density of galaxies in
a certain SFR interval.  From that we see that all models have a similar
functional form, but that the normalisation of the SFR shows differences of up
to a factor of three between models.  This is reflected in \Tab{tab:sumMstar}
where we list the total stellar mass formed (second column), the present-day
star formation rate (third column) and specific star formation rate (last
column, i.e.~the ratio between SFR and total $M_*$). We see that, for instance,
the \galacticus\ model produced more than three times as many stars as
\lgalaxy.

\begin{table}
  \caption{Total stellar mass, star formation rate, and global specific star
    formation rate in galaxies with $M_{*}>10^{9}\hMsun$ at redshift $z=0$
    (computed as total stellar mass divided by total SFR).}
\label{tab:sumMstar}
\begin{center}
\begin{tabular}{lccc}
\hline
code name	& $M_{*}$ & SFR	& sSFR \\
			& $[10^{14}\hMsun]$ & [$10^{4} $\hMsun yr$^{-1}$]	& Gyr$^{-1}$ \\
\hline
 \galacticus	&  $2.91$ & $2.22$ & $0.0761$ \\
 \galics		&  $2.73$ & $0.88$ & $0.0321$ \\
 \morgana 	&  $1.96$ & $1.21$ & $0.0614$ \\
 \sag			&  $2.37$ & $1.15$ & $0.0486$ \\
 \somerville 	&  $1.11$ & $0.53$ & $0.0475$ \\
 \ysam 		&  $1.14$ & $0.85$ & $0.0749$ \\
\\
\underline{Durham flavours:}\\
 \galformVGP	&  $0.98$ & $0.50$ & $0.0511$ \\
 \galformBOW	&  $1.16$ & $0.51$ & $0.0442$ \\
 \galformFONT	&  $1.06$ & $0.52$ & $0.0491$ \\
\\
\underline{Munich flavours:}\\
 \DLB 		&  $1.76$ & $0.99$ & $0.0563$ \\
 \lgalaxy		&  $0.87$ & $1.07$ & $0.1234$ \\
 \sage 		&  $1.01$ & $0.83$ & $0.0815$ \\
\\
\underline{HOD models:}\\
 \mice 		&  $1.77$ & $0.96$ & $0.0543$ \\
 \skibba		&  $1.49$ & n/a       &  n/a  \\
\hline
\end{tabular}
\end{center}
\end{table}

 \begin{figure*}
   \includegraphics[width=\columnwidth]{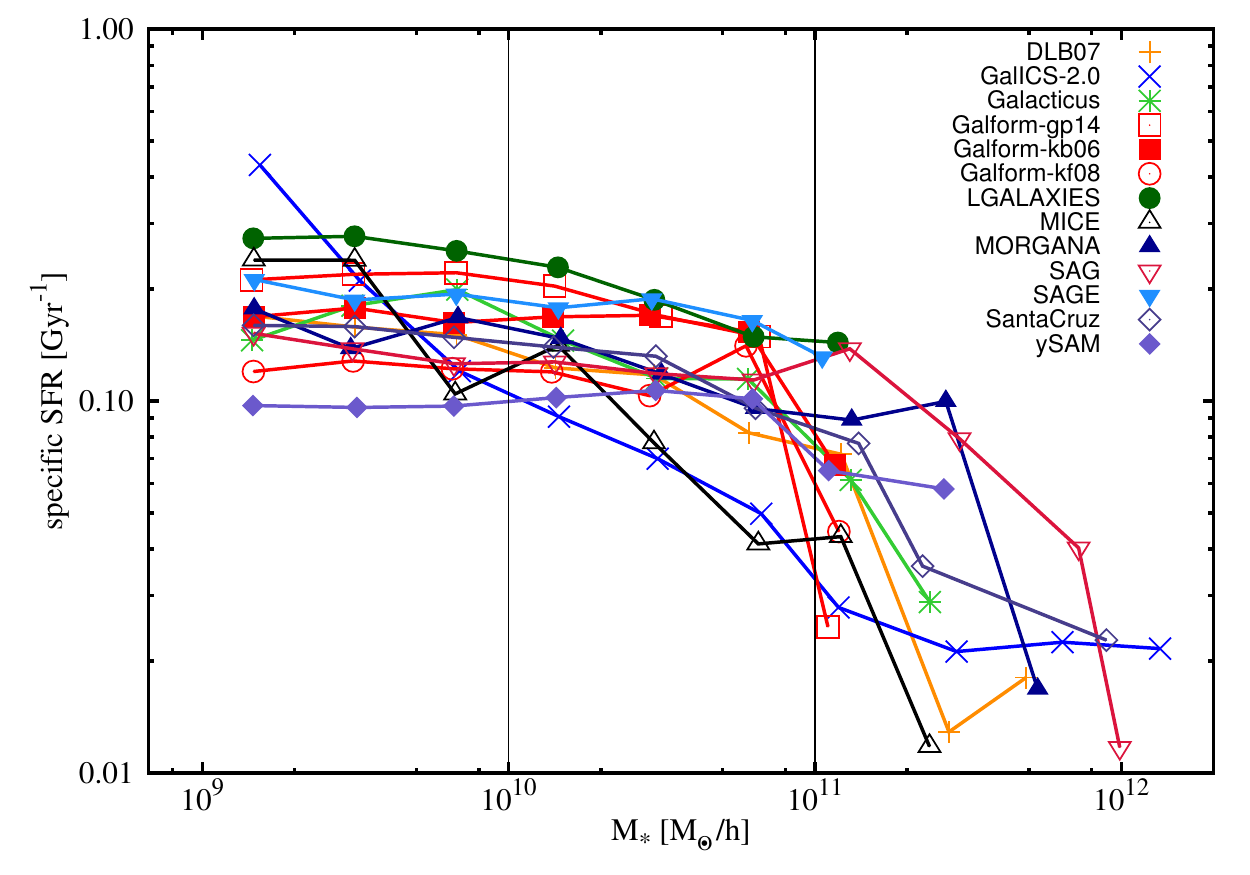}
   \includegraphics[width=\columnwidth]{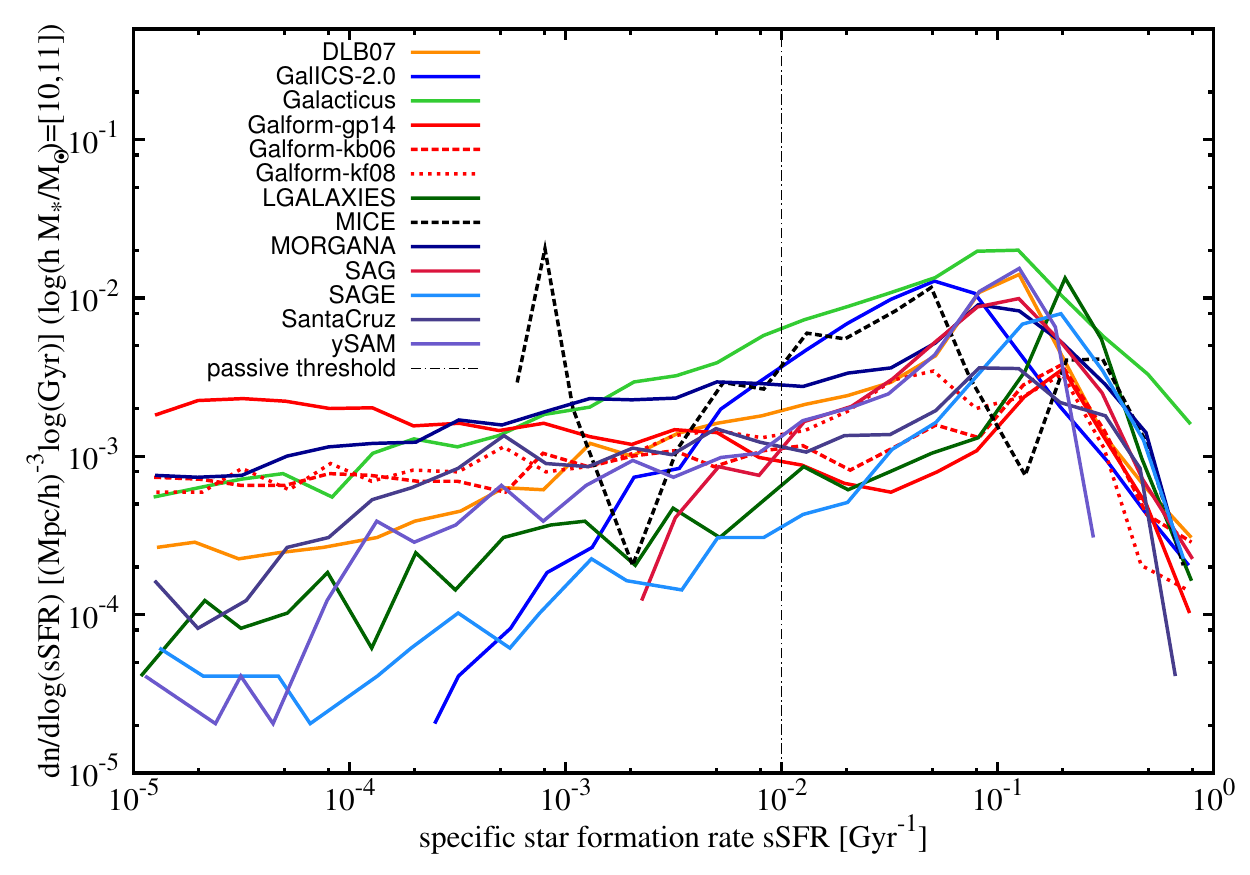}
   \caption{Specific star formation rate of star-forming galaxies at redshift
     $z=0$. The left panel shows the star formation rate per stellar mass as a
     function of stellar mass $M_{*}$; points represent mean values binned in
     both the $y$ and $x$ direction for the star forming sequence of
     galaxies. The right panel serves as a proxy for the (omitted) error bars:
     it shows the distribution of specific star formation rates for galaxies in
     the mass range $M_{*} \in [10^{10},10^{11}]\hMsun$ (indicated by the two
     vertical lines); the vertical dashed line indicates our choice for the
     passive galaxy fraction threshold.}
 \label{fig:sSFRMstar}
 \end{figure*}

The two previous figures showed the overall star formation rate, but
now we focus in \Fig{fig:sSFRMstar} on the {\em specific} star
formation rate (\sSFR) as a function of stellar mass $M_{*}$ at
redshift $z=0$. The left panel shows the \sSFR\, excluding passive
galaxies, i.e.~galaxies that are not considered `star-forming', which we define
as those with \sSFR~$ < 0.01$\ Gyr$^{-1}$. Points
shown are the mean values in the bin, both for the $y$- and $x$-axes,
(which explains why they do not start exactly at our mass threshold
of $10^{9}$\hMsun).\footnote{Although not shown we also reproduced the plot
using medians and 25 and 75 percentiles, but as these give very
similar results we decided to adopt mean values for this and all
subsequent plots.} Instead of error bars, the right panel of
\Fig{fig:sSFRMstar} shows the distribution of \sSFR\ values for a mass
bin $M_{*}\in [10^{10},10^{11}]\hMsun$ with our choice for the passive
threshold shown as a vertical dotted line. We are aware that the
choice of this mass bin for the right panel encompasses the `knee' of
the stellar mass function, but this right panel nevertheless shows
that our passive threshold value cuts the wing to the left
at approximately the same height as the right side. Please note that
the right panel does not substantially change when considering a
different mass range, although this is not explicitly shown here.

\Fig{fig:sSFRMstar} reflects what has already been seen in \Fig{fig:SFRzred},
i.e. there is a great diversity in star formation rates across the models
irrespective of the stellar mass of the galaxy. Bearing in mind the differences
in stellar mass functions visible here again on the $x$-axis, the curves
vary by a factor of about 3 at essentially all masses, with the
primary difference being in overall normalisation.

We would like to remark on the interplay between \Fig{fig:SFRzred},
\Tab{tab:sumMstar} and \Fig{fig:sSFRMstar} as at first sight the results seem to
be counterintuitive.  For instance, \galacticus\ has a much higher star
formation rate (at all times) than \lgalaxy, yet the specific star formation
rate is higher for \lgalaxy. This is readily explained by the presence of, on
average, more massive galaxies in \galacticus, which is confirmed by the stellar
mass function presented in \Fig{fig:NgalMstar}. One should also bear in mind
that the specific star formation rate \sSFR\ could be considered a proxy for the
(inverse of the) age of a galaxy. Therefore, \Fig{fig:sSFRMstar} indicates, for example, that
it took galaxies in \lgalaxy\ less time to assemble their stellar mass than
galaxies in \galacticus. 

We close our discussion of \Fig{fig:sSFRMstar} with the remark that it does
not change when considering only centrals: the differences across
models remain unaffected by restricting the analysis to this galaxy
population.  However, the specific star formation rate rises, with
approximately constant ratios between the curves, when moving to higher
redshifts.

\subsection{Stellar mass fractions} \label{sec:smfrac}

 \begin{figure*}
   \includegraphics[width=\columnwidth]{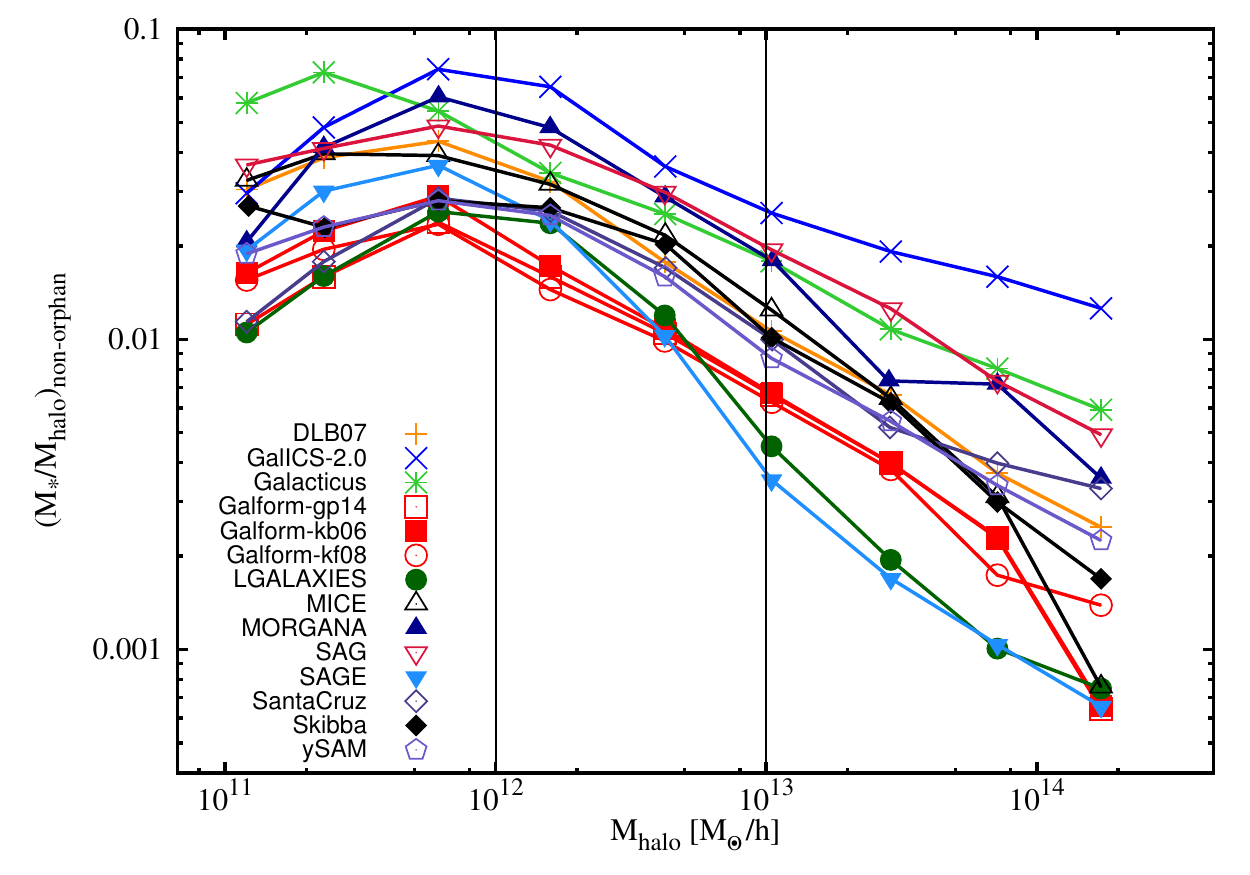}
   \includegraphics[width=\columnwidth]{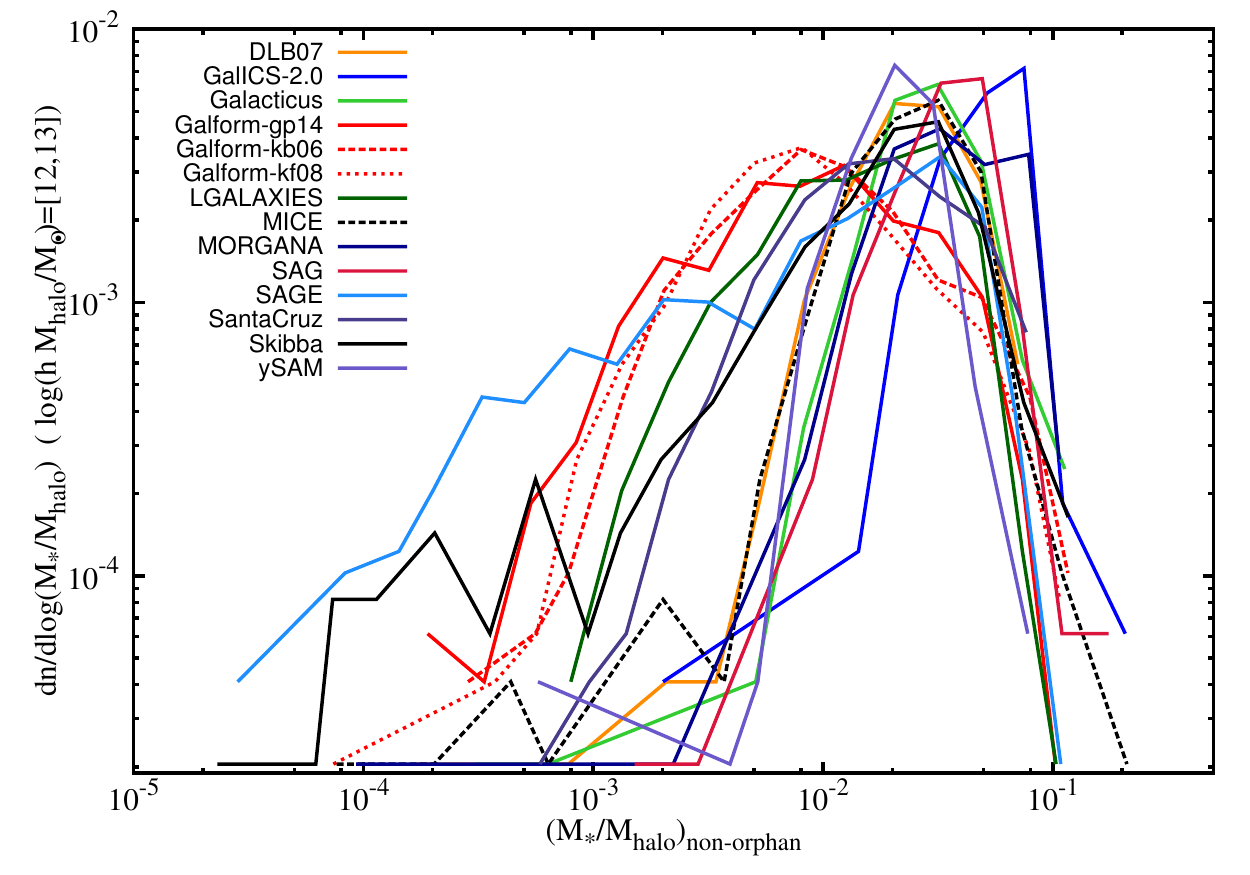}
   \caption{Stellar-to-halo mass ratio for all (non-orphan) galaxies at redshift
     $z=0$. The left panel shows mean values (again in both directions) as a
     function of galaxy host halo mass \Mhalo\ whereas the right panel indicates
     the (omitted) error bars: it shows the distribution of $M_{*}/M_{\rm halo}$
     for galaxy halo masses in the range $M_{\rm halo} \in
     [10^{12},10^{13}]\hMsun$.}
 \label{fig:MstarMhaloMhalo}
 \end{figure*}

In \Fig{fig:MstarMhaloMhalo} we show the stellar-to-halo mass ratio
$M_{*}/M_{\rm halo}$ as a function of galaxy host halo mass $M_{\rm halo}$ for
non-orphan galaxies.\footnote{We omit orphan galaxies as they lack an associated
  halo mass.} The layout of this figure is similar to the previous one, i.e.~the
left panel shows the actual mean of the ratios (omitting error bars), with the
distribution of the values in a galaxy host halo mass bin $M_{\rm halo}\in
[10^{12},10^{13}]\hMsun$ the right serving as a proxy for the missing error
bars.  All models have a similar maximum stellar mass fraction of about 0.1, but
there is a large spread in the modal value and this leads to a difference in
overall normalisation of approximately a factor of 3 -- where some of this
variation can be attributed to the different mass definitions applied (see
\App{app:halomassdefinition}). Note that the curves remain unaffected by
restricting the data to central galaxies only.  Further, the distributions shown
in the right panel are not influenced by the halo mass bin.

\subsection{Number density} \label{sec:ndens}

 \begin{figure}
   \includegraphics[width=\columnwidth]{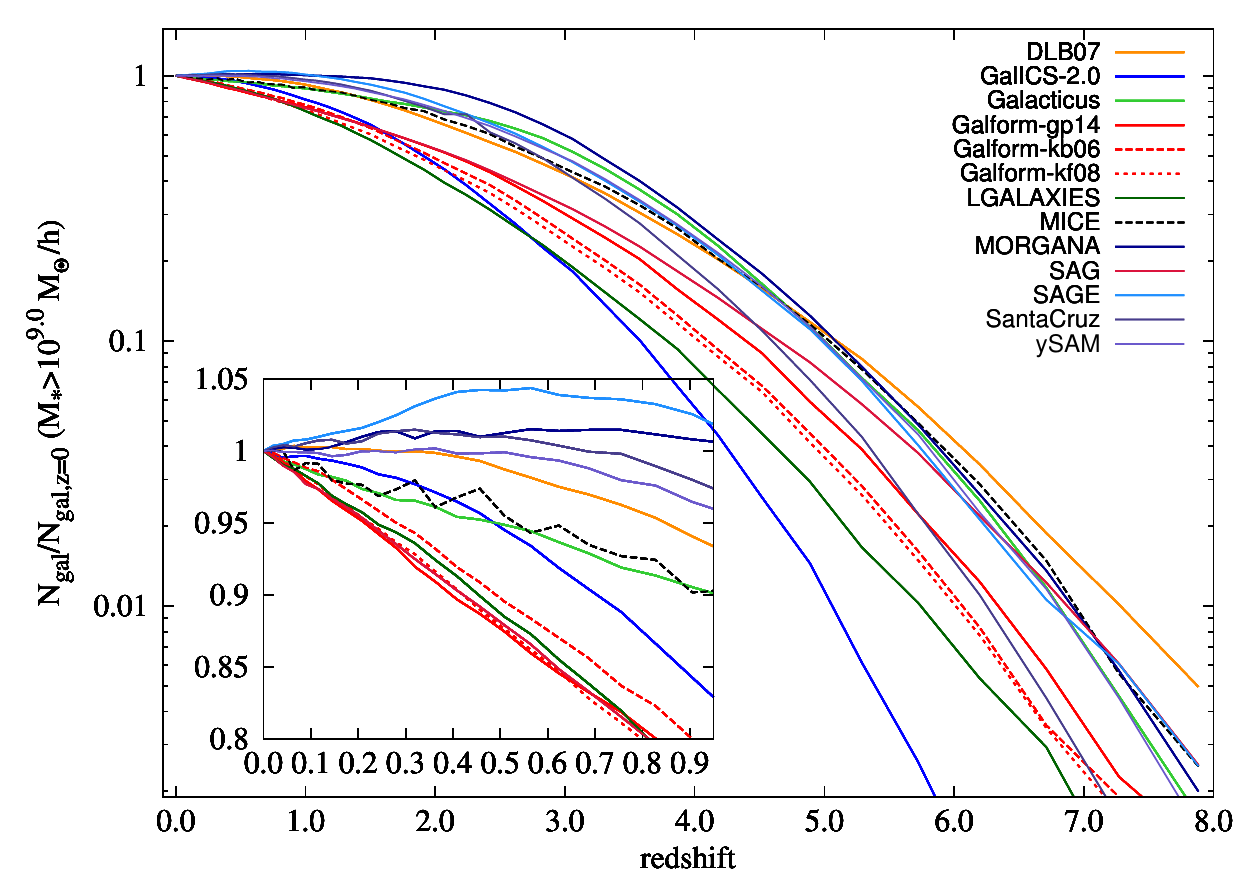}
   \caption{The number of all galaxies with stellar mass $\Mstar > 10^{9}\hMsun$
     (normalized to redshift $z=0$ values as listed in \Tab{tab:Ngal}) as a
     function of redshift. The inset panel shows a zoom (using a linear
     $y$-axis) into the range $z\in [0,1]$.}
 \label{fig:Ngalzred}
 \end{figure}

\begin{table}
  \caption{Number of galaxies at redshift $z=0$ with a stellar mass in excess of
    $M_{*}>10^{9}\hMsun$. (For the \morgana\ and \somerville\ models the number
    of orphans is in fact the number of satellite galaxies, see text.)}
\label{tab:Ngal}
\begin{center}
\begin{tabular}{lcccc}
\hline
code name	& $N_{\rm gal}$ & $N_{\rm central}$ & $N_{\rm non-orphan}$ & $N_{\rm orphan}$ \\
\hline
 \galacticus	& 14255 & 7825 & 10019 & 4236\\
 \galics		& 9310 & 7462 & 9310 & 0\\
 \morgana 	& 10008 & 6186 & 6186 & \textit{3822}\\
 \sag			& 19516 & 13571 & 16256 & 3260\\
 \somerville 	& 8901 & 6682 & 6682 & \textit{2219}\\ 
 \ysam 		& 11138 & 7423 & 9458 & 1680\\
\\
\underline{Durham flavours:}\\
 \galformVGP	& 8824 & 5097 & 6098 & 2726\\ 
 \galformBOW	& 11563 & 6669 & 7897 & 3666\\
 \galformFONT	& 12116 & 6430 & 7664 & 4452\\  
\\
\underline{Munich flavours:}\\
 \DLB 		& 15132 & 9420 & 11897 & 3235\\ 
 \lgalaxy		& 7499 & 4792 & 6287 & 1212\\ 
 \sage 		& 8437 & 6588 & 8437 & 0\\ 
\\
\underline{HOD models:}\\
 \mice 		& 12191 & 7286 & 10106 & 2085\\ 
 \skibba		& 9203 & 5088 & 7973 & 1230\\
\hline
\end{tabular}
\end{center}
\end{table}

Turning to the galaxies themselves we show in \Fig{fig:Ngalzred} the evolution
of the number density of galaxies (with stellar mass in excess of
$M_{*}>10^{9}\hMsun$) as a function of redshift -- normalized to the number of
galaxies at redshift $z=0$ (provided in \Tab{tab:Ngal}).\footnote{Note that as
  the \skibba\ model solely provides $z=0$ data it does not appear in the
  figure.} It is noteworthy that the various models display different
evolutionary trends of galaxy density. For instance, \DLB\ starts with the
largest fraction of galaxies at high redshift whereas \galics\ begins with the
lowest fraction -- with the difference being more than one order of magnitude at
redshift $z=6$ between these two models. We further note (in the inset panel)
that some of the models -- in particular \sage\ -- have a flat or falling galaxy
number density between a redshift $z=1$ and the present day -- this is perfectly
allowable as galaxy merging can reduce the galaxy density. Although not
explicitly shown here, the main features of the plot remain the same when
restricting the analysis to central galaxies only.

\Fig{fig:Ngalzred} should be viewed together with \Tab{tab:Ngal} as the former
provides the trend whereas the latter quantifies the normalization (at redshift
$z=0$); for a combination of both, i.e. an un-normalized version of
\Fig{fig:Ngalzred}, we refer the reader to \Fig{fig:Ngalzred2} in the Appendix.  The total
number of galaxies with $M_{*}>10^{9}\hMsun$ ranges from $\approx$ 7500 for the
\lgalaxy\ model to $\approx$15000 in the \DLB\ model.  
All models (apart from \skibba) populate all dark matter (sub-)haloes found in
the simulation down to at least $M_{\rm halo}\approx 10^{11}$\hMsun. Hence,
any differences seen here originate from lower mass objects. This is confirmed
by re-calculating $N_{\rm central}$ applying a halo mass threshold of $M_{\rm
  halo} > 2\times 10^{11}$\hMsun\ (instead of the galaxy stellar mass threshold
of $M_{*}>10^{9}$\hMsun). This process results in 3774 galaxies for all models
-- a number identical to the number of host haloes in the \subfind\ catalogue
above this mass limit.

\begin{figure}
   \includegraphics[width=\columnwidth]{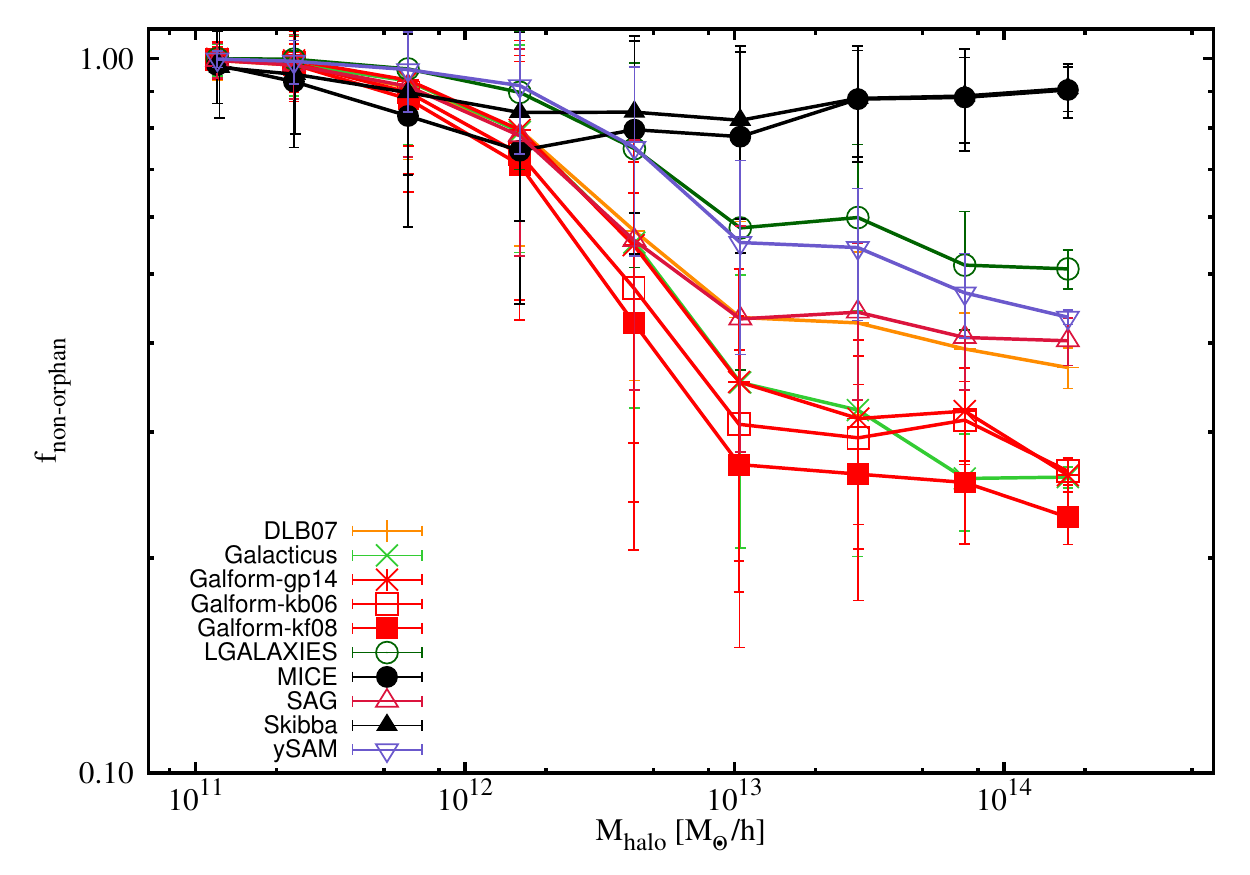}
   \includegraphics[width=\columnwidth]{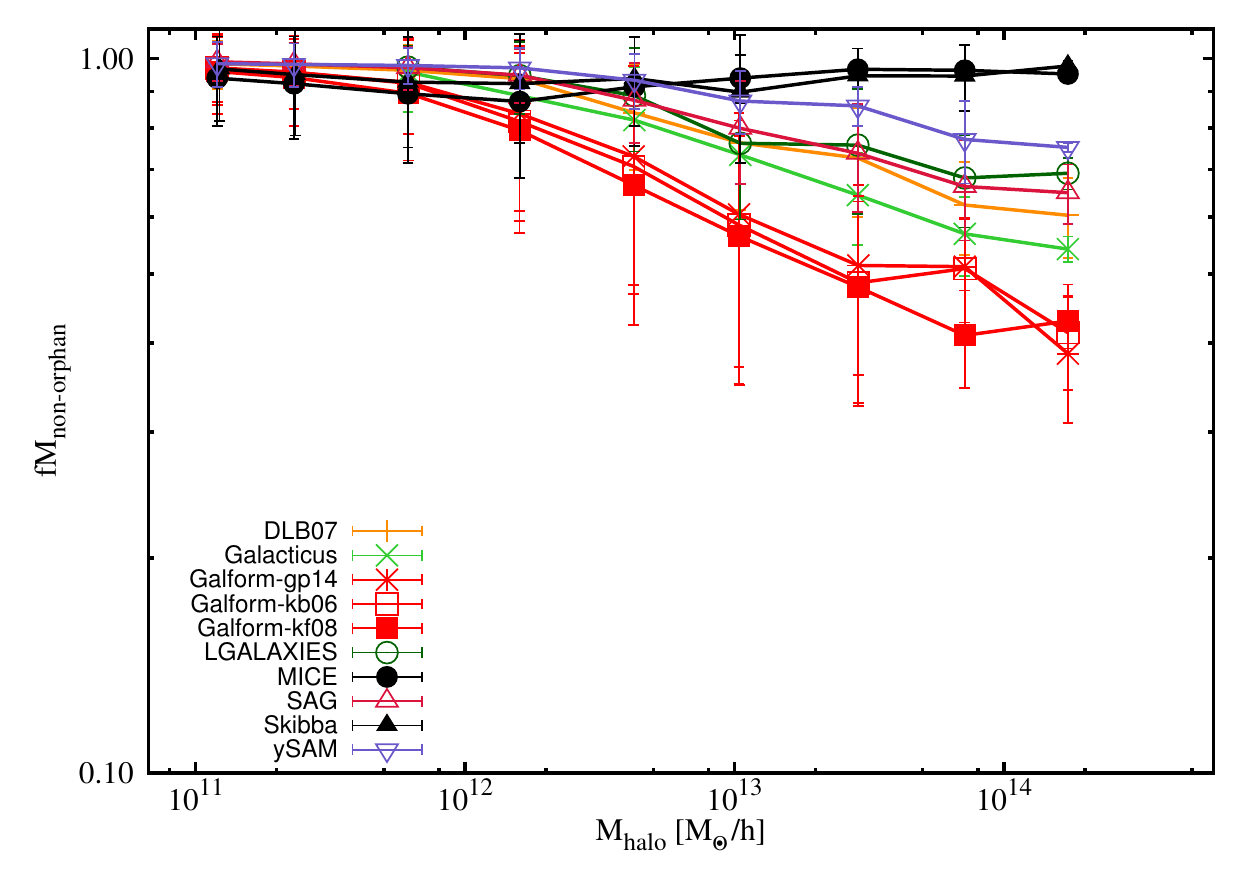}
   \caption{Number fraction (top panel) and stellar mass fraction (bottom panel) of non-orphan galaxies with stellar mass $\Mstar > 10^{9}\hMsun$ as a function of main halo mass \Mhalo\ at redshift $z=0$. Only models that actually feature orphans are shown here.}
 \label{fig:NxorphMhalo}
 \end{figure}

In \Tab{tab:Ngal} we further divide the galaxies into different populations,
i.e. centrals, non-orphans, and orphans.  The fraction of orphan galaxies also
shows a spread from a mere 13 per cent for \skibba\ to nearly 37 per cent for
the \galformFONT\ model.  Note that \galics, and \sage\ do not feature orphans
at all, whereas the \morgana\ and \somerville\ models, as previously mentioned,
do not make use of the $N$-body information for subhaloes and hence tag
satellite galaxies as orphans -- therefore all satellite galaxies in these
models are technically orphans as only central galaxies retain information about
their host halo; naturally, $N_{\rm central}=N_{\rm non-orphan}$ for these two
models.  To further explore the differences in the abundance of orphan galaxies
between models we show in the upper (lower) panel of \Fig{fig:NxorphMhalo} the
number (stellar mass) fraction of all galaxies that are classified as
non-orphans with stellar mass in excess of $M_{*}>10^{9}\hMsun$ orbiting inside
a main halo of given mass $M_{\rm halo}$.\footnote{As neither \galics, \morgana, \sage,
nor \somerville\ feature orphans, they have been omitted from the plot.} We can
observe some bimodality here: the two HOD models have the lowest fraction of
orphans in high-mass main haloes (cf. \Tab{tab:Ngal}) whereas for all other
models orphans form the dominant population, making up between 40 and 75 per
cent of all galaxies at $z=0$ within main haloes above $M_{\rm
  halo}>10^{13}\hMsun$. This trend is also true at higher redshifts although we
do not explicitly show it here. For those models which feature orphans, the
variation in the number of orphans is due to the various methods of dealing with
their eventual fate: over some timescale orphans are expected to suffer from
dynamical friction and merge into the central galaxy of the halo. This timescale
can be very long in some models \citep[see e.g.][]{DeLucia10}.  The basic
features of the plot do not change when applying a more strict threshold for
$M_{*}$ as can be verified in the lower panel of \Fig{fig:NxorphMhalo} where we
show the stellar mass weighted fraction of non-orphan satellite
galaxies. However, now the model variations are reduced. The difference from
unity is the fraction of stellar mass locked up in orphans, still as large as 60
per cent for the \galform\ models, yet substantially smaller than the number
fraction presented in the top panel. These differences might be ascribed to the
different treatment of merger times again. We also observe that for the HOD
models the orphan contribution has nearly vanished when weighing it by stellar
mass.

\subsection{Galaxy-Halo Connection} \label{sec:galaxyhaloconnection}

 \begin{figure}
   \includegraphics[width=\columnwidth]{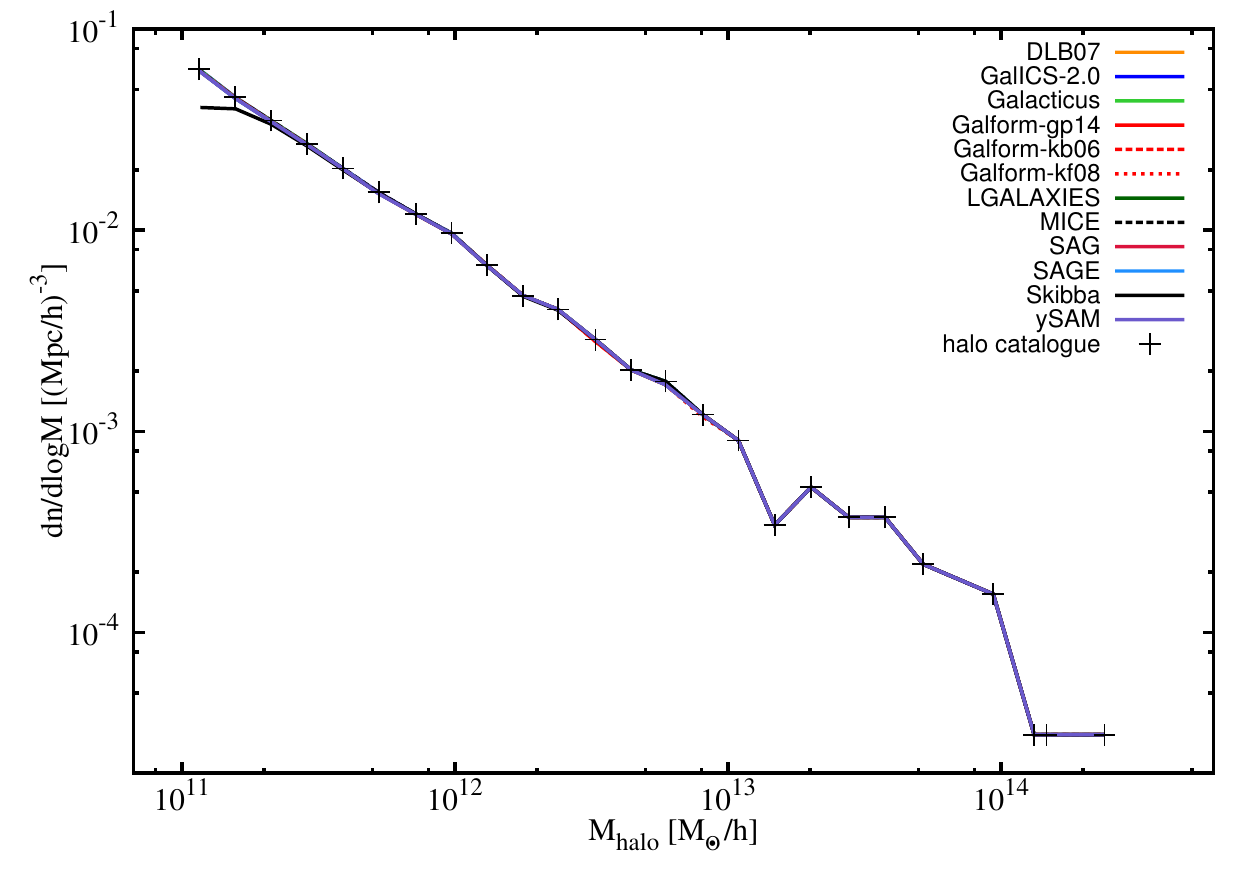}
   \includegraphics[width=\columnwidth]{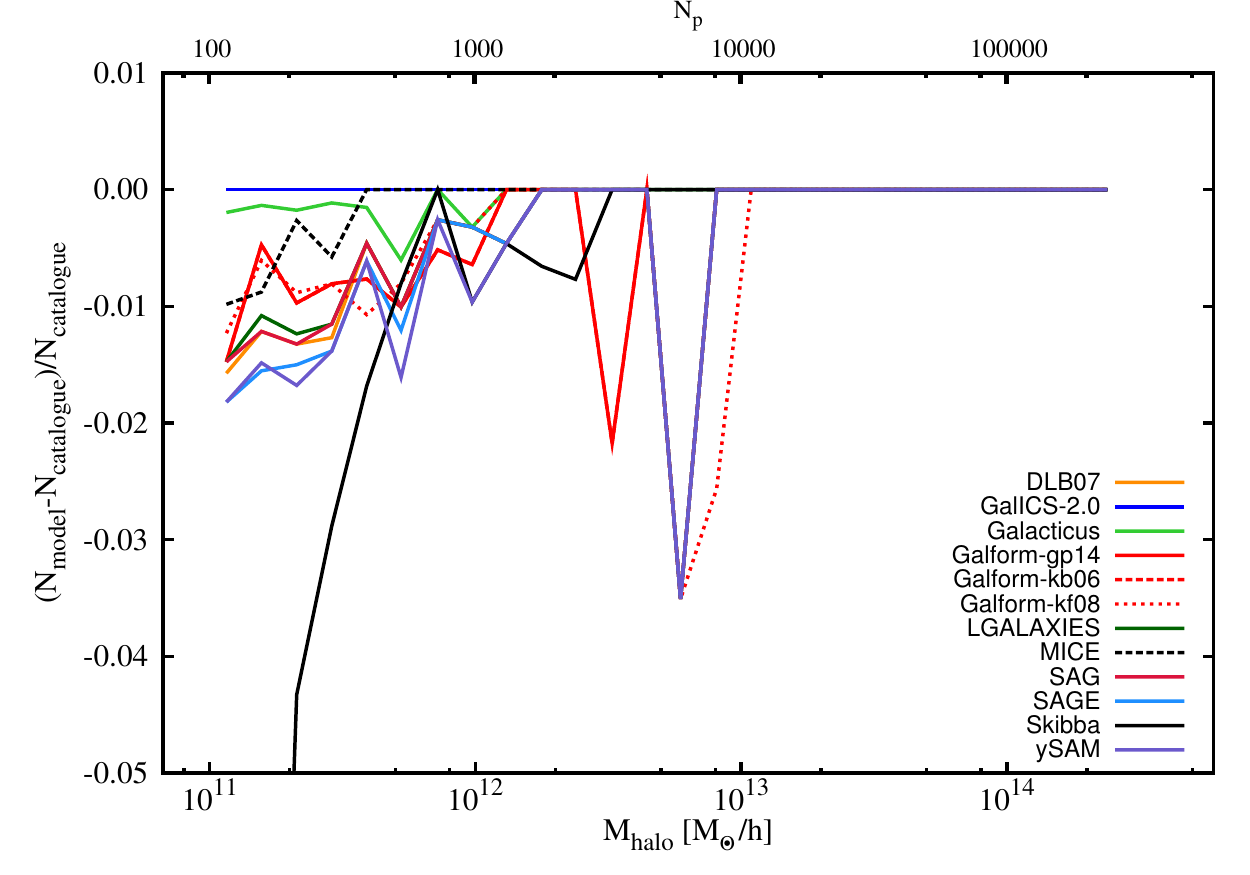}
   \caption{Mass function of all haloes at redshift $z=0$ as given in the input halo catalogue (crosses, all identified objects down to 20 particles including subhaloes) and as recovered from the non-orphan galaxy catalogues of each model. The upper panel shows the supplied mass function whereas the lower panel shows the fractional difference with respect to the input halo catalogue. The upper panel also gives the translation of $M_{\rm halo}$ to the number of particles in the halo as additional $x$-axis at the top.}
 \label{fig:NhaloMhalo}
 \end{figure}

 \begin{figure}
   \includegraphics[width=\columnwidth]{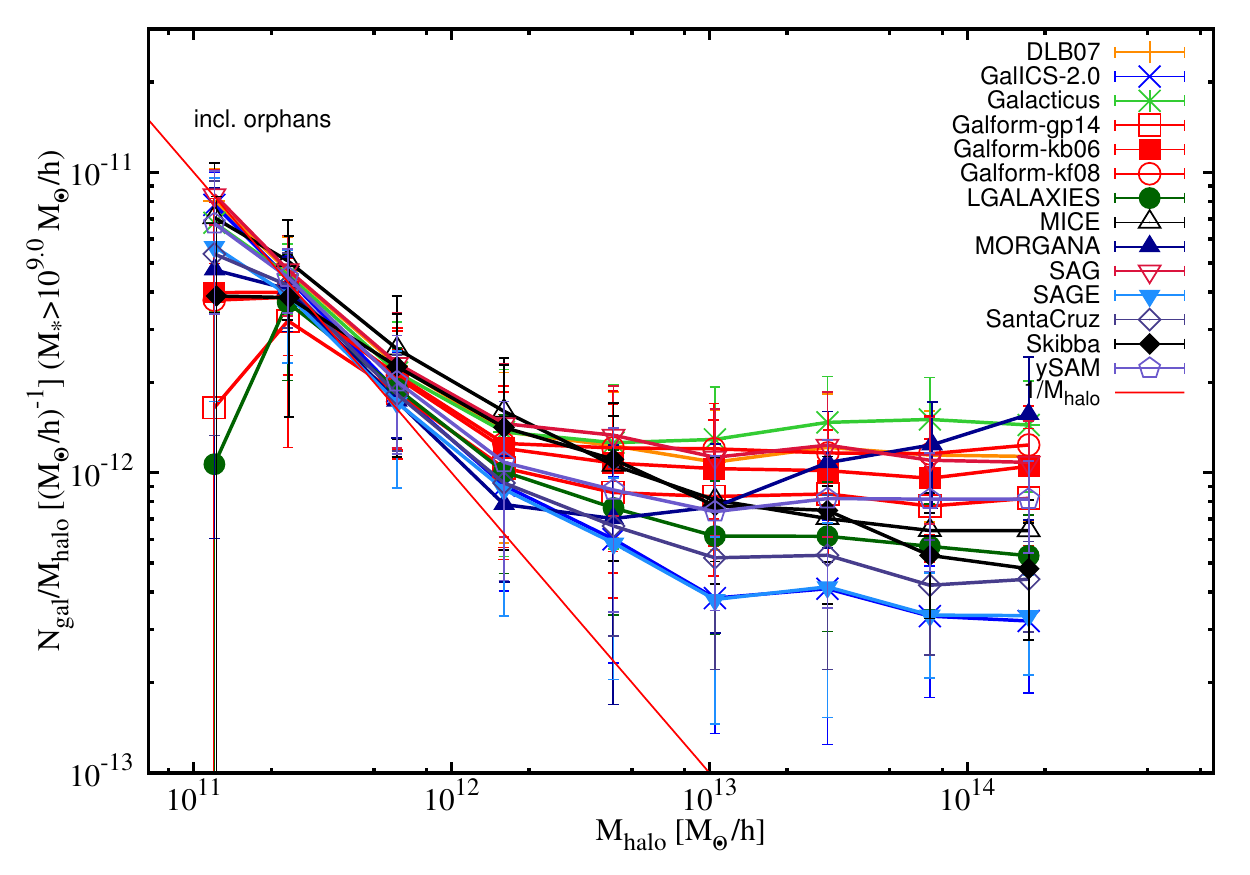}
   \includegraphics[width=\columnwidth]{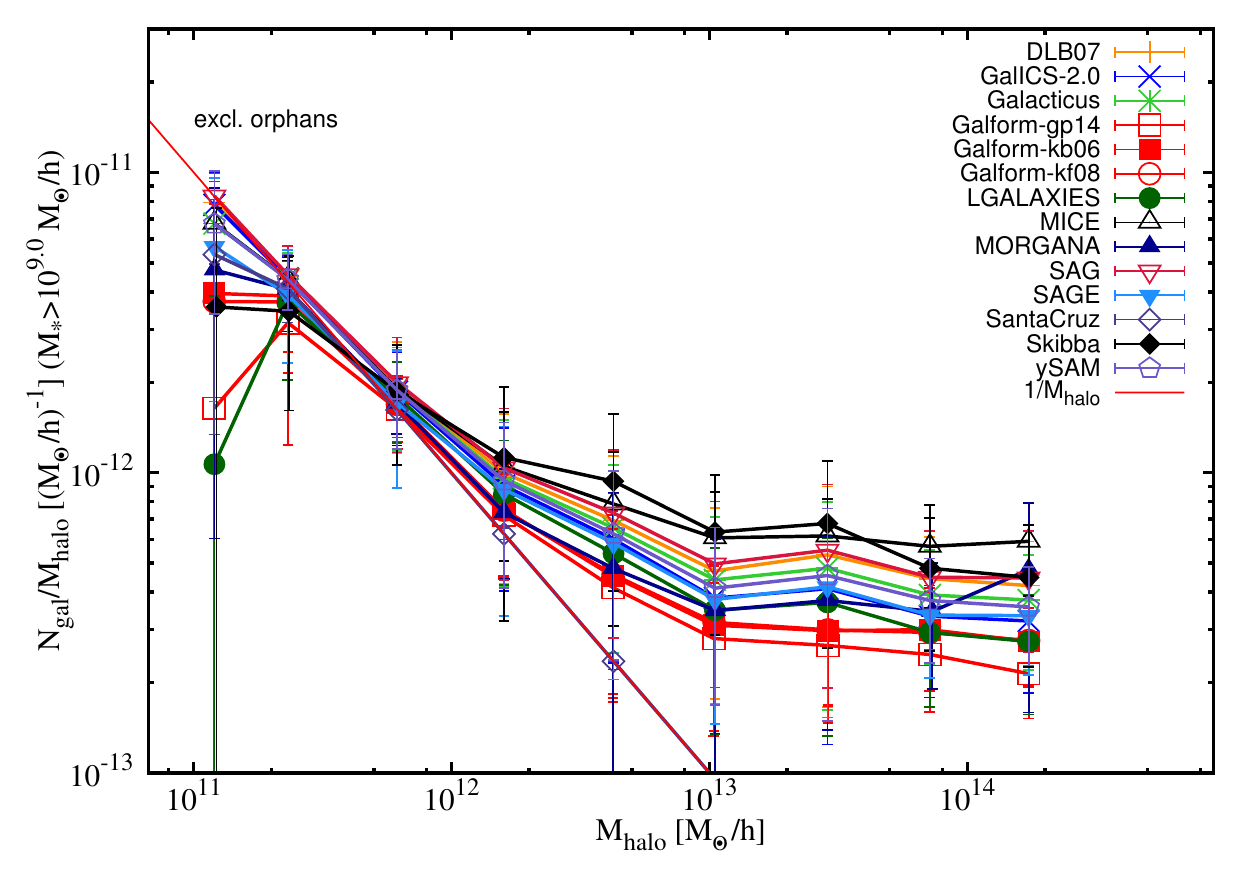}
   \caption{Number of galaxies \Ngal\ per halo mass (i.e. `specific frequency of galaxies') for galaxies more massive than $M_{*}>10^{9}\hMsun$ as a function of halo mass \Mhalo\ at redshift $z=0$. Points plotted are the mean values in the bin with respect to both axes and error bars are $1\sigma$. The solid line running from upper left to lower right represents one galaxy per halo. The upper panel shows all galaxies whereas the lower panel focuses on non-orphan galaxies.}
 \label{fig:NgalMhalo}
 \end{figure}

Lastly, we now turn to how the supplied tree hierarchy has been populated with
galaxies by each model. To this extent the upper panel of \Fig{fig:NhaloMhalo}
compares the supplied halo mass function (crosses, all identified objects down
to 100 particles including subhaloes) overplotted by the halo mass function as
derived from the galaxy catalogues returned by the models.  Note that the
\morgana\ and \somerville\ models have been omitted due to their treatment of
subhaloes.  
While the logarithmic scale of the upper panel masks any differences,
the lower panel -- showing the fractional difference of each model's halo mass
function with respect to the supplied input mass function -- indicates that
nearly every (sub-)halo found in the simulation contains a galaxy. The only
exception to this is the \skibba\ model, which has a high incompleteness
threshold that frequently leaves small haloes empty.
For the other models differences are all below 2 per cent.

In \Fig{fig:NgalMhalo} we relate the number of galaxies with stellar mass
$M_{*}>10^{9}\hMsun$ to the mass of the dark matter main halo they orbit
within. We normalize the average number of galaxies by the mass of the main
halo, i.e. presenting the `specific frequency of galaxies'. The solid line
running from the upper left to the lower right across the plot indicates a
frequency of `one galaxy per main halo'. Haloes to the right of this line
essentially always contain at least one galaxy while the values to the left are
indicating the fraction of haloes devoid of galaxies and are chiefly driven by
incompleteness -- and in retrospect justifying our threshold of $M_{\rm
  halo}>10^{11}$\hMsun\ for the previous plots. At high halo mass,
$M_\mathrm{halo}>10^{13}\hMsun$, the specific frequency of galaxies is roughly
constant for all models although the occupation number of haloes varies between
them by around an order of magnitude. While the upper panel of
\Fig{fig:NgalMhalo} shows all galaxies -- including orphans -- the lower panel
only shows non-orphan galaxies; we clearly see that the differences between
models are primarily due to the (treatment of) orphan galaxies, though
significant differences remain even for non-orphans. The difference between the
two panels also indicates that orphans are favourably found in higher mass
haloes while lower mass objects are practically devoid of them, as already seen in \Fig{fig:NxorphMhalo}.

The halo occupation distributions (as well as clustering properties) of all the
models will be further analyzed in a spin-off project of this collaboration
(Pujol et al., in prep.).

\section{Summary \& Discussion}\label{sec:discussion}

We have brought together \Nmodels\ models for galaxy formation in simulations of
cosmic structure formation, i.e. \NSAMs\ SAM and \NHODs\ HOD models.  In this
inaugural paper we presented the models and undertook the first comparison where
the models applied their published parameters (without any recalibration) to the
same small cosmological volume of $(62.5\hMsun)^3$ with halo merger trees
constructed using a single halo finder and tree building algorithm. Hence the
framework that underpins this study was designed to be the same for each
model. This approach allowed us to directly compare the galaxy formation models
themselves leaving aside concerns about cosmic variance, the influence of the
halo finder \citep{Avila14} or tree construction method
\citep{Srisawat13,Lee14}.  However, some teams had to slightly alter this
framework in order to make the catalogues compatible with the assumptions in
their methods (see \App{app:models}).

All of the contributing teams have been provided with a standardized
dark matter halo catalogue and merger tree; they were asked to
undertake their currently favoured model and were explicitly told the
underlying cosmology and mass resolution of the simulation to be
used. They supplied returns in a specified format and the analysis was
performed on these files using a single common analysis pipeline. This
approach has proved highly successful for other related scientific
issues in the past \citep[e.g.][]{Frenk99, Heitmann08cccp,
  Knebe11,Onions12,Srisawat13,Kim14}.

This paper should be viewed as the first in a series emerging out of the
nIFTy cosmology workshop\footnote{http://www.popia.ft.uam.es/nIFTyCosmology}.
A number of spin-off projects were also initiated at the meeting, including
more detailed studies of cold vs. hot gas properties, correlation functions, dust effects,
disk instabilities, and -- last but not least -- how to define a common calibration
framework. The results will be presented in future papers.

For the present paper each team applied their published calibration values to
the supplied cosmological model as specified in \App{app:models} which
also describes each model's specific choice of parameters. Each team
tends to use their own personal preference of which observables to
tune their model to, and these observables often require additional
processing to produce from the more physically fundamental quantities
studied here. For instance, the calculation of luminosities requires
the adoption of a particular stellar population synthesis model as
well as a certain dust model. Even the derivation of stellar mass
demands a choice for the stellar initial mass function.  We
deliberately deferred from studying magnitudes to avoid the
accompanying layer of complexity; and we left the preference for the
initial stellar mass function at the modeller's discretion.  This
choice of calibration freedom was made deliberately in order not to
favour any particular model if we happened to make similar (somewhat
arbitrary) post-processing choices.

We have explicitly chosen not to overlay our figures with observational
data for precisely the same reason: such data requires somewhat
arbitrary (reverse) conversion from the observed quantities and this
conversion may bias the reader in favour of a particular model that
happened to convert using the same approach (or happened to tune their
model to this particular observational quantity). We reserve such
comparisons to future work where we will consider the full range of
such conversions and include a careful review of the observational
literature on this point. But we nevertheless like to remind  the  reader that 
each  model has  been compared with a variety of 
observational data and that these comparisons are published. 

\bigskip

Given the variety in models and calibrations, the agreement found here
is gratifying
although a number of discrepancies exist, as summarized here (and discussed
below).

\paragraph*{Stellar Component}
The stellar mass functions at $z=0$ of all the models lie within a
range of around a factor of 3 in amplitude across the faint end of the curve and
have somewhat different effective breaks at the high mass end.
The star formation rate density and specific star formation rate
by halo mass is broadly similar for all the SAMs,
with the main difference being in normalisation, which can vary by a
little under an order of magnitude near the peak of the star formation
rate density curve and a factor of three elsewhere.

\paragraph*{Galaxies}
For most of the models considered here, galaxies without a surviving
dark matter halo (so-called `orphan galaxies') dominate the number
counts within each host halo, accounting for between 40 and 75 per
cent of all galaxies in main haloes above $10^{13}\hMsun$ at redshift $z=0$. The
treatment of these galaxies and their eventual fate differs
dramatically between the various models and is also expected to be
strongly dependent upon the resolution of the simulation.

\paragraph*{Galaxy-Halo Connection}
All the models populate the supplied trees adequately, i.e. all haloes found in the simulation contain a galaxy. We further found that
the specific frequency of galaxies, i.e. the number of galaxies per halo mass,
is constant above the completeness limit of the simulation although the average
number of galaxies per dark matter halo mass varies by around an order of magnitude across the
models, if orphans are included; otherwise the variation is reduced to a factor of about two.

\bigskip



When interpreting the results one needs to always bear
in mind that all models were used as originally tuned in the
respective reference paper, i.e. the way this comparison has been
designed might lead to scatter across models that is larger than the
scatter due to different implementations of the same physics within
them.  The factors entering into the model-to-model variations seen
here are differences due to a) models not being tuned to the same
observational constraints, b) models being tuned to different
cosmologies, c) the choice for the halo mass definition, d) the choice for the applied initial stellar mass function, and e) models not being
optimally tuned (for the merger tree structure at hand).  Elaborating on
these points:

\paragraph*{a) observational constraints:} Using different observational
constraints should not be a primary source of variation, at least as
long as constraints for all the relevant physical quantities are
included. For instance, the work of \citet[][using SMF+Bband+Kband
  constraints from $z=3$ to $z=0$]{Henriques13} and \citet[][using
  SMF+red fraction constraints]{Henriques15} lead to convergent
results, if a proper assessment of the observational uncertainties is
performed; the authors also state that they need the combination of
properties to arrive at converged likelihood regions in parameter
space.  However, the models included here show an even greater variety
of (potentially mutually exclusive) observational constraints and
hence we cannot exclude that those differences contribute
significantly to the scatter.

\paragraph*{b) cosmology:} Differences due to cosmology can often (but not
always) be absorbed by re-tuning the physical parameters of the
model. However, without re-tuning (as here), cosmology can make a big
difference; this can be seen, for instance, from the significant
changes in parameter values required to get the same stellar mass
function at $z=0$ for different cosmologies
\citep[see][]{Wang08b,Guo13,gp14}.

\paragraph*{c) halo mass definition:}  As can be verified in \Tab{tab:models} another difference across models
is the applied definition for halo masses. Furthermore these
differences can increase with redshift: for instance, \Mcrit\ only
depends on the evolution of $\rho_{c}(z)$ whereas \MBN\ has an
additional dependency on cosmology encoded in the overdensity parameter
$\Delta_{\rm BN98}(z)$. We confirmed that there are variations due to it, but these are not sufficiently large
to explain all of the scatter.

\paragraph*{d) initial mass function:} As for different cosmologies and halo mass definitions, the assumption
of different IMFs can be compensated for when calibrating the model: the
observational data set used for the calibration (and the initial mass function
assumed in its preparation) will determine the values of the model parameters --
whatever the assumption for the model IMF.  However, the model's stellar masses
(and other quantities not studied here such as the recycled fraction of gas, the
amount of energy available for supernovae, chemical enrichment, etc.) will
certainly be affected by the IMF choice. We can confirm that this has an
effect but is not the primary source of the variations between models.

\paragraph*{e) tuning:} This is by far the most decisive factor for the scatter
\citep{Henriques09,mutch_sam_2013}. It has been shown by, for instance,
\citet{Henriques09} that \DLB\ and an earlier version of the \sage\ model could
be brought into better agreement with the observational data by re-tuning their
parameters optimally. Further, \citet{Lee14} have shown that differences in
merger trees could be overcome by re-tuning the model parameters. This directly
applies to the comparison presented here: all models have been designed and
tested using different simulations and merger tree (structures), but were not
allowed to re-adjust their parameters for this initial project. We have seen
that this has a very strong influence on the stellar mass function and is potentially the
main source of the scatter seen in the plots throughout this paper.  Future
papers in this series will investigate the degree to which the models can be
brought into agreement, and the extent to which they still differ, once retuned
to the same set of observational constraints.
\bigskip

We deliberately did not include any comparison with luminosity-based
properties as their calculation involves another layer of complexity,
e.g. stellar population synthesis, dust extinction, etc. However -- as stated in  \App{app:models} -- some models are using
luminosity-related quantities to constrain their model parameters. Also,
one should not neglect the additional difficulties encountered when
moving from intrinsic galaxy properties such as mass to directly
observable quantities such as luminosity or colour. Conversely, the
stellar mass of a galaxy is not directly observable and hence any
derivation of it relies on modelling itself; therefore -- as
highlighted a couple of times before already -- all observational data
comes with its own error estimates that can be as large as 0.5 dex at
$z=0$. All of this will certainly leave its imprint on the models
presented here.

\section{Conclusions}\label{sec:conclusions}

We conclude that applying galaxy formation models without due consideration to
calibration with respect to cosmology, resolution, and -- most importantly --
merger tree prescription leads to scatter that could otherwise be avoided
\citep[see, e.g.][]{Fontanot09,DiazGimenez10,DeLucia11,Contreras13,Lu14,Lee14}.
But the need for re-calibration should not be viewed as a flaw of the models: it
is a necessary step required to match the model to the particular observational
data sets that are chosen to underpin the model.  The fact that a good match can
be obtained is itself a non-trivial success of the model which indicates that
the models capture the underlying key physical phenomena correctly. The
(adjusted) parameters then place bounds upon the relevant physics, and the
models can be used to test astrophysics outside that used in the calibration
step.

We close by mentioning again that this work only forms the initial step in a
wider and long overdue programme designed to inter-compare current SAM and HOD
models. The next stage is to calibrate all the models to a small, well
specified, set of training data, such as for instance the stellar mass function
at $z=0$ and $z=2$ before re-comparing the models on the other statistics shown
in this work. This approach will likely significantly narrow the spread of the
returned data on these physical quantities. It will also allow a more detailed comparison
on such observationally interesting measures as hot and cold gas fractions, gas
metallicity, galaxy sizes and morphologies, etc.  The work for this has been
started and will form the basis of a future workshop.

\section*{Acknowledgments}
The authors would like to express special thanks to the Instituto de Fisica Teorica (IFT-UAM/CSIC in Madrid) for its hospitality and support, via the Centro de Excelencia Severo Ochoa Program under Grant No. SEV-2012-0249, during the three week workshop `nIFTy Cosmology' where this work developed. We further acknowledge the financial support of the 2014 University of Western Australia Research Collaboration Award for `Fast Approximate Synthetic Universes for the SKA', the ARC Centre of Excellence for All Sky Astrophysics (CAASTRO) grant number CE110001020, and the two ARC Discovery Projects DP130100117 and DP140100198. We also recognize support from the Universidad Autonoma de Madrid (UAM) for the workshop infrastructure.

AK is supported by the {\it Ministerio de Econom\'ia y Competitividad} (MINECO) in Spain through grant AYA2012-31101 as well as the Consolider-Ingenio 2010 Programme of the {\it Spanish Ministerio de Ciencia e Innovaci\'on} (MICINN) under grant MultiDark CSD2009-00064. He also acknowledges support from the {\it Australian Research Council} (ARC) grants DP130100117 and DP140100198. He further thanks Nancy Sinatra for the last of the secret agents.
PAT acknowledges support from the Science and Technology Facilities Council (grant number ST/L000652/1).
FJC acknowledges support from the Spanish Ministerio de Econom\'ia y Competitividad project AYA2012-39620.
SAC acknowledges grants from CONICET (PIP-220), Argentina.
DJC acknowledges receipt of a QEII Fellowship from the Australian Government.
PJE is supported by the SSimPL programme and the Sydney Institute for Astronomy (SIfA), DP130100117.
FF acknowledges financial contribution from the grants PRIN MIUR 2009 `The Intergalactic Medium as a probe of the growth of cosmic structures' and PRIN INAF 2010 `From the dawn of galaxy formation'.
VGP acknowledges support from a European Research Council Starting Grant (DEGAS-259586). This work used the DiRAC Data Centric system at Durham University, operated by the Institute for Computational Cosmology on behalf of the STFC DiRAC HPC Facility (www.dirac.ac.uk). This equipment was funded by BIS National E-infrastructure capital grant ST/K00042X/1, STFC capital grant ST/H008519/1, and STFC DiRAC Operations grant ST/K003267/1 and Durham University. DiRAC is part of the National E-Infrastructure.    
The work of BH was supported by Advanced Grant 246797 "GALFORMOD" from the European Research Council.
MH acknowledges financial support from the European Research Council via an Advanced Grant under grant agreement no. 321323 NEOGAL.
PM has been supported by a FRA2012 grant of the University of Trieste, PRIN2010-2011 (J91J12000450001) from MIUR, and Consorzio per la Fisica di Trieste.
NDP was supported by BASAL PFB-06 CATA, and Fondecyt 1150300.  Part of the calculations presented here were run using the Geryon cluster at the Center for Astro-Engineering at U. Catolica, which received funding from QUIMAL 130008 and Fondequip AIC-57.
CP acknowledges support of the Australian Research Council (ARC) through Future Fellowship FT130100041 and Discovery Project DP140100198. WC and CP acknowledge support of ARC DP130100117.
AP was supported by beca FI and 2009-SGR-1398 from Generalitat de Catalunya and project AYA2012-39620 from MICINN.
RAS acknowledges support from the NSF grant AST-1055081.
RSS thanks the Downsbrough family for their generous support.
SKY acknowledges support from the National Research Foundation of Korea (Doyak 2014003730). Numerical simulations were performed using the KISTI supercomputer under the programme of KSC-2013-C3-015. 

The authors contributed to this paper in the following ways: AK \& FRP formed the core team and wrote the paper (with substantial help from PAT). They also organized week \#2 of the nIFTy workshop out of which this work emerged. CS supplied the simulation and halo catalogue for the work presented here. The authors listed in \Sec{app:models} performed the SAM or HOD modelling using their codes, in particular AB, FJC, AC, SC, DC, GDL, FF, VGP, BH, JL, PM, RAS, RS, CVM, and SY actively ran their models with the assistance of JH, MH, and CS. DC and RAS have written \Sec{sec:models}. WC, DC, PJE, CP, and JO assisted with the analysis and data format issues. All authors had the opportunity to proof read and comment on the paper.

This research has made use of NASA's Astrophysics Data System (ADS) and the arXiv preprint server.

\bibliographystyle{mn2e}
\bibliography{archive}

\appendix

\section{Galaxy Formation Models} \label{app:models}

\subsection{\galacticus\ (Benson)}
The \galacticus\ semi-analytic galaxy formation code \citep{Benson12} was designed to be highly modular. Every physical process is implemented through a simple and well-defined interface into which an alternative implementation of a calculation can easily be added. Similarly, the physical description of galaxies is extremely flexible. Each galaxy has a set of components (e.g. disk, dark matter halo, super-massive black hole, etc.) which can be created/destroyed as needed, each of which has a set of properties. The formation and evolution of galaxies is treated by simply defining a set of differential equations for each galaxy. These are all simply fed in to an ordinary differential equation (ODE) solver which evolves them to a specified accuracy, removing any need for fixed timesteps. When running on merger trees derived from $N$-body simulations, halo masses are interpolated linearly in time between available snapshots. In addition to this differential evolution galaxy components can define `interrupts' so that the ODE solver stops, allowing the creation of new components (e.g. the first time gas cools and infalls it needs to create a disk component) or to handle discrete events (e.g. if a merger occurs the ODE solver is interrupted, the merger processed, and then the ODE solver starts up again).

\paragraph*{\it Cooling}
Cooling rates from the hot halo are computed using the traditional cooling radius approach \citep{white91}, with a time available for cooling equal to the halo dynamical time, and assuming a $\beta$-model profile with isothermal temperature profile (at the virial temperature). Metallicity dependent cooling curves are computed using \textsc{CLOUDY} \citep[v13.01,][]{Ferland13} assuming collisional ionization equilbrium.

\paragraph*{\it Star formation}
Star formation in disks is modeled using the prescription of \citet{Krumholz09}, assuming that the cold gas of each galaxy is distributed with an exponential radial distribution. The scale length of this distribution is computed from the disks angular momentum by solving for the equilibrium radius within the gravitational potential of the disk+bulge+dark matter halo system \citep[accounting for adiabatic contraction using the algorithm of][]{Gnedin04}.

\paragraph*{\it Initial mass function}
A \citet{Chabrier03} IMF is used throughout.

\paragraph*{\it Metal treatment}
Metal enrichment is followed using the instantaneous recycling approximation, with a recycled fraction of 0.46 and yield of 0.035. Metals are assumed to be fully mixed in all phases, and so trace all mass flows between phases. Metals affect the cooling rates from the hot halo, and also the star formation rate in disks \citep{Krumholz09}.

\paragraph*{\it Supernova feedback and winds}
The wind mass loading factor, $\beta$, is computed as $\beta = (V_{\rm disk}/250 {\rm km/s})^{-3.5}$ where $V_{\rm disk}$ is the circular velocity at the disk scale radius. Winds move cold gas from the disk back into the hot halo, where it remains in an outflowed phase for some time before being reincorporated and possibly cooling once again. In the case of satellite galaxies, the ouflowing gas is added to the hot halo of the satellite's host.

\paragraph*{\it Gas ejection \& reincorporation}
Gas removed from galaxies by winds is retained in an outflowed reservoir. This reservoir gradually leaks mass back into the hot halo on a timescale of $t_{\rm dyn}/5$ where $t_{\rm dyn}$ is the dynamical time of the halo at the virial radius.

\paragraph*{\it Disk instability}
The \citet{Efstathiou82} criterion is used to judge when disks are unstable, with a stability threshold that depends on the gas fraction in the disk (0.7 for pure gas disks, 1.1 for pure stellar, linearly interpolated in between). When a disk is unstable, it begins to transfer stars and gas from the disk to the bulge on a timescale that equals the disk dynamical time for a maximally unstable disk, and increases to infinite timescale as the disk approaches the stability threshold.

\paragraph*{\it Starburst}
There is no special ``starburst'' mode in \galacticus. Instead, gas in the spheroid forms stars at a rate $\dot{M}_\star = 0.04 M_{\rm gas}/t_{\rm dyn} (V/200 {\rm km/s})^{-2}$, where $t_{\rm dyn}$ is the dynamical time of the spheroid at its half mass radius, and $V$ its circular velocity at the same radius. Starburst-level star formation rates are reached if enough gas is deposited into the spheroid, such as happens after a merger.

\paragraph*{\it AGN feedback}
The mass and spin of black holes are followed in detail, assuming black holes accrete from both the hot halo and spheroid gas. When black holes are accreting from an advection dominated accretion flow, we compute the power of the jet produced by the black hole using the method of \citet{Benson09}. This jet power is used to offset the cooling luminosity in the hot halo (if, and only if, the hot halo is in a hydrostatic phase), thereby reducing the cooling rate onto the galaxy. Additionally, a radiatively-driven wind is launched from the spheroid by the black hole, assuming that a fraction 0.0024 of its radiative output couples efficiently to the outflow.

\paragraph*{\it Merger treatment}
A merger between two galaxies is deemed to be ``major'' if their mass ratio exceeds 1:4. In major mergers, the stars and gas of the two merging galaxies are rearranged into a spheroidal remnant. In other, minor mergers, the merging galaxy is added to the spheroid of the galaxy that it merges with, while the disk of that galaxy is left unaffected.

\paragraph*{\it Substructures}
Substructures are traced using the subhalo information from the $N$-body simulation.

\paragraph*{\it Orphans}
When a subhalo can no longer be found in the $N$-body merger trees, a ``subresolution merging time'' is computed for the subhalo (based on its last known orbital properties and the algorithm of \citet{Boylan-Kolchin08} algorithm. The associated galaxy is then an orphan, which continues to evolve as normal (although we have no detailed knowledge of its position within its host halo) until the subresolution merging time has passed, as which point it is assumed to merge with the central galaxy of its host halo.

\paragraph*{\it Calibration method}
The parameters of galaxy formation physics in \galacticus\ have been chosen by manually searching parameter space and seeking models which provide a reasonable match to a variety of observational data, including the $z=0$ stellar mass function of galaxies \citep{li_distribution_2009}, $z=0$ K and b$_{\rm J}$-band luminosity functions \citep{cole_2df_2001,norberg_2df_2002}, the local Tully-Fisher relation \citep{pizagno_tully-fisher_2007}, the color-magnitude distribution of galaxies in the local Universe \citep{weinmann_properties_2006}, the distribution of disk sizes at $z=0$ \citep{de_jong_local_2000}, the black hole mass--bulge mass relation \citep{haring_black_2004}, and the star formation history of the Universe \citep{hopkins_evolution_2004}. {\sc Galacticus} has also been calibrated to the local stellar mass function using MCMC techniques \citep{benson_building_2014}, but so far only for a simplified implementation of galaxy formation physics. As such, we do not use this MCMC-calibrated version of \galacticus\ here.

\paragraph*{\it Model origin}
The parameters used were calibrated using Monte Carlo trees built using the algorithm of \citet{Parkinson08}.

\paragraph*{\it Modifications to the supplied data}
When importing merger trees for this project, \galacticus\ aims to make minimal changes to the tree structure. Two small modifications are required to ensure consistency of the merger trees. First, where halo $A$ is indicated as being the host of halo $B$, but $A$ is not present in the merger trees (i.e. is not listed as the progenitor of any other halo), then $A$ is assumed to be a progenitor of the same halo of which $B$ is a progenitor. Second, if two haloes are mutual hosts (i.e. $A$ is the host of $B$, while $B$ is the host of $A$), \galacticus\ resolves this inconsistency by reassigning the more massive of the two haloes to be unhosted (i.e. to no longer be classified as a subhalo). No further modifications are made. In particular, this means that subhalo-subhalo mergers are allowed, subhaloes are allowed to become non-subhaloes later in their evolution, halo masses are permitted to decrease with time if indicated by the $N$-body simulation, and subhaloes are permitted to jump between branches of merger trees (and between separate merger trees) if indicated.

\paragraph*{\it Halo finder properties used}
The standard incarnation of \galacticus\ uses \Mbnd\, but runs have been performed for all of the five supplied mass definitions. \galacticus\ further uses the following information from the provided halo catalogues: haloid, hosthaloid, number of particles, mass, radius, concentration, spin parameter, angular momentum, position, and velocity; if any of these values is not supplied (e.g. spin parameter), a random value is drawn from a distribution as measured by cosmological simulations. Further, the peak value of the circular velocity curve and the velocity dispersion are carried through the code, but not used for any calculation. 

\subsection{\galics\ (Cattaneo, Blaizot, Devriendt \& Mamon)} \label{app:galics}
\galics\ is not a simple development of \textsc{GalICS} \citep{Hatton03,cattaneo_etal06},
but a totally new and different code. Its main characteristics are presented here for the first time and hence described in more detail for than the other models. 

\paragraph*{\it Gas accretion}

 A baryonic mass $M_{\rm b} = f_{\rm b}M_{\rm h}$ is assigned to a halo of
  mass $M_{\rm h}$.  Let $T_{\rm reio}$ be the temperature at which the
  intergalactic medium is reionised; $f_{\rm b} = f_{\rm b}(M_{\rm h},z)$ is a function such
  that $f_{\rm b}\sim 0$ at $T_{\rm vir}\ll T_{\rm reio}$ and $f_{\rm b}\sim
  \Omega_{\rm b}/\Omega_{\rm M}$ at $T_{\rm vir}\gg T_{\rm reio}$. Its precise form is
  irrelevant for this article because the $N$-body simulation used for this
  comparison has such poor resolution that it can resolve only haloes with
  $T_{\rm vir}\gg T_{\rm reio}$.  The baryonic mass that accretes onto a halo
  between two timesteps is the maximum between $\Delta M_{\rm b}$ and zero.  A
  fraction $f_{\rm hot} = f_{\rm hot}(M_{\rm h},z)$ of this gas is shock heated and
  added to the hot halo. The rest is put into cold streams, which accrete
  onto the disk of the central galaxy on a dynamical timescale. 
  Cosmological hydrodynamic simulations show that $f_{\rm hot}\sim 0$ at $M_{\rm h}\lsim 3\times10^{10}\,M_\odot$ and that 
  $f_{\rm hot}\sim 1$ at $M_{\rm h}>3\times 10^{12}\,M_\odot$ \citep{ocvirk_etal08,nelson_etal13}. The halo mass $M_h$ at which
  $f_{\rm hot}=0.5$ increases with redshift (at least at $z>2$).
  In this paper, we assume that $f_h$ is a linear ramp between the halo
  mass $M_{\rm shock\,min}$  at which some gas begins to be shock heated and the halo mass $M_{\rm shock\,max }$ above which the infalling gas is entirely shock heated.
 We assume that  $M_{\rm shock\,max }= M_{\rm shock}+\alpha(z-z_c)$ for $z>z_{\rm c}$ and that $M_{\rm shock\,max }= M_{\rm shock}$ for
 $z\le z_{\rm c}$ where $z_{\rm c}$, $M_{\rm shock}$ and $M_{\rm shock\,min}$ are free parameters of the model to be determined by fitting observational
 data. On a theoretical ground, $M_{\rm shock}$ is the critical mass halo mass at which the cooling time equals the gravitational compression time and
 $z_{\rm c}$ is the critical redshift at which $M_{\rm shock}$ equals the non-linear mass \citep{dekel_birnboim06}.
 
 \paragraph*{\it Cooling}

Following \citet{cattaneo_etal06} and \citet{dekel_etal09}, we assume that galaxies are built through the accretion of cold gas and that hot
 gas never cools. This is an extreme assumption, but is in good agreement
 with the galaxy colour-magnitude distribution, while, if we let the gas cool, the predictions of the model are in complete disagreement with the
    observations \citep{cattaneo_etal06}.
 Introducing cooling makes sense
    only if one has a physical model of how AGN feedback suppresses
    it. Attempts in this direction have been made, starting from
    \citep{Croton06}, but the physics are uncertain. Hence it was considered that
    not much would be gained from
    implementing them in \galics.

\paragraph*{\it Star formation}

Following \citep{bigiel_etal08},
we assume a constant star formation
  timescale of $t_{\star{\rm\,d}} = 2.5\times 10^9{\rm\,yr}$ for disks and $t_{\star{\rm b}} =
  2.5\times 10^8{\rm\,yr}$ for bulges (merger-driven starbursts) for all gas. $\dot{M}_\star = M_{\rm gas}/t_\star$, where $M_{\rm gas}$ is the gas mass in the
  component.
  Star formation is suppressed when $\Sigma_{\rm gas}<1\,{\rm M}_\odot{\rm\,pc}^{-2}$. 
  Disks with $\Sigma_{\rm gas}>20\,{\rm M}_\odot{\rm\,pc}^{-2}$ are assumed to be in a starburst mode and are assigned the same star formation timescale as bulges.
  Not only are these assumptions observationally motivated, but also they have the practical 
   advantage that star formation rates are affected only mildly by errors in the modelling of disk
  sizes. 
  
\paragraph*{\it Initial mass function}

As in \citet{cattaneo_etal06}, we assume a \citet{kennicutt83} stellar initial mass function.
The IMF that our model uses is determined by our choice of input stellar evolution tables (see Metal treatment below).

\paragraph*{\it Metal treatment}

Stars are created
  with metallicity that equals that of the gas from which they form.  Stellar
  evolution is computed following \citet{pipino_etal09},  who tabulated the mass loss rate and the metal yield of a stellar
  population as a function of its age and metallicity. 
We can select different stellar evolution models/IMFs by replacing the input file with the stellar evolution tables.

\paragraph*{\it Supernova feedback and winds}

Matter ejected from  stars is mixed into the interstellar medium. The mass outflow rate through stellar feedback is assumed to be given by
$\dot{M}_{\rm w}=(M_{\rm h}/M_{\rm SN})^\beta\dot{M}_\star$, where $\dot{M}\star$ is the star formation rate, while
 $M_{\rm SN}$ (the halo mass at which outflow rate equals star
  formation rate) and $\beta<0$ are free parameters of the model.
  We consider a scaling with $M_h$ rather than with $v_{\rm c}$ because the latter would
  lead to weaker feedback at high redshift for a same halo mass.

\paragraph*{\it Gas ejection \& reincorporation}

As our model does not include cooling, it is irrelevant whether the gas that is blown out of the galaxy escapes from the halo or whether it is mixed to the
hot component, as we assume currently.

\paragraph*{\it Disk structure}

Gas that accretes onto a galaxy contributes to the
  disk's angular momentum ${\bf J}_d$ with specific angular momentum that
  equals that of the halo at the time of accretion. The disk radius $r_d$ is
  computed by assuming that the disk is exponential and by solving the
  equation
\begin{equation}
|{\bf J}_d| = {M_d\over r_{\rm d}^2} \int_0^\infty  {\rm e}^{-{r\over r_{\rm d}}}[v_{\rm d}^2(r)+v_{\rm b}^2(r)+v_{\rm h}^2(r)]^{1\over 2}r^2{\rm\,d}r,
\end{equation}
where the disk term $v_{\rm d}$, the bulge term $v_{\rm b}$, and the halo term $v_{\rm h}$ are
the three terms that contribute in quadrature to disk's rotation curve.  The
halo term $v_h$ is computed by assuming that the halo follows an NFW profile
modified by adiabatic contraction \citep{Blumenthal86}.

\paragraph*{\it Disk instability}

Disk instabilities are not enabled yet because we first want the test the properties of disks and bulges in a scenario in which mergers are the only mechanism for the formation of bulges.

\paragraph*{\it Starburst}

Gas is starbursting after a major merger or when $\Sigma_{\rm gas}>100\,M_\odot{\rm\,pc}^{-2}$. In these cases, we lower the star formation timescale from 
$t_{\star{\rm\,d}} = 2.5\times 10^9{\rm\,yr}$ to $t_{\star{\rm b}} = 2.5\times 10^8{\rm\,yr}$.

\paragraph*{\it AGN feedback}

Two types of AGN feedback have been suggested in the literature: one is related powerful AGN, the other is mainly linked to Fanaroff-Riley type I radio sources. The need for the former is unclear while the second is essential to explain the absence of cooling flows in massive systems \citep{cattaneo_etal09}.
Our code does not contain any explicit model of AGN feedback. However, the assumption that the hot gas never cools is an implicit model for AGN feedback.
It corresponds to the assumption that $P_{\rm jet}=L_{\rm X}$, i.e., that that jets  self-regulate so that the power they damp into the hot gas matches exactly that which is lost to X-rays (see, e.g., \citealp{cattaneo_teyssier07}). Observationally, these quantities are equal to $\sim 10\%$ (\citealp{cattaneo_etal09}, and references therein). Hence, this is a reasonable first approximation.

\paragraph*{\it Merger treatment}
In minor mergers (mass ratio $< 1:3$), the disk and the bulge
  of the smaller galaxy are added to the disk and the bulge of the larger
  galaxy, respectively. In major mergers, the galaxies are scrambled into one
  large bulge and a fraction $\epsilon_\bullet$ of their gas content feeds
  the growth of a central supermassive black hole
  \citep{cattaneo_etal09}. The size of the merger remnant is computed by
  applying an energy conservation argument that has been tested in hydrodynamic
  simulations \citep{covington_etal11,oser_etal12}. 
  Our calculation assumes that merging pairs start with zero energy of interaction at infinity
  in agreement with what we see in cosmological hydrodynamic simulations and with the constraints from the
  mass-size relation
  \citep{shankar_etal13,shankar_etal14}.

\paragraph*{\it Substructures}

Substructures are traced from the $N$-body simulation. No gas accretion is allowed onto them but
  cold gas already accreted keeps streaming onto satellite galaxies as long as the host halo mass is
  $M_h<M_{\rm shock\,max}$.

\paragraph*{\it Orphans}

Whenever the code encounters a halo with more than one progenitor, 
it computes the dynamical friction time for all progenitors bar the most massive one. The dynamical friction time is computed with \citet{jiang_etal08}'s formula as in \citet{Cattaneo11}. This formula is a modification of \citet{chandrasekhar43}'s (see \citealp{2008gady.book.....B}) that includes the
effects of orbital eccentricity and that has been calibrated on the results of cosmological simulations. Our calculation of the dynamical friction time contains a fudge factor $\epsilon_{\rm df}$ that is a free parameter of the model. A halo/subhalo merges with its descendent according the merger tree only after this time has elapsed. The halo catalogues are completed with the creation of a ghost halo at all timesteps between the one when the halo/subhalo was last detected and the elapsing of the dynamical friction time. If $\epsilon_{\rm df}=0$, the dynamical friction time is set equal to zero. Hence halo/subhalo mergers result into immediate galaxy mergers.
 
\paragraph*{\it Parameters and calibration method}

For this project, the code has been calibrated manually  by requiring that it fits the evolution of the galaxy stellar mass function at $0<z<2.5$ 
\citep{baldry_etal12,ilbert_etal13,Bernardi13}. 
The parameters that have been tuned and that are relevant for this comparison are:
\begin{itemize}
\item{The mass $M_{\rm shock\,min}$ above which shock heating begins and the three parameters $M_{\rm shock}$,
$\alpha$ and $z_{\rm c}$ that determine the critical mass above which shock heating is complete.}
\item{The star formation timescales for disks ($t_{\star{\rm\,d}}$) and bulges (merger-driven starbusts; $t_{\star{\rm\,d}}$), and the gas surface density threshold for star formation
$\Sigma_{\rm min}$.}
\item{The mass $M_{\rm SN}$ at which outflows rate equal star formation rate and the exponent $\beta$ of the scaling of mass-loading factor with halo mass.}
\item{The dynamical friction parameter $\epsilon_{\rm df}$.}
\item{The critical mass ratio $\mu$ that separates minor and major mergers.}
\end{itemize}
Here we have used $M_{\rm shock\,min} = 10^{10.5}\,M_\odot$,  $M_{\rm shock} = 10^{12.3}\,M_\odot$, $\alpha = 0.3$, $z_{\rm c}=1.5$ and
$M_{\rm SN} = 4\times 10^{11}\,M_\odot$ (for $h=0.678$). The best fit to the galaxy stellar mass function of \citet{baldry_etal12} is found for 
$\beta = -2$ and $\epsilon_{\rm df}>0$ (our best fit is for $\epsilon_{\rm df}\ll 1$; hence we apply Occam's razor and prefer a model with no ghost/orphans).
The star formation timescales for disks ($t_{\star{\rm\,d}} = 2.5\times 10^9{\rm\,yr}$) and merger-driven starbursts ($t_{\star{\rm\,d}} = 2.5\times 10^8{\rm\,yr}$)
and the gas surface density threshold for star formation
$\Sigma_{\rm min}=1\,M_\odot{\rm\,pc}^{-2}$ were not tuned but they were set to the values found observationally by \citet{bigiel_etal08}.
The critical mass ratio of $\mu = 1/3$ that separates minor and major mergers was not tuned either.

\paragraph*{\it Model origin}

The model was conceived to be run on merger trees from $N$-body simulations.

\paragraph*{\it Modifications to the supplied data}

None besides the creation of ghost haloes if $\epsilon_{\rm df}>0$.

\paragraph*{\it Halo finder properties used}

They halo properties that enter the \galics\ SAM are the halo mass $M_{\rm halo}$, its radius $R_{\rm halo}$ and
the halo angular momentum ${\bf J}_{\rm halo}$ (the virial velocity is computed with the formula $v_{\rm c}^2={\rm G}M_{\rm halo}/R_{\rm halo}$). Positions and velocities are used to compute dynamical friction and tidal stripping. Halo concentration is needed to compute dynamical friction, tidal stripping, and the radii of disks and bulges, but it is not provided in the \subfind\ catalogues. Hence, it has been computed with the fitting formulae of \citet{munoz_etal11}.
\galics\ flags certain haloes as bad. Bad haloes have positive total energy or they are tidal features that broke off a good halo, in which case we assume
that no galaxy is formed in them. Normally, there is no gas accretion onto haloes that are flagged as bad. We could not impose this condition here because halo kinetic and potential energies were not provided but this is unlikely to affect our results because the halo/subhalo finder \subfind\ automatically removes unbound particles.

\subsection{\morgana\ (Monaco \& Fontanot)}
In this paper we use the version of \morgana\  that has been presented in \citet{LoFaro09}. Its main properties are detailed below.

\paragraph*{\it Cooling}
Described in detail in \cite{Viola08}, the model
assumes that the gas is in hydrostatic equilibrium within an NFW halo
and polytropic. Its thermal state responds to the injection of
energy. The cooling radius is treated as a dynamical variable. Disk
sizes are computed using the \cite{Mo98} model, taking into account
the presence of a bulge.

\paragraph*{\it Star formation}
Star formation is treated following the results of the model
by \cite{Monaco04}.  

\paragraph*{\it Initial mass function}
A \citet{Chabrier03} IMF is assumed.

\paragraph*{\it Metal treatment}
An Instantaneous Recycling Approximation is
assumed, only global metallicities are followed.

\paragraph*{\it Supernova feedback and winds}
Feedback and winds are treated following the results of the model
by \cite{Monaco04}.  The ejection rate of gas into the halo is always
equal to the star formation rate.

\paragraph*{\it Gas ejection \& reincorporation}
The halo acts as a
buffer for feedback. Galaxies inject (hot or cold) mass and (thermal
or kinetic) energy in the halo, when the typical specific energy is
larger than the escape velocity of the halo the gas is ejected into
the IGM.  Half of this ejected gas is re-accreted whenever the circular
velocity of the halo grows larger than the velocity at ejection.

\paragraph*{\it Disk instability}
The \cite{Efstathiou82} criterion is used.

\paragraph*{\it Starburst}
The size of star-forming gas in bulges is
estimated by assuming that its velocity dispersion is determined by
turbulence injected by SN feedback. SFR is computed using a Kennicutt
law, the high gas surface densities guarantees short gas consumption
timescales.

\paragraph*{\it AGN feedback}
Star formation in the bulge
is responsible for the loss of angular momentum of a fraction of gas;
the reservoir of such low-angular momenum gas is accreted onto the BH
on a viscous time-scale. The accretion is Eddington-limited.  AGN
feedback can be in the quasar mode or in the `jet' (radio) mode if the
accretion rate in units of the Eddington one is higher or lower than
0.01.

\paragraph*{\it Merger treatment}
Major mergers transform the whole resulting
galaxy into a bulge, a minor merging satellite is put in the galaxy
bulge. A fraction of 80\% of stars in minor merging satellites is
positioned into the stellar halo component.

\paragraph*{\it Substructures}
The $N$-body simulation subhalo information is not used to follow the orbital evolution of substructures; they are rather tracked analytically. In detail, whenever a DM halo merges with a larger structure, the orbital decay of its galaxies is computed using an updated version of the fitting formulae provided by \citet{Taffoni03}, calibrated on the results of numerical simulations, which account for dynamical friction, tidal stripping and tidal disruption.

\paragraph*{\it Orphans}
All satellite galaxies are effectively treated as `orphans', i.e. the subhalo information is not explicitly used in modeling their evolution.

\paragraph*{\it Calibration method}
Parameters were manually to fit the
stellar mass function of galaxies at $z=0$ and the evolution of the
SFR density; see \citet{LoFaro09} for more details.

\paragraph*{\it Model origin}
\morgana\ has been designed to work with the
idealized merger trees obtained with the \pinocchio\ code 
\citep{Monaco02}, so the application to numerical merger trees
requires a significant amount of cleaning of the trees.

\paragraph*{\it Modifications to the supplied data}
It is assumed that the merger trees of haloes are mirrored by the evolution of their main haloes, so that main haloes either have no progenitors or have at least one main halo progenitor.  When this is not true, the merger tree is modified to adapt to this requirement. This includes the reabsorption of any substructure that descends into a main halo (a `backsplash' halo) or the exchange of the role of main halo between two haloes. 

\paragraph*{\it Halo finder properties used}
As a halo mass, our adopted default choice is the FOF mass \Mfof, but we run the model on all five mass definitions. However, the halo mass used to obtain the budget of baryons available to the halo is assumed never to decrease, so it is computed as the maximum of the total mass that the halo and all its progenitors have got in the past. In addition to this \morgana\ made use of the provided haloid, hosthaloid, number of substructures, number of particles, position, and velocity.

\subsection{\sag\ (Cora, Vega-Mart\'inez, Gargiulo \& Padilla)}
While the \sag\ model originates from a version of the Munich code \citep{springel_subfind_2001}, it has seen substantial development and improvement; we are going to explain its features (and derivations from the Munich and other models) here.

\paragraph*{\it Cooling}
A gaseous disk with an exponential density profile is formed from gas inflow generated as the result of the radiative gas cooling suffered by the hot gas in the halo \citep{Springel01}. The metal-dependent cooling function is estimated by considering the radiated power per chemical element given by \cite{foster_xray_2012} multiplied by the chemical abundances of each element in the hot gas, thus being completely consistent with the metallicity of this baryonic component.

\paragraph*{\it Star formation}
When the mass of the disk cold gas exceeds a critical limit, an event of quiescent star formation takes place, as in \citet{Croton06}.

\paragraph*{\it Initial mass function}
\sag\ assumes a Salpeter IMF.

\paragraph*{\it Metal treatment}
\sag\ includes a detailed chemical implementation \citep{cora_sam_2006}, which estimates the amount of metals contributed by stars in different mass ranges. Metals are recycled back to the cold gas taking into account stellar lifetimes \citep{Padovani93}. The code considers yields from low- and intermediate-mass stars \citep{Karakas2010}, mass loss of  pre-supernova stars \citep{Hirschi2005}, and core collapse supernovae \citep[SNe CC,][]{Kobayashi2006}. Ejecta from supernovae type Ia (SNe Ia) are also included \citep{Iwamoto99}; SNe Ia rates are estimated using the single degenerate model
\citep{Lia2002}.

\paragraph*{\it Supernova feedback and winds}
Feedback from SNe CC is modeled following \citet{DeLucia04b}. The amount of reheated gas is proportional to the energy released by SNe CC and inversely proportional to the square of the halo virial velocity. \citet{cora_sam_2006} adapted this prescription according to the chemical model implemented, such that the energy contribution of SNe CC occurs at the moment of their explosions, for which the lifetimes of massive stars are considered.  SNe feedback produces outflows of material that transfer the reheated cold gas with its metal content to the hot gas phase; this chemical enrichment has a strong influence on the amount of hot gas that can cool, since the cooling rate depends on the hot gas metallicity.

\paragraph*{\it Gas ejection \& reincorporation}
\sag\ assumes a `retention' scheme in which the cold gas reheated by SNe feedback is transfered to the hot gas phase, being available for further gas cooling that takes place only in central galaxies (`strangulation' scheme).

\paragraph*{\it Disk instability}
When a galactic disk is sufficiently massive that its self-gravity is dominant, it becomes unstable to small perturbations by satellite galaxies. The circular velocity of the disk involved in the stability criterion of \citet{Cole00} is approximated by the velocity calculated at three times the disk scale-length, where the rotation curve flattens \citep[see][for details concerning disk features]{tecce_sam_2010}. We model the influence of a perturber by computing the mean separation between galaxies in a group; the instability is triggered when this separation is smaller than a certain factor (a free parameter of the model) of the disk scale radius of the perturbed galaxy. When an unstable disk is perturbed, existing stars are transferred to the bulge component along with the cold gas that is consumed in a starburst.

\paragraph*{\it Starburst}
Starbursts occur in both mergers and triggered disk instabilities and are the only channel for bulge formation. During a starburst episode, we consider that the cold gas available in the bulge that has been transfered from the disk, referred to as bulge cold gas, is gradually consumed. The period for star formation is chosen to be the dynamical time-scale of the disk. However, as the starburst progresses, effects of supernovae feedback, recycling of gas from dying stars and black hole growth modify the reservoir of cold gas of both disk and bulge, thus also changing the time-scale of the starburst \citep{gargiulo_2014}.

\paragraph*{\it AGN feedback}
\sag\ includes radio-mode AGN feedback following \citet{Croton06} as described in \citet{lagos_sam_2008}, which reduces the amount of gas that can cool thus providing a mechanism for regulating star formation in massive galaxies. AGNs are produced from the growth of central black holes, for which two possible mechanisms are considered: i) infall of gas towards the galactic centre, induced by merger events or disk instabilities, and ii) the accretion of gas during the cooling process. The current version of the code \citep{gargiulo_2014} considers that mass accretion rates in the latter case depends on the square of the virial velocity \citep[being consistent with a Bondi-type accretion,][]{bondi_bh_1952}, instead of on the cube of the velocity as in \cite{lagos_sam_2008}, in order to prevent super massive black holes at the centre of cluster-dominant galaxies to grow unrealistically large, at the expense of the intracluster medium. 

\paragraph*{\it Merger treatment}
The galaxy inhabiting the subhalo is assumed to merge with the central galaxy  of their host subhalo on a dynamical friction time-scale \citep{Binney87}. The merger can be major or minor depending on the the baryonic mass ratio between the satellite galaxy and the central galaxy. If this mass ratio is larger than 0.3, then the merger is considered a major one. In this case, stars and cold gas in the disk of the remnant galaxy are transferred to the bulge, with the latter being consumed in a starburst. The presence of a starburst in a minor merger will depend on the fraction of cold gas present in the disk of the central galaxy, as implemented by \citet{lagos_sam_2008}. In minor mergers, only the stars of the merging satellite are transferred to the bulge component of the central galaxy.

\paragraph*{\it Substructures}
Substructures are followed from the $N$-body simulation. The supplied data has not been modified. Those branches of merger trees that start with subhaloes give place to spurious galaxies with neither cold gas nor stars since gas cooling does not take place in galaxies residing in subhaloes. Those merger trees are ignored and we discard those galaxies in the output of \sag. 

\paragraph*{\it Orphans}
Orphan galaxies are created when their subhaloes are no longer identified after merging with a larger one because they lose mass as a result of tides. The position and velocity of orphan galaxies are estimated  assuming a circular orbit with a velocity given by the virial velocity of the parent subhalo and a decaying radial distance determined by dynamical friction \citep{Binney87}.

\paragraph*{\it Calibration method}
Calibrations of SAG are performed using the `Particle Swarm Optimization' technique, which yields best-fitting values for the free parameters of the model allowing it to achieve good agreement with specific observational data \citep{ruiz2014}. The free parameters that have been tuned are those related with star formation efficiency, the SNe feedback efficiencies that control the amount of disk cold gas and bulge cold gas reheated by the energy generated by SNe formed in the disk and in the bulge, respectively, parameters involved in the AGN feedback, that is, the fraction of cold gas accreted onto the central supermassive black hole (SMBH) and the efficiency of accretion of hot gas onto the SMBH, and finally, the factor involved in the distance scale of perturbation to trigger disk instability.

We calibrate the free parameters of the SAG model considering a set of observational constraints that involve the $z=0$ luminosity function in the {\em r}-band ({\em r}-band LF), the relation between the mass of the central SMBH and the bulge mass (BHB relation), the redshift evolution of SNe Ia and SN CC rates, and the $[\alpha / {\rm Fe}]$-stellar mass relation of elliptical galaxies. The first two constraints help to tune the free parameters associated to the star formation efficiency and the SNe and AGN feedback. The third one allows to fix the fraction of binary stars that explode as SNe Ia, and therefore, the amount of iron recycled into the interstellar medium. The last one establishes more restriction to the efficiency of SNe feedback arising from stars formed in the bulge. The observational data used are the {\em r}-band LF of \citet{Blanton2005}, the BHB relation given by \citet{HaringRix2004} and \citet{Sani2010}, the compilation of rates for both SNe Ia and SNe CC given by \citet{Melinder2012}, and the $[\alpha/{\rm Fe}]$ ratio of elliptical galaxies presented in \citet{Thomas2010} and \citet{Arrigoni2010}.

\paragraph*{\it Model origin}
While the \sag\ model originates from a version of the Munich code \citep{Springel01}, based on $N$-body simulations, it has seen substantial development and has been improved with a detailed chemical implementation \citep{cora_sam_2006}, the inclusion of AGN feedback and disk instabilities \citep{lagos_sam_2008}, a detailed estimation of disk scale-lengths \citet{tecce_sam_2010}, and a gradual star formation during starbursts \citep{gargiulo_2014}, among other aspects that are not taken into account in the version used for the current comparison, like the effects of accretion with misaligned angular momenta on the properties of galactic disks \citep{padilla_flip_2013}, a star formation dependent top-heavy integrated galactic IMF \citep{gargiulo_2014}, estimation of nebular emission of star-forming galaxies \citep{orsi_2014}, and environmental effects such as tidal stripping and ram pressure stripping (Cora et al., in prep.). 

\paragraph*{\it Modifications to the supplied data}
As mentioned above, branches of the merger tree that start with subhaloes are ignored. However, they contribute to generate the galaxy population if gradual removal of hot gas is allowed since, in that case, satellites receive cooling flows (which, however, is not the case in the version of the model considered here). No other modifications have been made.

\paragraph*{\it Halo finder properties used}
\sag\ uses the bound mass \Mbnd\ to construct its galaxies. Further properties entering \sag\ are the haloid, hosthaloid, number of substructures, radius, position, velocity, and spin parameter.

\subsection{\somerville\ (Somerville \& Hirschmann)} \label{sec:somerville}
The \somerville\ SAM was first presented in \citet{Somerville99}
and significantly updated in \citet[S08][]{Somerville08},
\citet{Somerville12}, \citet[][H12]{Hirschmann12}, and recently
in \citet[][P14]{Porter14}. The \somerville\ SAM includes the
following physical processes: (1) shock heating and radiative cooling
of gas, (2) conversion of cold gas into stars via an empirical
`Kennicutt-Schmidt' relation, (3) starbursts, black hole feeding, and
morphological transformation due to galaxy mergers 4) bulge and black
hole growth via `disk instabilities' 5) metal enrichment of the
interstellar and intracluster media by supernovae using the
instantaneous recycling approximation 6) galactic outflows driven by
stars and supernovae (7) galactic outflows driven by `quasar mode'
black hole activity, and heating of the hot intracluster and
intragroup medium by `radio mode' black hole activity. The code used
here adopts the modifications relative to the S08 model suggested by
\citet{Hirschmann12} in order to match the observed luminosity
function of AGN: massive seed black holes, black hole feeding via disk
instabilities, and suppressed black hole feeding in mergers at low
redshift (see H12 for details). These modifications have almost no
impact on galaxy properties but do affect black hole growth.

\paragraph*{\it Cooling}

The cooling model is based on the spherically symmetric cooling flow
model originally presented in \citet{white91}, and is described in
detail in S08 and P14. We assume a singular isothermal profile for the
hot gas density distribution and adopt the metallicity dependent
cooling functions of \citet{sutherland93}. If the cooling radius is
larger than the virial radius, gas is accreted on a halo dynamical
time. Cooling gas is only accreted onto `central' galaxies. 

\paragraph*{\it Star formation}
As gas cools, we assume that it settles into a rotationally-supported
exponential disk. We use a prescription based on the work of
\citet{Mo98} and \citet{Somerville:2008b} to compute the radial
size of the gas disk.

We allow for two modes of star formation: `disk mode' star formation,
which occurs in disks at all times as long as cold gas above a
critical surface density is present, and `burst mode' star formation,
which occurs after two galaxies merge. In the `disk mode' the star
formation rate density is dependent on the surface density of cold gas
in the disk, following the empirical Schmidt-Kennicutt relation
\citep{Kennicutt98}.  Only gas that is above a
critical surface density threshold is allowed to form stars. See S08
and P14 for further details.

\paragraph*{\it Initial mass function}
We adopt a \cite{Chabrier03} initial mass function.

\paragraph*{\it Metal treatment}
We model chemical enrichment using the instantaneous recycling
approximation.  The chemical yield is treated as a free parameter.

\paragraph*{\it Supernova feedback and winds}

Massive stars and supernovae are assumed to produce winds that drive
cold gas back into the ICM and IGM, heating the gas in the process.
The mass outflow rate is proportional to the star formation rate,
\begin{equation}
\dot{m}_{\rm rh}=\epsilon_{\rm SN}\left(\frac{200\ \rm km \ s^{-1}}{V_{\rm disk}}\right)^{\alpha_{\rm rh}}\dot{m}_{*},
\end{equation}
where $\epsilon_{\rm SN}$ and $\alpha_{rh}$ are free parameters and
$V_{\rm disk}$ is the circular velocity of the disk.  

\paragraph*{\it Gas ejection \& reincorporation}
The proportion of the gas that is ejected from the halo entirely is a
decreasing function of the halo's virial velocity.  This gas can then
fall back into the hot halo, at a re-infall rate that is proportional
to the mass of the ejected gas and inversely proportional to the
dynamical time of the halo (see S08 and P14 for details). When gas is
ejected due to supernova feedback, these winds are assumed to have a
metallicity $Z_{\rm cold}$. Ejected metals are assumed to
``re-infall'' back into the hot halo on the same timescale as the gas.

\paragraph*{\it Disk instability}
We include bulge formation and black hole growth due to disk
instabilities as in the ``stars'' model of P14 (see also
H12). Following \citet{Efstathiou82} we define the stability
parameter
\begin{equation}
\epsilon_{\rm disk} =\frac{V_{\rm max}}{(G M_{\rm disk}/ r_{\rm
    disk})^{1/2}},
\label{eqn:epsilon_disk}
\end{equation}
where $V_{\rm max}$ is the maximum circular velocity of the halo,
$r_{\rm disk}$ is the scale length of the disk, and $M_{\rm disk}$ is
the mass of stars in the disk. Disks are deemed unstable if
$\epsilon_{\rm disk}< \epsilon_{\rm crit}$, where $\epsilon_{\rm
  crit}$ is a free parameter. In every timestep in which the disk
becomes unstable, we move just enough stars from the disk to the bulge
component to restore stability. 

\paragraph*{\it Starburst}
Following a merger with mass ratio $\mu > \mu_{\rm crit} \sim 0.1$, we
trigger a burst of star formation. The burst mode star formation is
added onto the `disk'-mode star formation described above. The
efficiency and timescale of the burst mode depends on the merger mass
ratio, the gas fraction of the progenitors, and the circular velocity
of the progenitors, as described in S08. These scalings were derived
from hydrodynamic simulations of binary galaxy-galaxy mergers (see S08
for details).

\paragraph*{\it AGN feedback}
In addition to triggering starbursts, mergers drive gas into galactic
nuclei, fueling black hole growth. Every galaxy is born with a small
``seed'' black hole (typically $\sim 10^{4}\, \Msun$ in our standard
models). Following a merger, any pre-existing black holes are assumed
to merge fairly quickly, and the resulting hole grows at its Eddington
rate until the energy being deposited into the ISM in the central
region of the galaxy is sufficient to significantly offset and
eventually halt accretion via a pressure-driven outflow. This results
in self-regulated accretion that leaves behind black holes that
naturally obey the observed correlation between BH mass and spheroid
mass or velocity dispersion (see S08 and H12 for more
details). Large-scale winds associated with this rapid BH growth can
also remove gas from the galaxy (see S08).

A second mode of black hole growth, termed ``radio mode'', is assumed
to couple very efficiently with the hot halo gas, and to provide a
heating term that can partially or completely offset cooling during
the ``hot flow'' mode (we assume that the jets cannot couple
efficiently to the cold, dense gas in the infall-limited or cold flow
regime).

\paragraph*{\it Merger treatment}
Once haloes cross the virial radius of a parent halo, the Santa Cruz
SAM tracks the timescale for the orbits of the sub-haloes (satellites)
to decay via dynamical friction using a refined version of the
Chandrasekhar formula (see S08). Dark matter is stripped from the
sub-haloes on each orbit. If the sub-halo is stripped below a critical
mass, it and the galaxy it contains are considered to be tidally
disrupted and the stars in the galaxy are added to the `diffuse
stellar halo'. Any cold gas is considered to be heated and added to
the hot gas halo of the host. Galaxies that are not tidally destroyed
before they reach the center are merged with the central
galaxy. Following a merger, a fraction $f_{\rm scat}$ of the stars
from the satellite are added to the diffuse interstellar
halo. Depending on the merger mass ratio and the gas fraction of the
progenitors, some of the disk stars are moved to the bulge following a
merger (see P14).

\paragraph*{\it Substructures}
Sub-structures are tracked analytically. One minor issue is that our
SAM cannot track haloes that become sub-haloes and then travel outside
the virial radius of the host \citep[so-called `backsplash galaxies', e.g.][]{Gill05}-- we continue to treat these as
sub-haloes.

\paragraph*{\it Orphans}
All substructures are effectively treated as `orphans' in our models. 

\paragraph*{\it Calibration method}
The SAM parameters used here are identical to those used in
\citet{Porter14}. These parameters were chosen by tuning to
observations of the galaxy stellar mass function, gas fraction and
stellar metallicity as a function of stellar mass, and black hole mass
versus bulge mass relationship, all at $z\sim 0$ (see S08 and P14 for
details). In addition, our ``disk instability'' parameter is tuned to
attempt to reproduce the morphological mix of nearby galaxies as a
function of stellar mass (see P14).

\paragraph*{\it Model origin}
The \somerville\ SAM was originally developed based on EPS merger trees
\citep{Somerville99}. Subsequently, the model has been implemented
within merger trees extracted from the Bolshoi simulations (P14). Our
model results are quite insensitive to whether we use EPS mergers
trees or high-quality $N$-body merger trees.

\paragraph*{\it Modifications to the supplied data}

Currently the \somerville\ SAM does not make use of positional
information from the $N$-body simulation for ``sub-haloes'', haloes that
have become subsumed in other virialized haloes. As a result sub-haloes
were stripped from the merger trees provided for this project. No
further modifications were made.

\paragraph*{\it Halo finder properties used}
The \somerville\ SAM uses the $M_{\rm BN98}$ `virial mass' definition by
default. The only halo finder properties that we use are the halo mass
and redshift, and progenitor and descendant relationships.

\subsection{\ysam\ (Lee \& Yi)}
Here we briefly summarize the main features of \ysam. It has been developed to calculate galaxy properties on halo merger trees extracted from $N$-body simulations. \ysam\ assumes that haloes newly identified in a volume have hot gas components proportional to their virial mass, following the universal baryonic fraction, $\Omega_{\rm b}/\Omega_{\rm m}$. The hot gas components are cooled (see below). Cold gas components form gas disks and stellar populations are newly born in the disks by the simple law in~\citet{Kauffmann93}. Gradual mass loss from stellar populations can enrich the metallicity of galaxies by recycling the ejecta in hot and cold gas reservoirs. \ysam\ also includes feedback processes (see below). In addition, environmental issues can affect the gas components of subhaloes: hot gas is stripped by tidal forces~\citep[see][]{kimm11} and ram pressure~\citep{mccarthy08,Font08}. Perhaps the most notable difference of \ysam\ is that it calculates stellar mass loss in all constituent stellar populations  in each galaxy step by step. This often involves tracking tens of thousands of separate populations in each galaxy, which  helps to trace the gas recycling more realistically than in the case of instant recycling assumption. Further details can be found in~\citet{lee13}. 

\paragraph*{\it Cooling}
\ysam\ calculates gas cooling rates by adopting the prescription proposed by \citet{white91}.

\paragraph*{\it Star formation}
\ysam\ follows a simple law suggested by \citet{Kauffmann93} for quiescent star formation.

\paragraph*{\it Initial mass function}
Chabrier, Salpeter, and Scalo IMFs are available in ySAM. Stellar populations formed in quiescent and bursty modes can have different IMFs.

\paragraph*{\it Metal treatment}
\ysam\ calculates stellar mass loss from every single stellar population at each time step. The amount of metals in the ejecta is computed based on a given IMF and chemical yields of Type Ia \citep{Nomoto84} and Type II SNe \citep{Portinari98}, and intermediate mass stars \citep{Marigo01}. The metals can be recycled by star formation or circulated between galaxies and environments via gas cooling or heating.

\paragraph*{\it Supernova feedback and winds}
\ysam\ follows the prescriptions described in \citet{Somerville08} for SN feedback and winds.

\paragraph*{\it Gas ejection \& reincorporation}
Gas components can be blown away by SN and QSO mode AGN feedback. Some of them can be re-accreted onto galaxies in the dynamical timescale of haloes.

\paragraph*{\it Disk instability}
In \ysam, disk instability can be estimated by using a formula derived by \citet{Efstathiou82}. Due to uncertainties of disk instability \citep[e.g.][]{Athanassoula07}, however, we turn it off in this study.

\paragraph*{\it Starburst}
The amount of stars born in bursty mode is evaluated when galaxy mergers ($M_2/M_1>0.1$) take place. It is calculated following the prescriptions formulated by \citet{Somerville08} from the numerical simulations performed by \citet{Cox08b}.

\paragraph*{\it AGN feedback}
QSO and radio modes AGN feedback has been implemented into \ysam\ by following \citet{Croton06}.

\paragraph*{\it Merger treatment}
Satellite galaxies merge into their centrals when subhaloes harbouring them reach very central region of host haloes ($<0.1R_{\rm vir}$). If a subhalo is deprived of mass below baryonic mass, then the galaxy at the centre of the subhalo is considered to be disrupted and their stellar components become diffuse stellar components of its host halo.

\paragraph*{\it Substructures}
\ysam\ traces substructures following the results from the $N$-body simulation.

\paragraph*{\it Orphans}
If a substructure disappears before reaching the central region of its host halo, \ysam\ calculates its mass \citep{battin87} and orbit \citep{2008gady.book.....B} analytically until approaching the very central regions. This has a large impact on the lifetime of subhaloes and galaxy merging timescale \citep{Yi13}.

\paragraph*{\it Calibration method}
\ysam\ has been manually calibrated to match galaxy mass functions \citep[mainly that in][]{Panter08}, BH-bulge relation \citep{HaringRix2004}, global star formation density evolution \citep{Panter08}, and stellar-to-halo mass relation \citep{Moster10}.

\paragraph*{\it Model origin}
\ysam\ has been developed to be run based on halo merger trees extracted from $N$-body simulations.

\paragraph*{\it Modifications to the supplied data}
We ignore halo merger trees that disappear as independent host haloes before merging into other haloes. We also remove merger trees identified as subhaloes at the beginning. If, however, haloes born as subhaloes come out of their hosts and remain as host haloes by $z=0$, then we do not discard them.

\paragraph*{\it Halo finder properties used}
\ysam\ adopts as its prime mass \Mcrit\ and also uses the following information from the supplied halo catalogues: haloid, hosthaloid, radius, position, and velocity; the number of particles, peak value and position of the circular rotation curve, and spin parameter are also read, but not used.

\subsection{Durham - \galform\ (Gonzalez-Perez, Bower \& Font)}
\subsubsection*{\galformVGP, \galformBOW\ \& \galformFONT}
For the study presented here we use the \citet{gp14} (thereafter \galformVGP) flavour of the \galform\ model \citep{Cole00}, which exploits a Millenium Simulation-class $N$-body run performed with WMAP7 cosmology. We also compare the results from two variations of the \galformVGP\ model, to which we refer to as \galformBOW\ and \galformFONT. These have been generated by running a modified version of the \galformVGP\ that accounts for the main differences with respect to the \citet{Bower06} and \citet{Font08} models. These two models were developed using merger trees derived from the Millennium simulation \citep{Springel05b}, which assumes a cosmology close to that from WMAP1. 

The model referred here as \galformBOW\ is the \galformVGP\ model run assuming a single star formation law. The model referred here as \galformFONT\ is the \galformVGP\ model run assuming a single star formation law and including a gradual stripping of hot gas in satellite galaxies as opposed to the strangulation assumed by default.

\paragraph*{\it Cooling} Cooling rates are estimated by defining a cooling radius, assuming that the shock-heated halo gas is in collisional ionization equilibrium. The gas density profile in the halo is kept fixed and is well fit by the $\beta$-model \citep{cavaliere76} generally used to model hot X-ray emitting gas in galaxy clusters. We use the cooling functions tabulated by \citet{sutherland93}, which are a function of both the metallicity and temperature of the gas.

\paragraph*{\it Star formation} The \galformVGP\ model assumes two different star formation laws depending on the star formation being quiescent, happening in disks, or happening in a burst. In the star burst mode, the star formation rate is assumed to be simply
proportional to the mass of cold gas present in the galaxy  and inversely
proportional to star formation time-scale. This is what is also assumed for the quiescent star formation in both the \galformBOW\ and \galformFONT\ model variants. The quiescent star formation in the \galformVGP\ model is obtained in a self consistent calculation in which  HI and H$_2$ are tracked explicitly and the star formation in disks is assumed to depend on the amount of molecular gas, H$_2$, rather than on the total mass of cold gas \citep{lagos11}.

\paragraph*{\it IMF} A \citeauthor{kennicutt83} IMF is assumed. 

\paragraph*{\it Metal treatment} The model uses the instantaneous recycling approximation, with a recycled fraction of 0.44 and a metal yield of 0.021. 

\paragraph*{\it Supernova feedback and winds} The supernova feedback efficiency is quantified in the model in terms of the rate at which cold gas is reheated and thus ejected into the halo, $\dot{M}_{\rm
  reheated}$, per unit mass of stars formed, $\psi$, which are computed as $\dot{M}_{\rm reheated}=\psi\Big(\frac{v_{\rm circ}}{425\, {\rm km/s}}\Big)^{-3.2}$.

\paragraph*{\it Gas ejection \& reincorporation} The gas  affected by stellar feedback is assumed to be heated to the virial temperature of the current halo and placed into a reservoir. The mass in this reservoir, $M_{\rm res}$, will return to the hot halo at a rate given by $1.26 \times M_{\rm res}/t_{\rm dyn}$, where $t_{\rm dyn}$ is the dynamical time-scale of the halo. The gas that has been reincorporated into the halo can then cool back on to the galaxy disk.

\paragraph*{\it Disk instability} If a disk is strongly self-gravitating it will be unstable with respect to the formation of a bar. This will happen in those disks satisfying the \citet{Efstathiou82} criterion, with a stability threshold of 0.8.

\paragraph*{\it Starburst} During star bursts episodes, the star formation law assumed in the \galformVGP\ model is different from the quiescent case (not so for the other two model variants). The available cold gas is assumed to be consumed during a starburst event with a finite duration.

\paragraph*{\it AGN feedback} The onset of the AGN supression of the cooling flow can only happen in the model in haloes undergoing quasi-hydrostatic cooling. This is assumed to happen for haloes hosting galaxies such that $t_{\rm cool}>t_{\rm ff}/0.6$, where $t_{cool}$ is the cooling time of the gas and $t_{ff}$ is the free-fall time for the gas to reach the centre of the halo.

\paragraph*{\it Merger treatment} Mergers such that the ratio between masses exceeds 0.1 will trigger a burst of star formation. Disks are transformed into spheroids when mergers happen.

\paragraph*{\it Substructures} The $N$-body simulation subhalo information is used to trace substructure.

\paragraph*{\it Orphans} When the subhalo hosting a satellite galaxy can no longer be followed with the N-body simulation information, the \citet{Lacey93} analytical expression is used to compute the merging timescale of this orphan galaxy.

\paragraph*{\it Calibration method} The free parameters in this model where chosen manually such that the predicted luminosity functions in both b$_J$ and K-band at redshift 0 and the predicted evolution of the rest-frame UV and V-band luminosity function were in reasonable agreement with observations \citep{norberg02,2mass,driver12}.

\paragraph*{\it Model origin} The   \galformVGP\ model uses dark matter halo trees derived from the MS-W7 $N$-body simulation \citep{Guo13}, with a simulation box of 500 Mpc/$h$ side.

\paragraph*{\it Modifications to the supplied data} In order to run \galform, we have remapped the given merger trees using \textsc{D-Haloes} \citep{Jiang14b}. This algorithm groups subhaloes in $N$-body cosmological simulations avoiding transient structures and losses in mass. From the provided list of subhaloes, there is a percentage smaller than 15 per cent that \textsc{D-Haloes} classifies as independent haloes. This happens for haloes that when becoming subhaloes either retain more than 75 per cent their mass or that are located away from the main halo more than two half mass radii. 

\paragraph*{\it Halo finder properties used} The standard \galform\ models use \Mbnd, but runs have been performed for all the five supplied mass definitions.

\subsection{Munich - \DLB\ (De Lucia \& Blaizot)}
The variant of the Munich model described in \citet{DeLucia07}, with its generalization to the 3-yr Wilkinson Microwave Anisotropy Probe (WMAP3) cosmology discussed in \citet{Wang08b}. The model includes prescriptions for gas cooling, star formation, stellar feedback, merger driven starburst, AGN feedback and chemical enrichment. The latter is based on an instantaneous recycling approximation. For more details on the physical models, we refer to \citet{Croton06}, \citet{DeLucia07}, and references therein.

\paragraph*{\it Cooling}
The rate of gas cooling is computed following the model originally proposed by \citet{white91}, and an implementation similar to that adopted in \citet{Springel01subfind}. Full details can be found in \citet{DeLucia10}.

\paragraph*{\it Star formation}
A Kennicutt-type prescription is adopted. Only gas above a critical surface density for star formation can be converted into stars. Details in \citet{Croton06}.

\paragraph*{\it Initial mass function}
A \citet{Chabrier03} IMF is used throughout.

\paragraph*{\it Metal treatment}
As detailed in \citet{DeLucia04b}, an instantaneous recycling approximation is adopted. Metals are ejected (and instantaneously mixed) into the cold gas component after each star formation event.

\paragraph*{\it Supernova feedback and winds}
Supernovae explosions are assumed to reheat a (cold) gas mass that is proportional to the mass of stars formed. The amount of gas that leaves the dark matter halo in a wind is determined by computing whether the excess supernova energy is available to drive the flow after reheating the material to the halo virial temperature. Details can be found in \citet{Croton06}.

\paragraph*{\it Gas ejection \& reincorporation}
Ejected gas is re-incorporated into the hot gas component on a time-scale that is related to the dynamical time-scale of the halo, as detailed in \citet{DeLucia04b}. 

\paragraph*{\it Disk instability}
The instability criterion adopted is that of \citet{Efstathiou82}. When the instability condition is verified, we transfer enough stellar mass from the disk to the bulge so as to restore stability. For details, see again \citet{Croton06} and \citet{DeLucia11}.

\paragraph*{\it Starburst}
Galaxy mergers are accompained by starbursts modelled using the `collisional starburst' prescription introduced by \citet{Somerville01} with updated numerical parameters so as to fit the numerical results by \citet{Cox04}.

\paragraph*{\it AGN feedback}
The model includes a distinction between a `quasar mode' and a `radio mode'. AGN feedback is implemented as detailed in \citet{Croton06}.

\paragraph*{\it Merger treatment}
The model explicitly follows dark matter substructures. This allows us to follow properly the motion of the galaxies at their centres, until tidal truncation and stripping disrupt the subhalo at the resolution  limit of the simulation. When this happens, a residual merger time is estimated using the current orbit and the classical dynamical friction formula of \citet{Binney87}. For details, see \citet{DeLucia07} and \citet{DeLucia10}.

\paragraph*{\it Substructures}
The model follows dark matter haloes after they are accreted onto a larger system, i.e. substructures are explicitly followed from the $N$-body simulation.

\paragraph*{\it Orphans}
An `orphan' galaxy is created each time a substructure falls below the resolution limit of the simulation. The stellar mass of the galaxy is unaffected. The galaxy is assigned a residual merger time as detailed above. Its position and velocity are traced by following the most bound particle of the substructure at the last time it was present. As this information could not be reconstructed for this project  positions and velocities of orphan galaxies are kept fixed to those of the substructures at the last time they were identified.

\paragraph*{\it Calibration method}
The model has been calibrated `by-hand' (note that in practice this means that a grid of model parameters was considered in order to evaluate the influence of each of them on model predictions). The main constrain is the K-band luminosity function at $z=0$.

\paragraph*{\it Model origin}
The model is designed to work with merger trees from $N$-body simulations.

\paragraph*{\it Modifications to the supplied data}
No significant modification. Merger trees are reconstructed from original files provided, and using each halo's uniquely assigned descendant.

\paragraph*{\it Halo finder properties used}
The following quantities from the supplied halo catalogue are used: snapnum, positions and velocities of each
halo, its mass \Mcrit\ and spin parameter (this latter quantity is used to model the disk radius).

\subsection{Munich - \lgalaxy\ (Henriques, Srisawat \& Thomas)}
The `Munich' model of galaxy formation is a semi-analytic scheme for
simulating the evolution of the galaxy population as a whole and has
been continually developed over the last quarter century
\citep{white91, Kauffmann93, Kauffmann99, springel_subfind_2001,
Springel05b}. Recent updates to the baryonic physics have resulted
in a model that is capable of reasonably describing the observed
population of galaxies in the local Universe. These include a detailed
treatment of gas reheating, ejection and reincorporation by supernova
\citep{delucia_sfr_2004}, updated in \citet{Guo11}, black hole growth
during mergers \citep{kauffmann00} and feedback from quiescent
accretion \citep{Croton06}, continuous environmental effects acting
on satellite galaxies \citep{Guo11}, dust extinction from the ISM
and molecular clouds \citep{DeLucia07}.  The model used in this
paper, \citet{Henriques13}, includes all previous developments and
aims at better representing the observed evolution of stellar mass
across most of the age of the Universe. This is done by modifying the
timescales for gas to be reincorporated after ejection from supernova
in order to avoid an excessive build up of low mass galaxies at early
times. As in \citet{Guo13}, a WMAP7 cosmology is adopted.

In the latest major release of the Munich model,
\citep{Henriques15}, the modifications implemented in
\citet{Henriques13} were combined with a less efficient ram-pressure
striping implementation in low mass groups and a lower threshold for
star formation. These ensure that low mass galaxies are predominately
star forming down to $z=0$ and that the model can simultaneously match
the evolution of the stellar mass function and the fraction of red
galaxies. In addition, the AGN feedback and dust model were adjusted
in order to better follow the properties of intermediate and high-z
galaxies, respectively. Although the \citet{Henriques15} model
provides a significantly better representation of the observable
universe, we are unable to run it for this project since it requires
information on the trajectories of most-bound particles for haloes
striped below resolution. These are used to follow the dynamics of
orphan galaxies and are crucially in order to track their
properties. Without it, merger times and the disruption efficiency of
satellites will be significantly different, changing the black-hole
growth, star-burst efficiency and morphology evolution of all central
galaxies. We therefore use \citet{Henriques13} for which the
combination of physics is less sensitive to these effects.

\paragraph*{\it Cooling}
The cooling follows the implementation of \citet{white91}. Infalling
diffuse gas is expected to shock-heat as it joins a halo. At early
times and for low-mass haloes the post-shock cooling time is short and
new material is assumed to settle onto the cold gas disk in a
dynamical time. At later times and for higher mass haloes the shocked
heated gas forms a quasi-static hot atmosphere from which it can
gradually accrete to the centre via an element and temperature.

\paragraph*{\it Star formation}
Following \citet{Kauffmann96}, star formation is assumed to be
proportional to the mass in cold gas above a given threshold. 
\paragraph*{\it Initial mass function}
A \citet{Chabrier03} IMF is assumed and 43\% of the total mass of
stars formed is assumed to be in massive, short lived stars, and
immediately returned to the cold gas. 

\paragraph*{\it Metal treatment}
Following \citet{DeLucia04b}, these stars are enrich the surrounding medium with a fixed yield of
metals. \citet{Fu2013} and \citet{Yates2013} implemented,
respectively, more detailed models of star formation and chemical
enrichment in the Munich model. This will be incorporated in future
releases.

\paragraph*{\it Supernova feedback and winds}
Supernova feedback is treated following \citet{DeLucia04b} and
\citet{Guo11}. The fraction of the total available energy used in
feedback is parametrised in a virial velocity dependent way. 

\paragraph*{\it Gas ejection \& reincorporation}
Part of the energy released during supernovae is used to reheat gas from the cold gas to the hot phase
and the left-over used to eject material into an external reservoir.
A new reincorporation model returns gas back into the hot phase at a
rate inversely proportional to the halo mass \citep{Henriques13}.

\paragraph*{\it Disk instability}
Disk instabilities are followed as in \citet{Guo11} and transport
material inwards to the bulge and they occur in galaxies where
self-gravity of the disk dominates the gravitational effects of the
bulge and halo. When the instability criteria is met, we transfer
sufficient stellar mass from the inner parts of the disk to the bulge
to make it marginally stable again.

\paragraph*{\it Starburst}
The stellar mass formed during a merger (see below) is modelled using
the "collisional starburst" formulation of \citet{Somerville01}.

\paragraph*{\it AGN feedback}
\citet{Henriques13} includes two black-hole related processes: a
``quasar'' and a ``radio'' mode.  Following \citet{kauffmann00},
black holes are taken to form and grow when cold gas is driven to
the centre of merging systems. Black holes are also allowed
to accrete gas from the hot gas atmospheres of their galaxies. This
is assumed to generate jets and bubbles which produce "radio mode"
feedback, suppressing cooling onto the galaxy and so eliminating the
supply of cold gas and quenching star formation \citep{Croton06}.

\paragraph*{\it Merger treatment}
When a satellite galaxy finally merges with the object at the centre of the
main halo, the outcome is different for major and minor mergers.  In a
major merger, the disks of the two progenitors are destroyed and all
their stars become part of of the bulge of the descendent, along with
any stars formed during the merger. In a minor merger, the disk of the
larger progenitor survives and accretes the cold gas component of the
smaller galaxy, while its bulge accretes all the stars of the
victim. Stars formed during the merger stay in the disk of the
descendent. 

\paragraph*{\it Substructures}
When the host halo of a galaxy enters the virial radius of a larger
system it becomes a satellite and its properties are strongly affected
by a number of processes collectively called environmental
effects. Satellite galaxies do not receive primordial infall, they are
stripped of their hot and ejected gas components due to ram-pressure
and tidal striping, and they might merge with other galaxies
\citep{Guo11}.

\paragraph*{\it Orphans}
Once a satellite subhalo is disrupted, its central galaxy becomes an
orphan and a merging clock is started. This will estimate how long the
satellite will take to spiral into the central object due to dynamical
friction. Since our implementation of tidal stripping of hot gas is
directly connected to the stripping of dark matter, orphans have no
hot or ejected gas. In addition, any cold gas reheated by supernovae
is added to the central halo and tidal forces might completely strip
the stars and cold gas into the intracluster medium.

In the default version of our model, the dynamical properties of
orphan galaxies are given by those of the most bound particle
identified at the time at which the halo falls below resolution.  The
vector offset is decayed due to dynamical friction. This has been
shown to be crucial in order to correctly trace satellite
distributions and achieve convergence for simulations of different
resolution \citep{Guo11}. Since the most bound particle information
is not provided, the positions and velocities of orphans are frozen at
the time of disruption and should be ignored. These do not have a
significant impact on the physics in \citet{Henriques13} which
depend mostly on the independently calculated merger times but stop us
from running the latest \citet{Henriques15}.

\paragraph*{\it Calibration method}
The best-fit model was chosen by fully sampling the allowed regions in
parameter space using the MCMC methodology described in
\citet{Henriques09,Henriques13}. The stellar mass, $K$-band
and $B$-band Luminosity functions at $z=0,1,2$ and 3 were used as
observational constraints.
	
\paragraph*{\it Model origin}
\citet{Henriques13} is built on $N$-body merger trees following the
method introduced by \citet{Springel01subfind}.  The substructure are
followed directly from the dark-matter simulation and the supplied
data was used in unmodified form.  The following quantities from the
supplied halo catalogue are used: snapnum, positions and velocities of
each halo, its mass (\Mcrit200) and spin.

\paragraph*{\it Modifications to the supplied data}
The supplied data was used in unmodified form.  We note that insufficient
information was supplied to allow use of the latest version of L-Galaxies
\citep{HWT14} which better models the stripping of satellite galaxies within
groups and clusters.

\paragraph*{\it Halo finder properties used}
The following quantities from the supplied halo catalogue are used: snapnum, positions and velocities of each
halo, its mass (\Mcrit200) and spin.

\subsection{Munich - \sage\ (Croton)} \label{sec:sage}
The new Semi-Analytic Galaxy Evolution (\sage) model is an updated version of that first presented in \citet{Croton06}.  We only highlight the significant changes here and point to the 2006 paper and Croton et al. (in prep.) for a full description of the rest of the model.

\sage\ is publicly available through the Theoretical Astrophysical Observatory\footnote{https://tao.asvo.org.au} \citep[TAO;][]{Bernyk15}, an online virtual laboratory that includes tools to add hundreds of magnitude filters to the galaxy output (with or without dust), construct custom light-cones, build images, and then download the mock data to your local machine, all without any requisite programming knowledge.

\paragraph*{\it Cooling}
Cooling is handled as in \citet{Croton06}. An isothermal sphere is assumed and a cooling rate estimated from a simple continuity argument.

\paragraph*{\it Star formation}
The \sage\ model calculates the mass of cold gas in the disk that is above a critical surface density for star formation. New stars then form from this gas using a Kennicutt-type prescription.

\paragraph*{\it Initial mass function}
\sage\ assumes a \citet{Chabrier03} IMF.

\paragraph*{\it Metal treatment}
\sage\ follows the simplistic metal treatment introduced in \citet{DeLucia04b}. A yield of metals is produced from each star formation event and is recycled instantly back to the cold gas from very short-lived stars.

\paragraph*{\it Supernova feedback and winds}
Feedback from supernova in \sage\ is a two step process. First, an assumed mass loading factor pushes cold gas out of the disk into the hot halo. Second, if enough energy from supernova has been added to the hot halo carried by this gas, some of the hot gas becomes unbound and is removed to an ejected reservoir.

\paragraph*{\it Gas ejection \& reincorporation}
Gas can be ejected from the halo from supernova or quasar winds. Ejected gas can be reincorporated back into the hot halo at a rate in proportion to the dynamical time of the dark matter halo.

\paragraph*{\it Disk instability}
\sage\ uses the \citet{Mo98} approximation to determining when a disk becomes unstable. When so, enough existing stars are transferred to the bulge to make the disk stable, along with any new stars as a result of a starburst.

\paragraph*{\it Starburst}
The SAGE model applies the collisional starburst model introduced in \citet{Somerville01} to determine the mass of cold gas that becomes new stars during a merger.

\paragraph*{\it AGN feedback}
\sage\ uses the radio-mode AGN heating model introduced in \citet{Croton06}, and a new quasar-mode wind model introduced in Croton et al. (in prep.).

\paragraph*{\it Merger treatment}
Mergers are treated using the method described in Croton et al. (in prep.). Satellites are either merged with the central galaxy or added to the halo's intra-cluster stars, depending on the subhalo survival time relative to an average expected based on its infall properties.

\paragraph*{\it Substructures}
Substructures are explicitly followed from the $N$-body simulation.

\paragraph*{\it Orphans}
No orphans are used in \sage. A decision as to the fate of a satellite galaxy has already been made and executed before its subhalo is lost below the resolution limit of the simulation.

\paragraph*{\it Calibration method}
\sage\ is calibrated by hand using the $z=0$ stellar mass function, cold gas fraction, stellar metallicity--stellar mass relation, baryonic Tully-Fisher relation, and black hole--bulge mass relation.

\paragraph*{\it Model origin}
\sage\ is an evolution of the \lgalaxy\ semi-analytic code which is based solely on $N$-body simulations.

\paragraph*{\it Modifications to the supplied data}
No modification were made to the supplied data.

\paragraph*{\it Halo finder properties used}
The primary mass used was \Mcrit. Additional halo finder properties used to build galaxies are  the peak value of the circular rotation curve, the position, and the spin parameter.

\subsection{HOD -- \mice\ (Castander \& Carretero)}
The MICE project\footnote{http://www.ice.cat/mice} is producing large simulations to help the design and interpretation of large scale cosmological observational projects. In this paper, we use the galaxy mock generation code that has been developed within MICE. The galaxy mock code populates dark matter haloes with a hybrid Halo Occupation Distribution (HOD; \citealt[e.g.,][]{Jing98,Scoccimarro01,Berlind02}) and SubHalo Abundance Matching (SHAM; \citealt[e.g.,][]{Vale04,Tasitsiomi04,Conroy06}) technique \citep{Carretero14,Castander14,Crocce13}. Following the HOD philosophy, we assume that haloes are populated by central and satellite galaxies. We assign luminosities to central galaxies based on abundance matching taking into account the scatter between halo mass and luminosity. The HOD gives us the number of satellites in each halo. The satellite luminosities are drawn from the satellite luminosity function.  We distribute satellites inside the haloes following a triaxial `modified' NFW, tweaked to match the observed clustering. The HOD parameters are also varied until we find an acceptable fit to the galaxy clustering as a function of luminosity. We assign velocities to the galaxies assuming a Gaussian velocity dispersion distribution given by the halo mass~\citep{Bryan98}. Lastly, we assign colours and SEDs with recipes that fit the clustering as a function of colour. We calibrate our method with local constraints given by the Sloan Digital Sky Survey (SDSS; ~\citealt{York00}) using the MICE Grand Challenge simulation~\citep{Fosalba13a,Crocce13,Fosalba13b,Hoffmann:14} as starting halo catalogue. In particular we reproduce the galaxy luminosity function~\citep{Blanton03,Blanton05b}, the colour-magnitude diagram~\citep{Blanton05a} and the SDSS clustering properties as a function of luminosity and color~\citep{Zehavi:11}. We extend our recipes to higher redshift applying evolutionary corrections to the galaxy colours and then resampling from the cosmos catalogue~\citep{Ilbert09} galaxies with compatible luminosities and colours at the given redshift.

\paragraph*{\it Cooling}
---

\paragraph*{\it Star formation}
We obtained the star formation rate from the dust-corrected UV flux of the galaxy SED.

\paragraph*{\it Initial mass function}
We assume a `diet' Salpeter IMF ~\citep{Bell:01}.

\paragraph*{\it Metal treatment}
We compute the metallicity from the absolute magnitude using empirically determined relations.

\paragraph*{\it Supernova feedback and winds}
---

\paragraph*{\it Gas ejection \& reincorporation}
---

\paragraph*{\it Disk instability}
---

\paragraph*{\it Starburst}
---

\paragraph*{\it AGN feedback}
---

\paragraph*{\it Merger treatment}
---

\paragraph*{\it Substructures}
We use a hybrid method to treat substructures. We can use the substructures provided by $N$-body, but if they are not available we can generate them analytically.

\paragraph*{\it Orphans}
We compute the expected number of satellites for each halo following an HOD prescription. If the halo contains fewer sub-haloes, we generate as many new satellites as the HOD predicts. We call these new satellites orphans in this context. We place them in the halo following a NFW profile with a concentration index expected for its halo mass.

\paragraph*{\it Calibration method}
The method has been calibrated to reproduce the galaxy luminosity function, the colour magnitude-diagram and the galaxy clustering as a function of luminosity and colour. The calibration is performed at low redshift and extrapolated at higher redshifts. The calibration has been done minimizing a $\chi^2$, where we have altered the input parameters manually.  Note that
our method has only been calibrated out to redshift z=1.5 and although
we have computed quantities at higher redshifts in this paper, they
are just an extrapolation of our recipies that we have not calibrated. 
So, take that in mind when trying to interpret the MICE
results beyond that redshift.

\paragraph*{\it Model origin}
The model stems from $N$-body simulations, namely the MICE Grand Challenge $N$-body simulation.

\paragraph*{\it Modifications to the supplied data}
The supplied data has not been modified

\paragraph*{\it Halo finder properties used}
The base for our method are the halo masses and we adopted \Mfof\ as the choice for the data presented here. We further use the following properties from the input catalogue: haloid, hosthaloid, number of substructures, number of particles, position, velocity, radius, peak value and position of the circular rotation curve, velocity dispersion, and concentration.

\subsection{HOD -- \skibba\ (Skibba)} 
The model used for this work is based on the halo model of galaxy clustering developed in \citet{Skibba06} and \citet{Skibba09}.  `Central' and `satellite' galaxy luminosities are modelled such that the luminosity function and luminosity dependence of clustering are the same as that observed.  All galaxy properties and their occupation distributions are determined by halo mass, concentration, and halo-centric position; therefore, all environmental correlations are entirely a consequence of the fact that massive haloes tend to populate dense regions (i.e., no `assembly bias' is assumed).  Satellites are assigned to subhaloes in the simulation, assuming an abundance matching-like procedure that includes scatter between stellar mass, subhalo mass, and $V_{\rm max}$.  To include colours, we develop a prescription for how the colour-luminosity distribution of central and satellite galaxies depends on halo mass, and this model is consistent with the observed colour mark correlation function.  I assume that the fraction of satellite galaxies which populate the red sequence increases with luminosity, and this fraction is in agreement with galaxy group catalogues \citep{Skibba09} and with the gradual quenching of satellites' star formation \citet{Font08}.  Stellar masses are based on luminosities and colors; since these are consistent with observations, it is not surprising that the stellar mass function and clustering are consistent with observations as well.  It is important to note that the model's observational constraints are robust only at $M_\ast>10^{9}$\hMsun\ and $M_\mathrm{halo}>10^{11}$\hMsun; therefore, I have applied a mass threshold, below which this model should not be extrapolated or compared to other models (or to observations).

We have made many updates and improvements to the model, which will be described in Skibba (in prep.). We are including colour and stellar mass gradients within haloes \citep{Hansen09,vandenBosch08} and a dependence of the colour distribution on halo mass at fixed luminosity \citep{More11,Hearin13}.  In addition, the model includes a treatment of dynamically unrelaxed systems, including some non-central brightest halo galaxies, the fraction of which is constrained by SDSS and mock group catalogues \citep{Skibba11a,Skibba11b}.

The model provided here is a sort of HOD-SHAM hybrid, in which I populated subhaloes when possible and when an insufficient number of resolved subhaloes were found, I distributed the satellites with my model's standard prescription.  I generously populated low-mass subhaloes in this model, and as a result the orphan fraction is relatively low.

\paragraph*{\it Cooling}
---

\paragraph*{\it Star formation}
The model currently only includes optical colours; a subsequent version will include a model of star formation rates. The model's stellar masses apply a calibration from \citet{Zibetti2009}.

\paragraph*{\it Initial mass function}
A \citet{Chabrier03} IMF is used throughout.

\paragraph*{\it Metal treatment}
Gas-phase and stellar metallicities are based on scaling relations of \citet{Tremonti04} and \citet{Gallazzi05}.

\paragraph*{\it Supernova feedback and winds}
---

\paragraph*{\it Gas ejection \& reincorporation}
---

\paragraph*{\it Disk instability}
Disk instabilities are not included in the model, but spiral and elliptical morphologies are included based on clustering and other constraints from \citet{Skibba09}.

\paragraph*{\it Starburst}
---

\paragraph*{\it AGN feedback}
AGN feedback is not modelled here, but a simple black hole mass scaling relation basted on \citet{Tundo07} is applied.

\paragraph*{\it Merger treatment}
---

\paragraph*{\it Substructures}
Substructure properties are taken from the simulation at a given snapshot, but subsequent subhalo evolution is not modelled.

\paragraph*{\it Orphans}
When a sufficient number of substructures are not resolved to match them to satellites, haloes are populated with the remaining satellites (`orphans') in order to reproduce the model's occupation distributions. Orphans are spatially distributed with a \citet[][NFW]{Navarro97} profile with the \citet{Maccio08b} mass-concentration profile is assumed, while accounting for the fact that galaxies and subhaloes are typically less concentrated than dark matter \citet{Munari13}.

\paragraph*{\it Calibration method}
The model has been designed to reproduce real- and redshift-space luminosity and colour-dependent galaxy clustering \citep{Zehavi:11} and mark clustering statistics \citep{Skibba06}.  The model is also constrained by the SDSS luminosity function \citep{Blanton03,Yang09} and colour-luminosity distribution \citep{Skibba09}. The model is consistent with the Moustakas et al. (2013) stellar mass function, which is not used as a constraint. The model is consistent with the \citet{Moustakas13} stellar mass function, which is not used as a constraint though. 

\paragraph*{\it Model origin}
The model originates from an analytic halo-model formalism \citep{Cooray02}.

\paragraph*{\it Modifications to the supplied data}
No modification were made to the supplied data.

\paragraph*{\it Halo finder properties used}
The model uses the 3-D positions, 3-D velocities, halo mass and radius (200c and bound), $V_{max}$, velocity dispersion (200c), and the substructure abundances and properties. 

\bigskip

Because the observational constraints are less robust at low masses and luminosities, we have applied a halo mass threshold near $10^{11}\hMsun$.  The \skibba\ model is complete only above this mass.

\section{Halo-Mass Definition} \label{app:halomassdefinition}
\begin{figure}
   \includegraphics[width=\columnwidth]{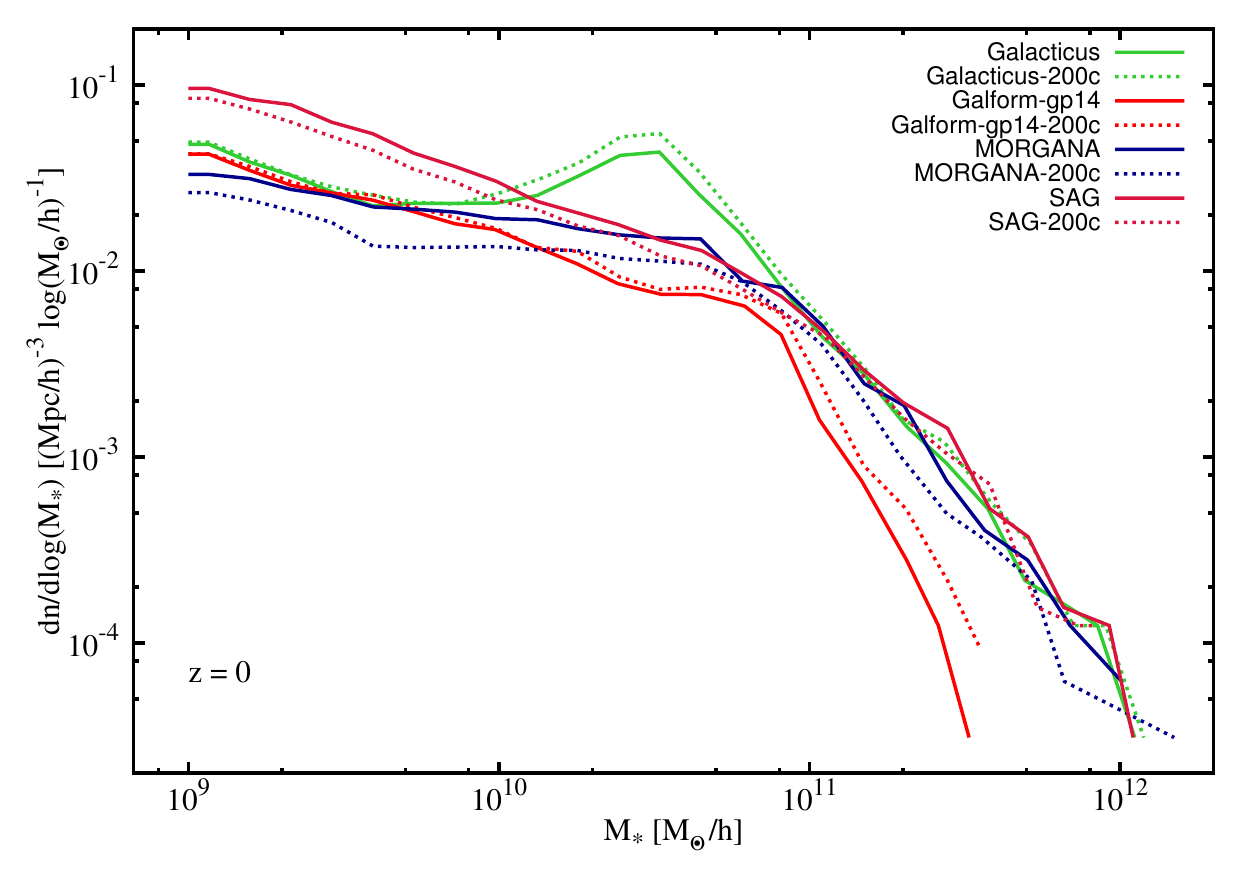}
   \caption{Stellar mass function at redshift $z=0$ for models that also provided data for \Mcrit\ as the mass definition. To be compared against \Fig{fig:NgalMstar}.}
 \label{fig:NgalMstar200c}
 \end{figure}

While we have already discussed the influence of the applied mass definition on the model-to-model variation seen in \Sec{sec:SMF} we like to extend this here a bit more by directly comparing two mass definitions on a model-by-model basis. Namely, for those models that provided both an \Mcrit\ galaxy catalogue and a catalogue based upon their own mass definition (different from \Mcrit, of course) we show in \Fig{fig:NgalMstar200c} the stellar mass functions: solid lines are for the model mass definition whereas dotted lines are for \Mcrit. Some models appear to be sensitive to the choice of the definition for halo mass, but the overall level of scatter across various models seems to stay similar.
We append some more plots that focus on the redshift evolution as this is where the mass definition leaves its largest imprint. The plots are accompanied by \Tab{tab:Ngal200c} where we list the number of galaxies in various populations. This table should be compared against \Tab{tab:Ngal}.

 \begin{figure}
   \includegraphics[width=\columnwidth]{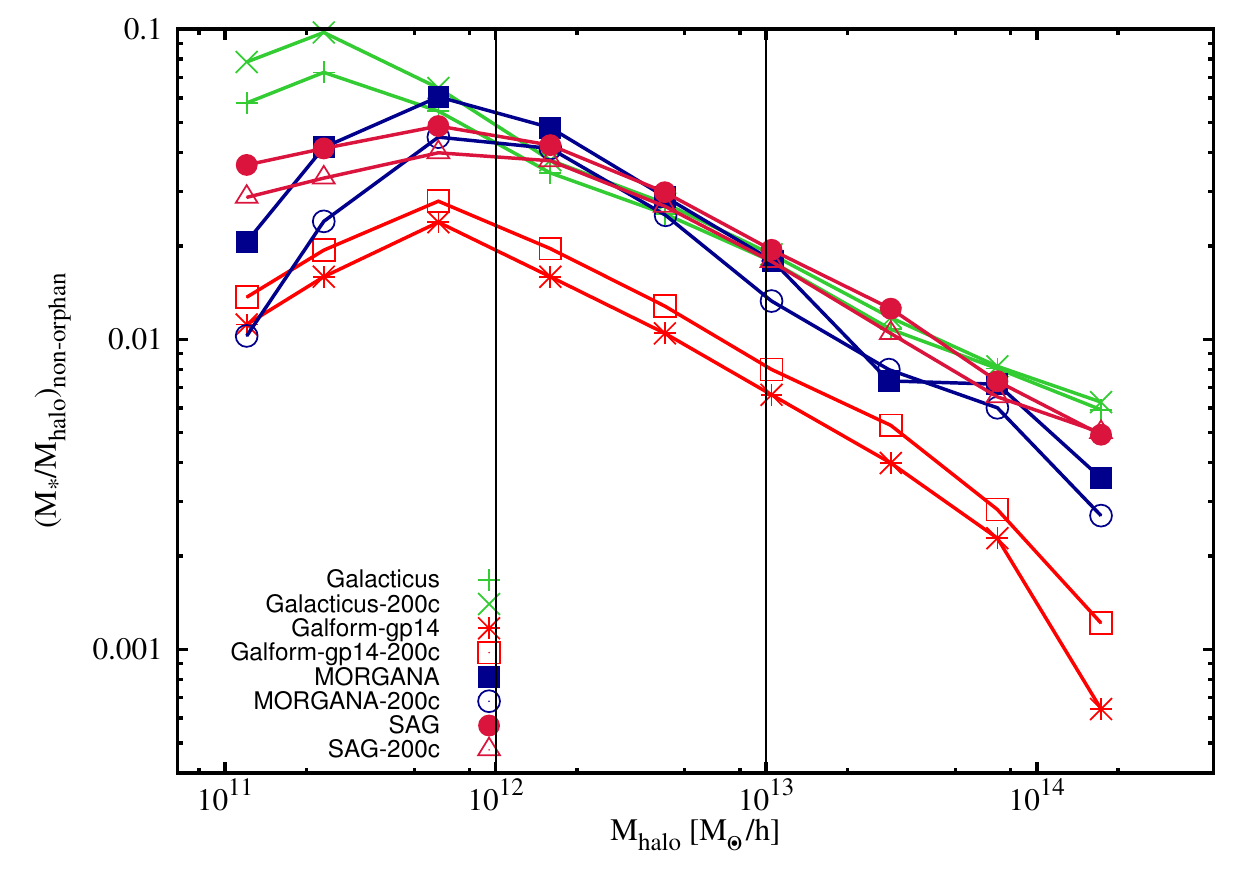}
   \caption{Stellar-to-halo mass ratio for all (non-orphan) galaxies at redshift $z=0$. To be compared against \Fig{fig:MstarMhaloMhalo}.}
 \label{fig:MstarMhaloMhalo200c}
 \end{figure}

 \begin{figure}
   \includegraphics[width=\columnwidth]{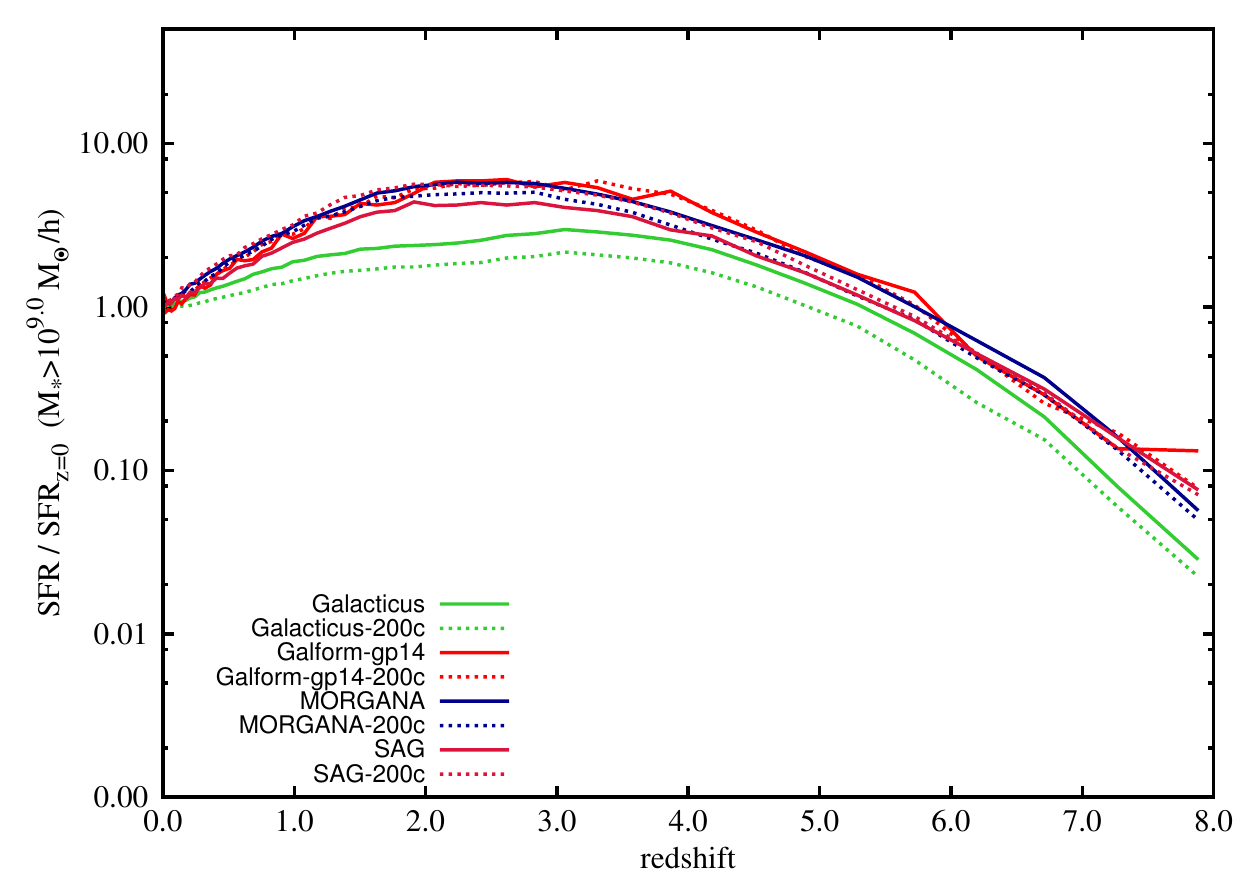}
   \caption{Star formation rate density as a function of redshift for models that also provided data for \Mcrit\ as the mass definition. To be compared against \Fig{fig:SFRzred}.}
 \label{fig:SFRzred200c}
 \end{figure}

 \begin{figure}
   \includegraphics[width=\columnwidth]{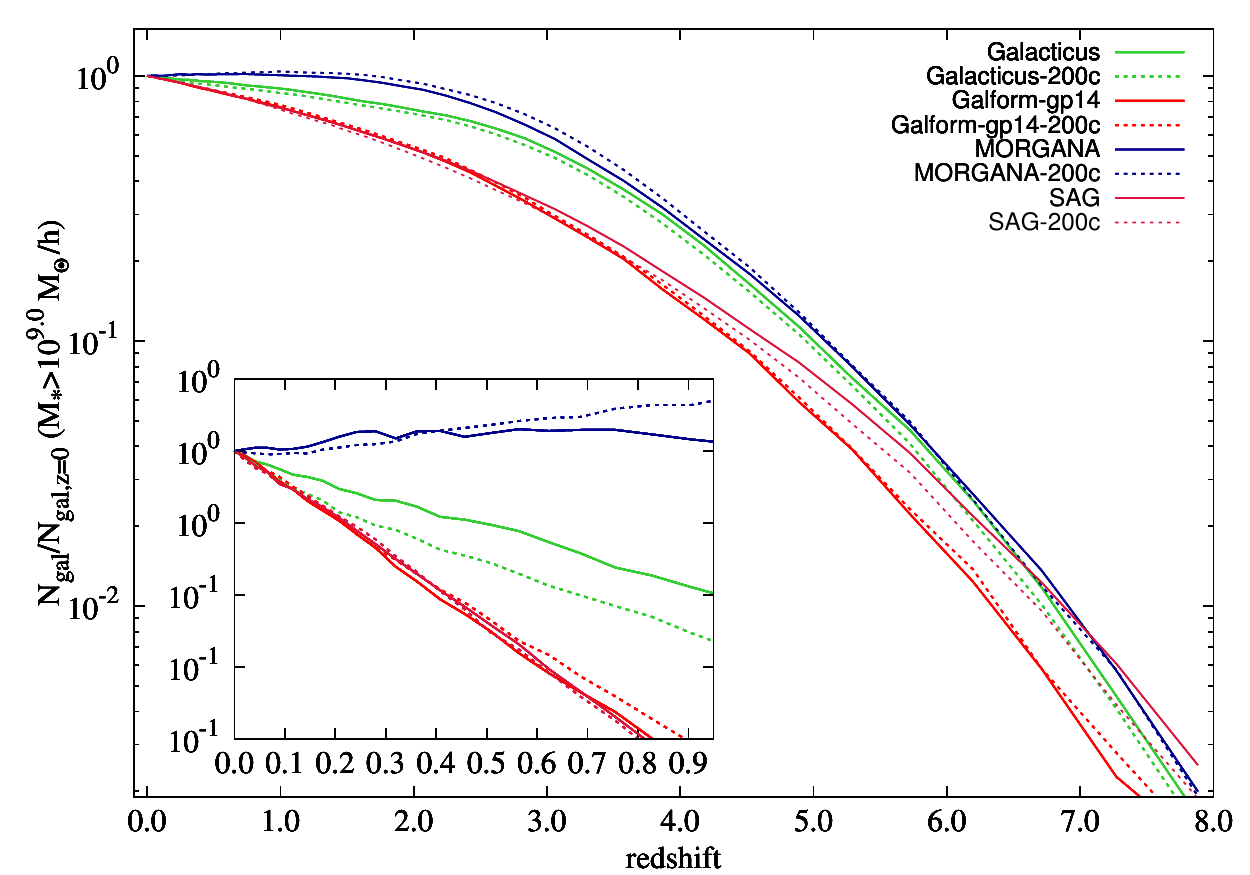}
   \caption{The number density of all galaxies with stellar mass $\Mstar > 10^{9}\hMsun$ as a function of redshift for models that also provided data for \Mcrit\ as the mass definition. To be compared against \Fig{fig:Ngalzred}.}
 \label{fig:Ngalzred200c}
 \end{figure}

\begin{table}
  \caption{Number of galaxies at redshift $z=0$ with a stellar mass in excess of $M_{*}>10^{9}\hMsun$ for models when applying \Mcrit\ as the mass definition ($-200c$ extension) and when using their favourite mass definition. Note again that for \morgana\ all satellites have been tagged as `orphan'.}
\label{tab:Ngal200c}
\begin{center}
\begin{tabular}{lcccc}
\hline
code name			& $N_{\rm gal}^{z=0}$ & $N_{\rm central}$ & $N_{\rm non-orphan}^{z=0}$ & $N_{\rm orphan}^{z=0}$ \\
\hline
 \galacticus			& 14255 & 7825 & 10019 & 4236\\
 \galacticus$-200c$		& 16123 & 9026 & 11393 & 4730\\
 \galformVGP			& 8824 & 5097 & 6098 & 2726\\
 \galformVGP$-200c$	& 9320 & 5595 & 6666 & 2654\\
 \morgana	 			& 10008 & 6186 & 6186 & \textit{3822}\\
 \morgana$-200c$ 		& 7316 & 4925 & 4925 & \textit{7316}\\ 
 \sag					& 19516 & 13571 & 16256 & 3260\\ 
 \sag$-200c$			& 16505 & 11332 & 13773 & 2732\\
\\
\hline
\end{tabular}
\end{center}
\end{table}

\section{Initial Stellar Mass Function} \label{app:IMF}
As the change in stellar mass will also influence the number of galaxies above our usual threshold $10^{9}$\hMsun\ we also list the (change in) numbers in \Tab{tab:NgalIMF}, only showing the affected models. This table should be compared against \Tab{tab:Ngal} again.

\begin{table}
  \caption{Number of galaxies at redshift $z=0$ with a stellar mass in excess of $M_{*}>10^{9}\hMsun$ for those models that have been affected by the transformation to a Chabrier IMF. To be compared against \Tab{tab:Ngal}.}
\label{tab:NgalIMF}
\begin{center}
\begin{tabular}{lcccc}
\hline
code name	& $N_{\rm gal}$ & $N_{\rm central}$ & $N_{\rm non-orphan}$ & $N_{\rm orphan}$ \\
\hline
 \galics		& \multicolumn{4}{c}{tuned parameters to observations w/ Chabrier IMF}\\
 \sag			& 14025 & 9565 & 11758 & 2267\\
\\
\underline{Durham flavours:}\\
 \galformVGP	& 9842 & 5680 & 6770 & 3072\\ 
 \galformBOW	& 10467 & 6060 & 7225 & 3242\\
 \galformFONT	& 13340 & 7054 & 8371 & 4969\\
\\
\underline{HOD models:}\\
 \mice 		& 9510 & 5638 & 8067 & 1443\\
\hline
\end{tabular}
\end{center}
\end{table}

\section{Un-normalized Redshift Evolution} \label{app:unnormalized}
In \Sec{sec:galaxies} we discussed the redshift evolution of both the number (density) of galaxies and the star formation rate (density), normalizing the respective curves to their redshift $z=0$ values. The normalizations have been provided in \Tab{tab:sumMstar} \& \Tab{tab:NgalIMF}  and hence we separated `trends' (as shown in the figures) from `absolute' differences (as listed in the tables). Here we now provide the un-normalized plots directly showing the different evolutions for both these quantities.

 \begin{figure}
   \includegraphics[width=\columnwidth]{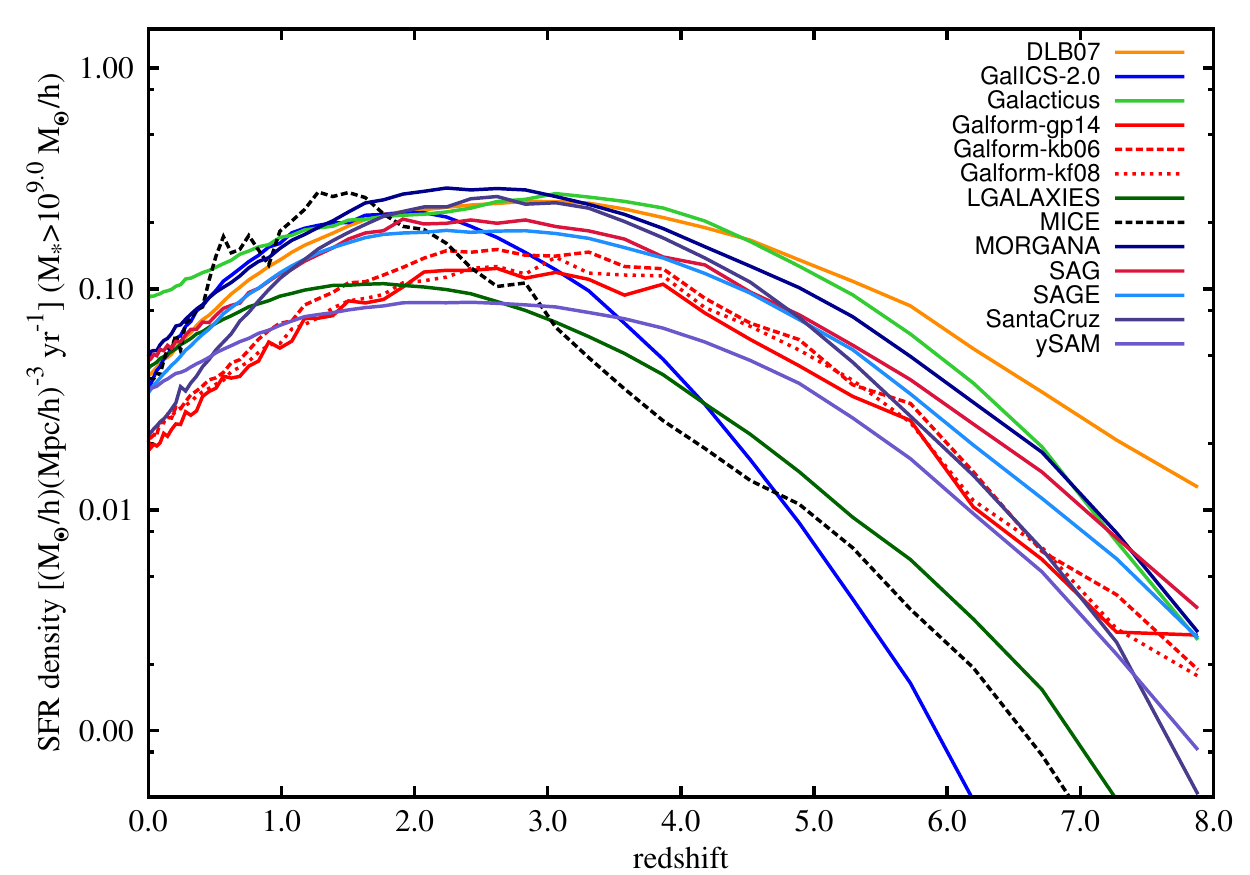}
   \caption{Star formation rate density as a function of redshift.}
 \label{fig:SFRzred2}
 \end{figure}

 \begin{figure}
   \includegraphics[width=\columnwidth]{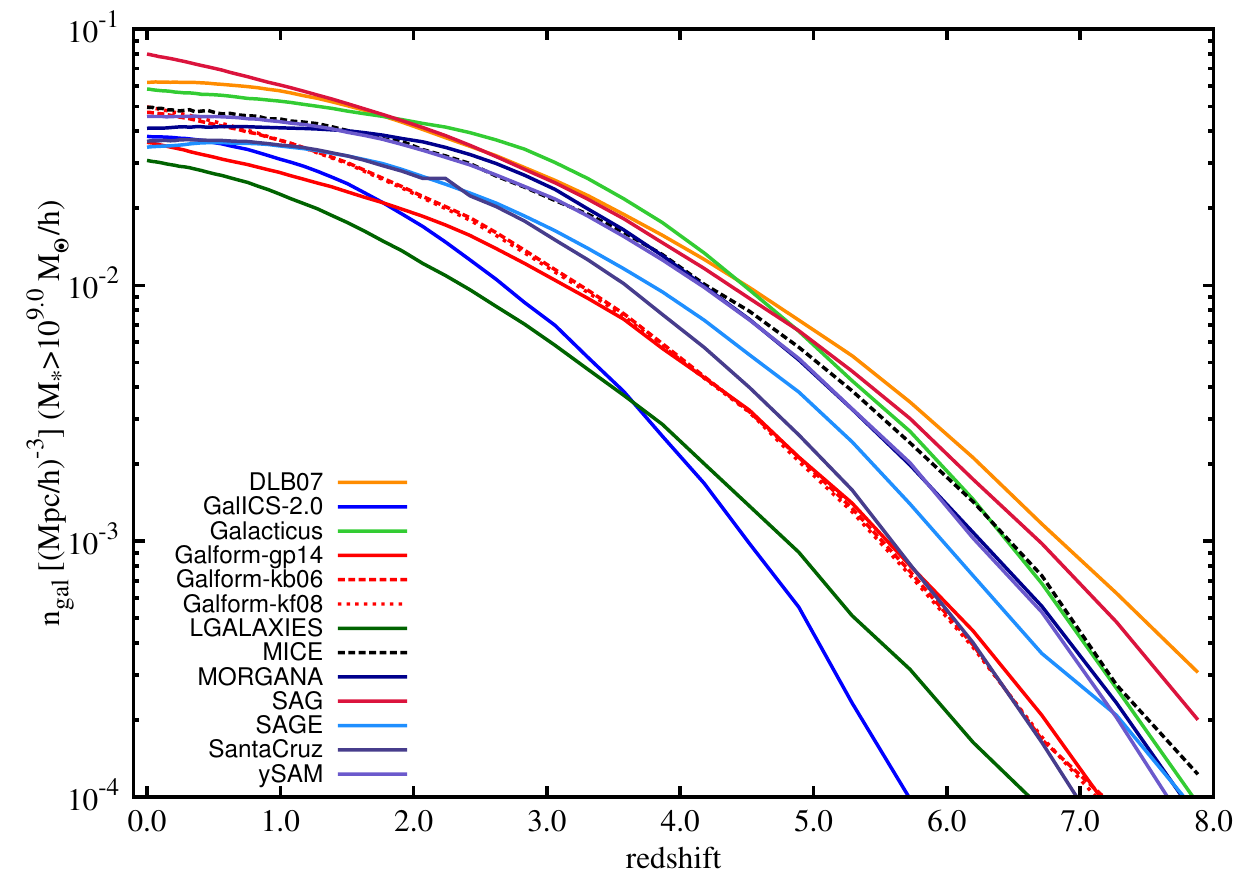}
   \caption{The number  of all galaxies with stellar mass $\Mstar > 10^{9}\hMsun$ as a function of redshift.}
 \label{fig:Ngalzred2}
 \end{figure}

\bsp

\label{lastpage}

\end{document}